\documentclass[lineno]{jfm}

\usepackage{subfiles}
\usepackage{graphicx}
\usepackage{newtxtext}
\usepackage{newtxmath}
\usepackage{natbib}
\usepackage{hyperref}
\hypersetup{
    colorlinks = true,
    urlcolor   = red,
    linkcolor=blue,
    citecolor  =blue,
    filecolor=black,
}
\usepackage{amsfonts}
\usepackage{amsmath}
\usepackage{amssymb,color,mathtools}
\usepackage{epstopdf,epsfig,subfloat,subfig,floatrow}
\usepackage{bm}

\newcommand{\RomanNumeralCaps}[1]
\linenumbers

\title{Rarefaction effects in head-on collision of two identical droplets}

\author{Tao Chen\aff{1},
	Lei Wu\aff{1}
	\corresp{\email{wul@sustech.edu.cn}},
	Lian-Ping Wang\aff{2,3,4}
	\and 
	Shiyi Chen\aff{2,3,4}
}

\affiliation{
	\aff{1} Department of Mechanics and Aerospace Engineering, Southern University of Science and Technology, Shenzhen 518055, China
	\aff{2} Southern Marine Science and Engineering Guangdong Laboratory (Guangzhou), 1119 Haibin Road, Nansha District, Guangzhou,
	511458, China
	\aff{3} Guangdong Provincial Key Laboratory of Turbulence Research and Applications, Center for Complex Flows and Soft Matter Research and Department of Mechanics and Aerospace Engineering, Southern University of Science and Technology, Shenzhen 518055, Guangdong, China
	\aff{4} Guangdong-Hong Kong-Macao Joint Laboratory for Data-Driven Fluid Mechanics and Engineering Applications, Southern University of Science and Technology, Shenzhen 518055, China
}
\begin{document}
\maketitle

\begin{abstract}
	The head-on collision of two identical droplets is investigated based on the BGK-Boltzmann equation. Gauss-Hermite quadratures with different degree of precision are used to solve the kinetic equation, so that the continuum (solution truncated at the Navier-Stokes order) and  non-continuum (rarefied gas dynamics) solutions can be compared. When the kinetic equation is solved with adequate accuracy, prominent variations of the vertical velocity (the collision is in the horizontal direction), the viscous stress components, and droplet morphology are observed during the formation of liquid bridge, which demonstrates the importance of the rarefaction effects and the failure of the Navier-Stokes equation.
	The rarefaction effects change the topology of  streamlines near the droplet surface, suppress the high-magnitude vorticity concentration inside the interdroplet region, and promote the vorticity diffusion around outer droplet surface. 
	Two physical mechanisms responsible for the local energy conversion between the free and kinetic energies are identified, namely, the total pressure-dilatation coupling effect and the interaction between the density gradient and strain rate tensor.
	An energy conversion analysis is performed to show that the rarefaction effects can enhance the conversion from free energy to kinetic energy and facilitate the discharge of interdroplet gas film along the vertical direction, thereby boosting droplet coalescence. 
\end{abstract}
%\begin{keywords}
%\end{keywords}

\section{Introduction}

Studying the droplet collision dynamics is of fundamental importance in understanding a variety of natural phenomena and engineering problems, such as atmospheric raindrop formation, ink-jet printing, dense spray system, and so on. Undoubtedly, complex interactions among a large number of droplets impose great difficulties in quantitative measurements and analysis. Therefore, the binary droplet collision problem is usually used as a canonical case for studying the physics of droplet collisions. 

%The physical dimensionless parameters for this problem are conventionally introduced as follows~\citep{Ashgriz1990,QianLaw1997,HZ2020PRF}.
%The collision Weber number  $We\equiv\rho_{l}D_{l}U_r^2/\sigma_s$ (where $\rho_l$ is the density of the liquid, $D_l$ is the droplet diameter, $U_r$ is the relative collision speed, and $\sigma_s$ is the surface tension coefficient) measures the relative importance of the droplet inertia to the surface tension. The impact parameter $B\equiv\chi/D_l$ ranges from 0 to 1 with $\chi$ being the projection of two mass centers connection line in the direction perpendicular to the droplet velocity. In particular, $B=0$ represents the head-on collision, while $B=1$ represents the grazing collision.
%In addition, the collision outcome is also influenced by the Ohnesorge number $Oh\equiv\mu_{l}/\sqrt{\rho_{l}D_{l}\sigma_{s}}$ ($\mu_l$ is the dynamic viscosity of the liquid) or the Reynolds number $Re\equiv\rho_lD_lU_r/\mu_l=\sqrt{We}/Oh$, the droplet diameter ratio $\Delta\equiv D_{l}^{(1)}/D_{l}^{(2)}$, the density ratio $r_{\rho}\equiv\rho_l/\rho_g$, and the kinematic viscosity ratio $r_\nu\equiv\nu_l/\nu_g$. 

The droplet collision is affected by many dimensionless parameters.
Previous studies are devoted to identifying and interpreting the outcomes of binary collision in the continuum regime where the Navier-Stokes equation is valid, such as the coalescence, separation, bouncing, and shattering.
Different collision outcomes are well summarized in a $We$-$B$ regime diagram ($We$ is the Weber number and $B$ is the impact parameter), where some phenomenological criteria are also proposed to predict their transition boundaries for both the water and hydrocarbon droplets~\citep{Ashgriz1990,JiangYJ1992,QianLaw1997,Orme1997,PanKL2008,PanKL2009}. 
In this paper, we are interested in the role of gas cushion in the droplet collision, since it has been observed that increasing (decreasing) the ambient gas pressure promotes droplet bouncing (coalescence)~\citep{JiangYJ1992,QianLaw1997,Zhangzhenyu2016}. 
%droplet coalescence could also be facilitated when the fuel sprays contain the fuel vapour~\citep{QianLaw1997}. 

%However, the present paper focuses on the rarefaction effects beyond the Navier-Stokes order during the head-on collision of two identical droplets.

%Effects of the Reynolds number on the binary droplet collision outcomes and the transition boundaries between different regimes were also investigated in addition to the droplet collision dynamics~\citep{Willis2003,DaiM2005,Gotaas2007,Sommerfeld2016,Finotello2017}. 
%Recently, from the vortex-dynamical perspective, a helicity analysis identified a strong interaction between the ring-shaped vortices in the droplet interior and the line-shaped shear layer in the droplet interaction region during the stretching separation~\citep{HXZ2019,HXZ2020}. 
%Recent studies also showed that the spinning motions of droplets could induce nonaxisymmetric flow features and a delayed separation after temporary coalescence~\citep{PanKL2019,HZ2020PRF}. For the head-on collision between two nonspinning droplets, the internal flow in the interdroplet gas film is approximately parallel to the collision plane; for the head-on collision between a spinning droplet and a nonspinning droplet, the interdroplet gas film is curved with its thickness varying prominently along the lubrication layer.

When the thickness of gas film between the colliding droplets is small, the rarefied (or non-continuum) gas dynamics beyond the description of Navier-Stokes equations should be considered. However, they are relatively less investigated in
problems of binary droplet collision, probably due to the complexity in the Boltzmann-type kinetic equation. Therefore, several simplified methods are used to account for the rarefaction effects.
\citet{ZhangWW1999} first studied the van der Waals rupture of a thin film on a solid substrate using the disjoint pressure model which was later adopted by several researchers to explore the binary droplet collision dynamics~\citep{JiangX2007,Yoon2007}.
Starting from the linearised Boltzmann equation, \citet{ZhangPeng2011} derived a modified lubrication force expression to account for the rarefaction effects inside the interdroplet gas film. It was found that the lubrication pressure was only modified by a correction factor for the shear viscosity of the gas.
\citet{LiJiePRL2016} made two corrections to the axisymmetric Navier-Stokes equations to account for the rarefied nature of the interdroplet gas film. On one hand, the disjoint pressure model of~\cite{ZhangWW1999} was used to incorporate the intermolecular van de Waals forces in a sharp-interface description. On the other hand, the modified shear viscosity model of~\cite{ZhangPeng2011} was adopted to describe the rarefaction effects. 
It is reported that the disjoint pressure model provides a leading-order approximation for relatively flat interfaces, but fails to capture the flow physics involved during the final stage of gas film rupture and therefore is not robust in predicting droplet coalescence~\citep{Yoon2007,ChenXiaodong2020}.
For low-Reynolds-number collision of two rigid spheres moving in an ideal isothermal gas, it is found that either the non-continuum or compressible effect allows the particles to contact, while a continuum incompressible lubrication would prevent particle contact~\citep{Gopinath1997}. This indicates that both the compressible and non-continuum nature of the gas should be included in an accurate macroscopic description of the collision process. There also exist some studies relevant to the rarefaction effects in the gas film, which includes a high-speed liquid splash on a dry solid surface~\citep{Mandre2012}, air entrainment in dynamic wetting~\citep{Sprittles2015,Sprittles2017}, and the creation of anti-bubbles from air films~\citep{Beilharz2015}.

The aforementioned studies indicate that when the characteristic film thickness is comparable to the mean free path of gas molecules, the no-slip boundary condition breaks down and hence the rarefaction effects should be considered in the physical modelling.
The slip effect decreases the lubrication resistance in the gap flow, which facilitates the drainage of the gas and the rupture of the film, therefore boosting the liquid-solid or liquid-liquid contact.
In addition, by reducing the ambient gas pressure,  the mean free path of the gas increases, which implies the enhancement of the rarefaction effect within the gas film.

As the characteristic Knudsen number (i.e., the ratio of the mean free path to the characteristic length scale) increases, the Navier-Stokes equation is no longer adequate to capture the fundamental physics: not only the slip velocity condition arises, but also the constitutive relation breaks down. Although the modified macroscopic equations can deal with certain problems, there is no self-consistent way macroscopically to include higher-order non-continuum effects in an a priori manner~\citep{ShanXW2006}.
On the other hand, although the 13-moment and 26-moment equations can go beyond the Navier-Stokes equations and capture the flow physics up to a certain Knudsen-number order~\citep{Rana2021,Struchtrup2022}, their derivation and numerical simulation are rather complicated, and their validity region is limited.
For such scenarios, the Boltzmann-type kinetic equation is more fundamental to include the higher-order non-continuum effects beyond the Navier-Stokes level, which provides an unified kinetic description for different flow regimes.
To the authors' knowledge, rarefaction effects in head-on collision of two droplets are not yet investigated by solving the kinetic equation, which becomes the main topic of the present study.

The rest of this paper is organized as follows. In Section~\ref{Section2}, we provide a brief introduction to the free-energy-based multiphase flow model. In Section~\ref{Section3}, we briefly review the progresses on kinetic modelling of single-component liquid-vapour flows and then introduce the present thermodynamically consistent mesoscopic model.
In addition, we describe the details of the numerical method and the construction of Gauss-Hermite quadratures. Then, numerical simulations are performed in Section~\ref{Section4}, where we discuss the rarefaction effects on macroscopic quantities, droplet morphology, streamline topology and energy conversion. Finally, conclusions are drawn in Section~\ref{Section5}.

\section{Free-energy-based multiphase flow model}\label{Section2}

The quasi-local thermodynamics of a two-phase flow system at equilibrium can be described by the second-gradient theory through the density and its gradient~\citep{CahnHilliard1958,Rowlinson1982,Anderson1998,YuePengtao2004}.
The corresponding free energy functional is given by
\begin{eqnarray}
	\mathcal{F}(\rho,\bm{\nabla}\rho)
	=\int_{V}\left(\psi(\rho)+\frac{1}{2}\kappa\lVert\bm{\nabla}\rho\rVert^2\right)dV,
\end{eqnarray}
where $\rho$ is the density, $\kappa$ is the interfacial free energy coefficient associated with the surface tension coefficient $\sigma_s$, and $V$ is the integral domain occupied by the fluids. The first term represents the bulk free energy density and the second term is the interfacial free energy density caused by the non-local molecular interactions.

In the present paper, the double-well formulation of the bulk free energy density is employed~\citep{Jacqmin1999,Jamet2000NED}:
\begin{eqnarray}\label{bulk_free_energy_density}
	\psi(\rho)=\beta(\rho-\rho_l)^2(\rho-\rho_g)^2,
\end{eqnarray}
where $\rho_l$ and $\rho_g$ represent the densities of liquid and vapour phases at saturation, respectively, and $\beta$ is a positive constant coefficient. Equation \eqref{bulk_free_energy_density} works perfectly in the vicinity of the critical point of the equation of state while may cease to be valid away from the critical point, namely, at a large density ratio or an equivalently low temperature~\citep{LeeTaehun2006,YuePengtao2004}.
Therefore, the present paper only considers the near-critical fluids with small density ratio.

The first-order variation of the free energy functional with respect to the density defines the chemical potential $\mu_\rho$ as follows:
\begin{eqnarray}\label{chemical_potential}
	\mu_\rho\equiv\frac{\delta\mathcal{F}}{\delta \rho}=\frac{\partial\psi}{\partial\rho}-\kappa\nabla^2\rho.
\end{eqnarray}
Moreover, for a given $\psi$, the thermodynamic pressure $p_{0}$ (i.e., the equation of state) is determined by
\begin{eqnarray}\label{pth}
	p_{0}=\rho\frac{\partial\psi}{\partial\rho}-\psi.
\end{eqnarray}
Using (\ref{bulk_free_energy_density}), the bulk chemical potential is evaluated as
$\mu_0\equiv{\partial\psi}/{\partial\rho}=4\beta(\rho-\rho_l)(\rho-\rho_g)[\rho-(\rho_l+\rho_g)/2]$ and the corresponding thermodynamic pressure is  $p_0=\beta(\rho-\rho_l)(\rho-\rho_g)[3\rho^2-\rho_{l}\rho_{g}-\rho(\rho_{l}+\rho_{g})]$.

For a flat surface at equilibrium, the density profile across the interface can be obtained by solving the energy-minimized equation $\mu_\rho=0$, which results in 
\begin{eqnarray}\label{density1D}
	\rho(\zeta)=\frac{\rho_l+\rho_g}{2}+\frac{\rho_l-\rho_g}{2}\tanh\left(\frac{2\zeta}{W}\right),
\end{eqnarray}
where $\zeta$ is the surface normal coordinate measured from the point satisfying $\rho=(\rho_{l}+\rho_{g})/2$  and $W=(\rho_l-\rho_g)^{-1}\sqrt{8\kappa/\beta}$ is a length parameter with the same order of magnitude as the interfacial thickness. 
The surface tension coefficient $\sigma_s$ is equal to the integral of free energy density through the interface per unit area~\citep{Jacqmin1999,Jacqmin2000}.
From (\ref{density1D}), it is explicitly expressed as
\begin{eqnarray}
 \sigma_s=\kappa\int_{-\infty}^{+\infty}\left(\frac{d\rho}{d\zeta}\right)^2d\zeta=\frac{1}{6}(\rho_l-\rho_g)^3\sqrt{2\kappa\beta}.
\end{eqnarray}
Conversely, a simple calculation will give
\begin{eqnarray}\label{betakappa}
	\beta=\frac{12\sigma_s}{W(\rho_l-\rho_g)^4},
	\quad
	\kappa=\frac{3\sigma_s W}{2(\rho_l-\rho_g)^2},
\end{eqnarray}
which can be used to determine $\beta$ and $\kappa$ when $\sigma_s$, $W$ and saturation densities are given.

\section{Mesoscopic model and numerical method}\label{Section3}

A brief summary of the progresses on kinetic modelling of single-component liquid-vapour flows is given in order to provide a research background for the present study. 
An approximate kinetic consideration leads to the Enskog-Vlasov equation, which can be viewed as an extension of the Boltzmann equation for dilute gases to single-component multiphase flows with interfacial dynamics. The Enskog-Vlasov equation handles the short-range repulsive interaction among molecules using the Enskog kinetic theory for hard-sphere dense fluids~\citep{Chapman1970} and incorporates the long-range attractive interaction by the mean-field approximation~\citep{Vlasov1978}. Together with its modified versions, they have been applied to investigate
evaporation of a liquid slab into near vacuum~\citep{Frezzotti2005},
net evaporation and condensation~\citep{Kon2014},
fluid-solid and vapour-solid interactions~\citep{Gibelli2015}, thermodynamics of noble gases~\citep{Benilov2018,Benilov2019}, evaporation of multicomponent substances into vapour and
vacuum~\citep{Kobayashi2017,Frezzotti2018,Busuioc2020}.
Similar kinetic models were also proposed to deal with surface-confined strongly inhomogeneous flows in nanoscale shale gas transport~\citep{GZS2005}.

%Kinetic modelling of the liquid-vapour interface and relevant applications of the Enskog-Vlasov equation were comprehensively reviewed by~\citet{FrezzottiBarbante2017}. It is also noted that evaluating the mean-field force can be performed by solving a screened-Poisson equation~\citep{Sadr2019}.

However, the collision and mean-field force terms of the original Enskog-Vlasov equation are too complex for practical applications.
In order to save the computational costs, simplifying the Enskog collision operator was also performed by applying the first-order Taylor-series expansion in terms of the molecular diameter~\citep{Chapman1970,KremerRosa1988}, the Bhatnagar-Gross-Krook (BGK) or Shakhov approximation~\citep{Luo2000,HeDoolen2002,WangPeng2020,HuangRongzong2021}, as well as the Fokker-Planck approximation~\citep{Sadr2017}. In addition, a simplified mean-field force expression was also derived using the Taylor-series expansion (namely, the smooth density approximation)~\citep{HeDoolen2002}.

Although the (simplified) Enskog-Vlasov equation has inherent kinetic advantages for accurate interfacial description, the macroscopic or simplified mesoscopic models (methods) are preferred by taking into account both the model complexity and computational costs. So far, the continuum methods are extensively used. 
Numerous computational fluid dynamics (CFD) methods have been proposed to solve the Navier-Stokes equation, which include the front-tracking method, the level set method, and the volume-of-fluid  method~\citep{Scardovelli1999,Sethian2003}. The front-tracking method is usually not able to simulate interface coalescence and break-up phenomena. For the level set and volume-of-fluid methods, the interface reconstruction or reinitialization may introduce some non-physical numerical artifacts.
Over the past few decades, the lattice Boltzmann method (LBM) has been developed as an efficient mesoscopic CFD method, which allows a direct mesoscopic modelling of many complex physical problems covering both the single-phase and multiphase flows~\citep{ChenDoolen1998,ShanXW2006}.
Prevalent multiphase LBM models include the colour-gradient model~\citep{Gunstensen1991}, Shan-Chen model~\citep{ShanXW1993, ShanXW1994,LiQing2020}, free-energy model~\citep{Swift1995,Swift1996}, and phase field model~\citep{HCZ1999,YuePengtao2004,ZhangChunhua2018}, which have been successfully applied to simulate a vast majority of multiphase flows at the continuum regime, with the superior advantage that
the interfacial fluid dynamics can be automatically captured by incorporating a non-ideal equation of state and the intermolecular force during the particle collision and streaming processes.

The aforementioned LBM models with on-lattice discrete particle velocity sets are originally designed for the continuum flows where the rarefaction effects are not considered. As the Knudsen number increases, higher-order non-continuum effects become more and more significant. Recently, part of the rarefaction effects are considered by the 13- and 26-moment equations~\citep{Struchtrup2019,Struchtrup2022}, which are derived from the Enskog-Vlasov equation. However, the derivation and numerical computation are rather complicated. In this paper, we aim to solve the kinetic equation for multiphase flows by a multiscale mesoscopic method which, through the proper discretization of molecular velocity space, can provide solutions either equivalent to the 13- and 26-moment equations or beyond.

\subsection{Mesoscopic model}
This paper focuses on the higher-order non-continuum effects beyond the Navier-Stokes level during the head-on collision of two identical droplets; the simplified kinetic model equation with the Bhatnagar-Gross-Krook (BGK) collision operator~\citep{BGK1954} is adopted:
\begin{eqnarray}\label{Boltzmann}
	\frac{\partial f}{\partial t}+\bm{\xi}\cdot\bm{\nabla}f=\frac{f^{eq}-f}{\tau}+F_{f}\equiv\bar{\Omega}_{f},
\end{eqnarray}
where $f(\bm{x},\bm{\xi},t)$ is the particle distribution function, $\bm{x}$ is the spatial location, $t$ is the time, $\bm{\xi}$ is the $D$-dimensional particle velocity and $\tau$ is the dimensional relaxation time. The kinematic viscosity is $\nu=\mu/\rho=c_s^2\tau$, where $\mu$ is the dynamic viscosity and $c_s=\sqrt{RT}$ is the speed of sound. $R$ is the gas constant and $T$ is the temperature.
The local Maxwellian equilibrium distribution function is given by
\begin{eqnarray}\label{Maxwellian}
	f^{eq}=\frac{\rho}{(2\pi RT)^{D/2}}\exp\left(-\frac{c^2}{2RT}\right),
\end{eqnarray}
where $\bm{c}=\bm{\xi}-\bm{u}$ is the thermal fluctuating velocity.
The forcing term can be modelled as
\begin{eqnarray}\label{F_f}
	F_f=\frac{\rho\bm{b}\bm{\cdot}\bm{c}}{\rho RT}f^{eq},
\end{eqnarray}
where $\rho\bm{b}$ should be determined by the interfacial force. Once the distribution functions are obtained, the density $\rho$, the momentum $\rho\bm{u}$ and the viscous stress tensor $\bm{\sigma}$ are evaluated through their velocity moments:
\begin{eqnarray}
	\rho=\int fd\bm{\xi},
	\quad
	\rho\bm{u}=\int \bm{\xi}fd\bm{\xi},
	\quad
	\bm{\sigma}=-\int\bm{cc}(f-f^{eq})d\bm{\xi}.
\end{eqnarray}

The hydrodynamic limiting equation should be recovered from  the Chapman-Enskog analysis~\citep{Chapman1970} of~\eqref{Boltzmann}.
The zeroth-order moment of~\eqref{Boltzmann} gives the continuity equation,
\begin{eqnarray}\label{hy1}
	\frac{\partial\rho}{\partial t}+\bm{\nabla}\bm{\cdot}\left(\rho\bm{u}\right)=0.
\end{eqnarray}
At the order $O(\tau^0)$, by assuming that $f=f^{eq}+O(\tau)$, the Euler momentum equation is obtained as
\begin{eqnarray}\label{EM}
	\frac{\partial(\rho\bm{u})}{\partial t}+\bm{\nabla}\bm{\cdot}(\rho\bm{uu})=-\bm{\nabla}(\rho RT)+\rho\bm{b}+\bm{O}(\tau).
\end{eqnarray}
The first-order moment of~\eqref{Boltzmann} gives the momentum conservation law,
\begin{eqnarray}\label{hy2}
	\frac{\partial(\rho\bm{u})}{\partial t}+\bm{\nabla}\bm{\cdot}(\rho\bm{uu})=-\bm{\nabla}(\rho RT)+\rho\bm{b}+\bm{\nabla}\bm{\cdot}\bm{\sigma},
\end{eqnarray}
which is not closed due to the presence of the viscous stress tensor $\bm{\sigma}$.
The remaining task is to find the explicit expression for $\bm{\sigma}$ in the continuum limit.
To this end, we perform the first-order Chapman-Enskog expansion up to $O(\tau)$ as
\begin{eqnarray}\label{CEf}
	f=f^{eq}-\tau\left(\frac{\partial f^{eq}}{\partial t}+\bm{\xi}\bm{\cdot}\bm{\nabla}f^{eq}\right)+\tau F_f+O(\tau^2).
\end{eqnarray}
By using~\eqref{Boltzmann} and~\eqref{CEf}, we obtain
\begin{eqnarray}\label{sigmaNS}
\bm{\sigma}=\tau\int\bm{cc}\left(\frac{\partial f^{eq}}{\partial t}+\bm{\xi}\bm{\cdot}\bm{\nabla}f^{eq}-F_f\right)d\bm{\xi}=2\mu\bm{S}+\bm{O}(\tau^2).
\end{eqnarray}
Substituting~\eqref{sigmaNS} into~\eqref{hy2} gives the momentum equation at the Navier-Stokes level:
\begin{eqnarray}\label{hy3}
	\frac{\partial(\rho\bm{u})}{\partial t}+\bm{\nabla}\bm{\cdot}(\rho\bm{uu})=-\bm{\nabla}(\rho RT)+\rho\bm{b}+\bm{\nabla}\bm{\cdot}\bm{\sigma}^{(NS)}+\bm{O}(\tau^2),
\end{eqnarray}
where $\bm{\sigma}^{(NS)}=2\mu\bm{S}$ is the viscous stress tensor with $\bm{S}=\left(\bm{\nabla}\bm{u}+\bm{\nabla}\bm{u}^T\right)/2$ being the strain rate tensor.

For isothermal flows, by selecting a proper density-dependent pair correlation function $\chi(\rho)$, a general formulation for the thermodynamic pressure can be expressed as~\citep{Chapman1970,HCZ1999,HeDoolen2002}
\begin{eqnarray}\label{thermo_pressure}
	p_{0}=\rho RT(1+b\rho\chi)-a\rho^2=\rho RT+b\rho^2RT\chi-a\rho^2,
\end{eqnarray}
where $b=2\pi d^3/3m$, with $d$ being the molecular diameter, $m$ is the molecular mass;
$a$ is a positive coefficient determined by the attractive intermolecular potential.
In order to be physically consistent with the simplified Enskog-Vlasov model introduced in~\citet{Chapman1970} and~\citet{HeDoolen2002}, the sum of the Enskog correction for short-range molecular interaction (namely, $-\bm{\nabla}(b\rho^2RT\chi)$) and the mean field force for the long-range molecular correction (namely, $-\rho\bm{\nabla}V_m=\bm{\nabla}(a\rho^2)+\kappa\rho\bm{\nabla}\nabla^2\rho$) should be used to model the force $\rho\bm{b}$ in~\eqref{F_f}, which gives 
\begin{eqnarray}\label{rhob}
	\rho\bm{b}=-\bm{\nabla}(b\rho^2RT\chi-a\rho^2)+\kappa\rho\bm{\nabla}\nabla^2\rho.
\end{eqnarray}
Using~\eqref{thermo_pressure},~\eqref{rhob} can be equivalently written as
\begin{eqnarray}\label{pressure_from}
\rho\bm{b}=\bm{\nabla}(\rho RT)-\bm{\nabla}p_{0}+\kappa\rho\bm{\nabla}\nabla^2\rho,
\end{eqnarray}
which is just the pressure form of the interfacial force~\citep{HCZ1999,LeeTaehun2006}. 

We note that~\eqref{pressure_from} can be connected to the free-energy-based description using the following two identities. First, a thermodynamic identity can be easily obtained using~\eqref{pth}, namely,
\begin{eqnarray}\label{ll0}
	\rho\bm{\nabla}\mu_0
	=\nabla\left(\rho\frac{\partial\psi}{\partial\rho}\right)-\frac{\partial\psi}{\partial\rho}\bm{\nabla}\rho
	=\bm{\nabla}(p_{0}+\psi)-\bm{\nabla}\psi
	=\bm{\nabla}p_{0}.
\end{eqnarray}
Using~\eqref{chemical_potential},~\eqref{pressure_from} and~\eqref{ll0}, we obtain the potential form of the interfacial force as
\begin{eqnarray}\label{ll1}
	\rho\bm{b}=\bm{\nabla}(\rho RT)-\rho\bm{\nabla}\mu_{\rho},
\end{eqnarray}
where the second term in~\eqref{ll1} is determined by the chemical potential gradient.
Second, by applying the following vector identity
\begin{eqnarray}
	\rho\bm{\nabla}\nabla^2\rho=\bm{\nabla}\cdot\left[\left(\rho\nabla^2\rho+\frac{1}{2}\lVert\bm{\nabla}\rho\rVert^2\right)\bm{I}-\bm{\nabla}\rho\bm{\nabla}\rho\right],
\end{eqnarray}
the divergence form of the interfacial force can derived from~\eqref{pressure_from}, namely,
\begin{eqnarray}\label{ll2}
	\rho\bm{b}=\bm{\nabla}(\rho RT)-\bm{\nabla}\cdot\bm{P},
\end{eqnarray}
with $\bm{P}$ being the Korteweg stress tensor~\citep{Rowlinson1982,Anderson1998}:
\begin{eqnarray}\label{Pressure_tensor}
	\bm{P}=\left(p_{0}-\kappa\rho\nabla^2\rho-\frac{1}{2}\kappa\lVert\bm{\nabla}\rho\rVert^2\right)\bm{I}+\kappa\bm{\nabla}\rho\bm{\nabla}\rho\equiv p\bm{I}+\kappa\bm{\nabla}\rho\bm{\nabla}\rho,
\end{eqnarray}
where $p$ is the nonlocal total pressure, serving as the sum of the thermodynamic pressure and two capillary contributions due to the density gradients. It is worth noting that~\eqref{Pressure_tensor} is formally consistent with the pressure tensor derived from the Enskog-Vlasov equation which combines the Enskog kinetic theory for short-range molecular interaction and the mean-field theory for long-range molecular interaction~\citep{Chapman1970,HCZ1999,HeDoolen2002}. In addition, if the bulk free energy density $\psi(\rho)$ is properly selected, other realistic equations of state can also be recovered including those of van der Waals, Peng-Robinson, Redlich-Kwong and Carnahan-Starling~\citep{CS1969,Rowlinson1982}.

Although these three forms of the interfacial force, i.e., the pressure form~\eqref{pressure_from}, the potential form~\eqref{ll1}, and the divergence form~\eqref{ll2}, are mathematically equivalent, numerical performances of their discrete versions could be different due to different discretization errors. Previous numerical studies~\citep{Jamet2000NED,Wagner2003,LeeTaehun2006,Guo2021POF} have shown that the potential form~\eqref{ll1} can greatly reduce the possible spurious currents, which therefore will be adopted in the present work.

\subsection{Numerical method}

All the following simulations are performed using the recently developed multiscale mesoscopic approach known as the discrete unified gas kinetic scheme (DUGKS)~\citep{Guo2013PRE}. As a finite volume method, DUGKS combines the advantages of the LBM~\citep{ChenDoolen1998} and the unified gas kinetic scheme~\citep{XuKun2010}, which can capture the flow physics at all kinetic regimes.
Instead of using the analytical solution of the Boltzmann equation, the particle distribution function at the cell interface is reconstructed by the numerical solution along the characteristic line, such that the particle transport and collision processes are naturally coupled and accumulated in a numerical time step scale. Therefore, the numerical dissipation is similar to that of LBM. 
The implementation details of the DUGKS approach that solves the kinetic model~\eqref{Boltzmann} are described below.

The computational domain is divided into many control volumes $V_j$ (the $j$-th control volume) with their cell centres denoted by $\bm{x}_{j}$. Integrating~\eqref{Boltzmann} over the control volume $V_j$ from $t_n$ to $t_{n+1}\equiv t_{n}+\Delta{t}$, and using the midpoint rule for the convection term and the trapezoidal rule for the collision operator $\bar{\Omega}_{f}$, we obtain
\begin{eqnarray}\label{DUGKS2}
	\tilde{f}_{j}^{n+1}(\bm{\xi})=\tilde{f}_{j}^{+,n}(\bm{\xi})-\frac{\Delta t}{\lvert V_{j}\rvert}J_{f}^{n+1/2}(\bm{\xi}),
\end{eqnarray}
where
\begin{eqnarray}\label{DUGKS3}
	J_{f}^{n+1/2}(\bm{\xi})=\oint_{\partial V_{j}}(\bm{\xi}\bm{\cdot}\bm{n})f(\bm{x},\bm{\xi},t_{n+1/2})dS
\end{eqnarray}
represents the mesoscopic flux across the cell interface at $t_{n+1/2}$, $\lvert V_{j}\rvert$ and $\partial V_{j}$ denotes the volume and interface of $V_j$. $\bm{n}$ is the outward unit normal vector.
$t_{n+1/2}=t_n+s$ represents the half time step with $s=\Delta{t}/2$. In order to remove the time implicity, two new particle distribution functions are introduced as $\tilde{f}\equiv f-(\Delta{t}/2)\bar{\Omega}_{f}$ and $\tilde{f}^{+}\equiv f+(\Delta{t}/2)\bar{\Omega}_{f}$, respectively. It is noted that $\tilde{f}_{j}^{n+1}(\bm{\xi})$ and $\tilde{f}_{j}^{+,n}(\bm{\xi})$ are the cell averaged values of $\tilde{f}(\bm{x},\bm{\xi},t_{n+1})$ and $\tilde{f}^{+}(\bm{x},\bm{\xi},t_n)$, respectively.   In the numerical simulation, we track the evolution of $\tilde{f}$ instead of the original distribution function $f$. 

Once the distribution function $\tilde{f}$ is obtained, the macroscopic flow variables at the cell centres are updated using
\begin{eqnarray}
\rho=\int\tilde{f}d\bm{\xi},~\rho\bm{u}=\int \bm{\xi}\tilde{f}d\bm{\xi}+\frac{\Delta{t}}{2}\rho\bm{b},~\bm{\sigma}=-\frac{2\tau}{2\tau+\Delta t}\int\bm{cc}(\tilde{f}-{f}^{eq})d\bm{\xi}.
\end{eqnarray}

The key to evaluate the cell interface flux $J_{f}^{n+1/2}(\bm{\xi})$ is to determine the original distribution function $f(\bm{x}_b,\bm{\xi},t_{n+1/2})$. This can be realized by integrating~\eqref{Boltzmann} along the characteristic line for a half time-step interval $s=\Delta{t}/2$ with the ending point $\bm{x}_{b}$ located at the centre of the cell interface. By using the trapezoidal rule for the collision term, we obtain
\begin{eqnarray}
f(\bm{x}_b,\bm{\xi},t_n+s)-f(\bm{x}_b-\bm{\xi}s,\bm{\xi},t_n)=\frac{s}{2}\left[\bar{\Omega}_f(\bm{x}_b,\bm{\xi},t_n+s)+\bar{\Omega}_f(\bm{x}_b-\bm{\xi}s,\bm{\xi},t_n)\right].
\end{eqnarray}
Once again, two new particle distribution functions $\bar{f}\equiv f-(s/2)\bar{\Omega}_f$ and $\bar{f}^{+}\equiv f+(s/2)\bar{\Omega}_f$ are introduced to remove the time implicity, which results in
\begin{eqnarray}\label{DUGKS4}
	\bar{f}(\bm{x}_{b},\bm{\xi},t_{n}+s)=\bar{f}^{+}(\bm{x}_{b}-\bm{\xi}s,\bm{\xi},t_n).
\end{eqnarray}
By applying the first-order Taylor-series expansion to the right hand side of~\eqref{DUGKS4}, we obtain
\begin{eqnarray}
	\bar{f}(\bm{x}_{b},\bm{\xi},t_{n}+s)\approx\bar{f}^{+}(\bm{x}_{b},\bm{\xi},t_n)-\bm{\xi}s\bm{\cdot}\bm{\nabla}\bar{f}^{+}(\bm{x}_{b},\bm{\xi},t_n),
\end{eqnarray}
where $\bar{f}(\bm{x}_{b},\bm{\xi},t_{n}+s)$ and its gradient $\bm{\nabla}\bar{f}^{+}(\bm{x}_{b},\bm{\xi},t_n)$ at the cell interface can be approximated by linear interpolations. In order to transform back to the original particle distribution functions, the conservative variables at the cell interface at the half time step are evaluated as
\begin{eqnarray}\label{half}
\rho=\int\bar{f}d\bm{\xi},
\quad
\rho\bm{u}=\int \bm{\xi}\bar{f}d\bm{\xi}+\frac{s}{2}\rho\bm{b}.
\end{eqnarray}
Therefore, the equilibrium distribution function $f^{eq}(\bm{x}_b,\bm{\xi},t_{n}+s)$ can be computed using the macroscopic variables obtained through~\eqref{half}. Then, the original distribution function at the cell interface at the half time step is updated as
\begin{eqnarray}
f(\bm{x}_b,\bm{\xi},t_{n}+s)
&=&\frac{2\tau}{2\tau+s}\bar{f}(\bm{x}_b,\bm{\xi},t_{n}+s)
+\frac{s}{2\tau+s}f^{eq}(\bm{x}_b,\bm{\xi},t_{n}+s)\nonumber\\
& &+\frac{\tau s}{2\tau+s}F_{f}(\bm{x}_b,\bm{\xi},t_{n}+s).
\end{eqnarray}
Finally, $J_{f}^{n+1/2}(\bm{\xi})$ can be obtained using~\eqref{DUGKS3}, followed by the update of $\tilde{f}$ through~\eqref{DUGKS2}. It is noted that two useful relations are used in the implementation:
\begin{equation}
\begin{aligned}[b]
\tilde{f}^{+}=\frac{4}{3}\bar{f}^{+}-\frac{1}{3}\tilde{f},\\
\bar{f}^{+}=\frac{2\tau-s}{2\tau+\Delta t}\tilde{f}+\frac{3s}{2\tau+\Delta t}f^{eq}+\frac{3\tau s}{2\tau+\Delta t}F_{f}.
\end{aligned}
\end{equation}

The time step $\Delta{t}$ is determined by the Courant-Friedrichs-Lewy (CFL) condition: $\Delta{t}=\alpha\Delta{x}_{min}/(\lVert\bm{u}\rVert_{max}+\lVert\bm{\xi}\rVert_{max})$, where $\alpha$ is the CFL number, $\Delta{x}_{min}$ is the minimal grid spacing, $\lVert\bm{u}\rVert_{max}$ is the maximum macroscopic flow velocity, and $\lVert\bm{\xi}\rVert_{max}$ is the maximum discrete particle velocity. For the convenience of comparison, the time step is set as a constant in the simulations.

\subsection{Construction of Gauss-Hermite quadratures}\label{GH33}
Different from Grad 13-moment method, the discrete particle distribution functions are used as the fundamental evolution variables in the DUGKS instead of the macroscopic variables and their fluxes.
When evaluating the particle velocity moments in the discrete particle velocity space, properly-designed Gauss-Hermite quadrature rules are needed to calculate the integral as accurately as possible. For example, in one-dimensional case, we have to consider the integral of the following type:
\begin{eqnarray}
\int\omega(\xi)p(\xi)d\xi=\sum_{\alpha=1}^{d}W_{\alpha}p(\xi_\alpha),
\end{eqnarray}
where $\{(\xi_\alpha, W_\alpha): \alpha=1, 2,\cdots, d\}$ are the abscissae and weights of a quadrature of degree of precision $q\leq 2d-1$ with  $p(\xi)$ representing a polynomial of an order not exceeding $q$. The dimensionless abscissae $\{\xi_\alpha: \alpha=1, 2,\cdots, d\}$ have been normalized by $\sqrt{RT}$. The weighting function $\omega(\xi)$ is given by
\begin{eqnarray}
\omega(\xi)=\frac{1}{\sqrt{2\pi}}\exp\left(-\frac{\xi^2}{2}\right).
\end{eqnarray}

By choosing the Hermite polynomials~\citep{Grad1949} as the base functions, a sufficient and necessary condition for $\{(\xi_\alpha, W_\alpha): \alpha=1, 2,\cdots, d\}$ to be a quadrature of degree of precision $q$ is given by~\citep{ShanXW2006,ShanXW2016}
\begin{equation}\label{Walpha}
\sum_{i=1}^{d}W_{\alpha}\mathcal{H}^{(n)}(\xi_{\alpha})
 = \left\{
\begin{array}{ll}
1, & n=0 \\[2pt]
0,         & n=1,2,\cdots,q.
\end{array} \right.
\end{equation}
The choice of the abscissae can be made to maximize the algebraic degree of precision for the given number of points.
A special choice leads to the $d$-point Gauss-Hermite quadrature of degree of precision $q=2d-1$, because 
$2d$ constraints are satisfied according to~\eqref{Walpha}.
The abscissae $\xi_{\alpha}$ of the Gauss-Hermite quadrature are specifically chosen as the zero points of the $d$-th order Hermite polynomial $\mathcal{H}^{(d)}(\xi)$ and the corresponding weights are given by
\begin{eqnarray}
W_{\alpha}=\frac{d!}{\left[d\mathcal{H}^{(d-1)}(\xi_{\alpha})\right]^2}, \alpha=1, 2,\cdots, d.
\end{eqnarray}
Several representative one-dimensional Gauss-Hermite quadratures used in this paper are listed in Tables~\ref{table1}, where the abscissae are normalized by $\sqrt{RT}$.
For two or higher dimensions, the Gauss-Hermite quadrature $\{(\bm{\xi}_\alpha, W_\alpha): \alpha=1, 2,\cdots, d\}$ can be constructed by generalizing its one-dimensional version with the production formulae.
\begin{table}[h!]
	\centering
	\begin{tabular}{ccc}
		\hline
		Quadrature& ${\xi}_{\alpha}$   & $W_{\alpha}$ \\
		D1Q3A5  &0   & 0.666666666666667\\	
		&$\pm$1.732050807568877 & 0.166666666666667\\
		D1Q5A9  &0   & 0.533333333333333\\	
		&$\pm$1.355626179974266 & 0.222075922005613\\	
		&$\pm$2.856970013872806&0.011257411327721\\
		D1Q11A21  &0   &0.369408369408369\\	
		&$\pm$0.928868997381064 &0.242240299873970\\	
		&$\pm$1.876035020154846&0.066138746071058\\
		&$\pm$2.865123160643646&0.006720285235537\\
		&$\pm$ 3.936166607129978& 0.000195671930271\\
		&$\pm$5.188001224374871&0.000000812184979\\	
		D1Q15A29  &0   &  0.318259518259518\\		
		&$\pm$0.799129068324548&0.232462293609732\\
		&$\pm$1.606710069028730& 0.089417795399844\\
		&$\pm$2.432436827009758&0.017365774492138\\
		&$\pm$3.289082424398766&0.001567357503550\\
		&$\pm$4.196207711269016& 0.000056421464052\\
		&$\pm$5.190093591304782&0.000000597541960\\
		&$\pm$6.363947888829840&0.000000000858965\\			D1Q19A37  &0   &   0.283773192751521\\	
		&$\pm$ 0.712085044042380 & 0.220941712199144\\	
		&$\pm$ 1.428876676078373 & 0.103603657276144\\
		&$\pm$ 2.155502761316935 & 0.028666691030118\\	
		&$\pm$ 2.898051276515754 & 0.004507235420342\\	
		&$\pm$ 3.664416547450639 & 0.000378502109414\\
		&$\pm$ 4.465872626831032 & 0.000015351145955\\
		&$\pm$ 5.320536377336039 &  0.000000253222003\\	
		&$\pm$ 6.262891156513252 &  0.000000001220371\\
		&$\pm$ 7.382579024030432 &  0.000000000000748\\									
		\hline
	\end{tabular}
	\caption{One-dimensional Gauss-Hermite quadratures. 
		$\xi_{\alpha}$ are the abscissae and $W_{\alpha}$ are the corresponding weights. The abscissae have been normalized by $\sqrt{RT}$.}
	\label{table1} 
\end{table}

For any square integrable function $f({\xi})$, it can be expanded in terms of the Hermite polynomials as~\citep{Grad1949,ShanXW2006}
\begin{eqnarray}\label{fH}
f({\xi})=\omega({\xi})\sum_{n=0}^{+\infty}\frac{1}{n!}a^{(n)}\mathcal{H}^{(n)}(\xi),
\quad
a^{(n)}=\int_{-\infty}^{+\infty} f(\xi)\mathcal{H}^{(n)}(\xi)d\xi.
\end{eqnarray}
From~\eqref{fH} and using the orthogonality of the Hermite polynomials, the $M$-th order truncated Hermite expansion $f^{M}$ preserves all the moments of $f$ up to the $M$-th order, which at least requires a Gauss-Hermite quadrature of degree of precision $q\geq2M$. Therefore,
according to the Chapman-Enskog expansion, a sufficient condition for the truncated Hermite expansion can be obtained for certain Knudsen number orders~\citep{ShanXW2006}.
The $3^{rd}$-order truncated Hermite expansion for the Maxwellian equilibrium ($s=3$) and a Gauss-Hermite quadrature with at least $6^{th}$-order degree of precision ($r\geq6$) are needed to evaluate the viscous stress tensor accurately at the truncated order $O(\tau)$ (namely, the solution truncated at the Navier-Stokes order). 
It is noted that most of the existing isothermal LBM models adopt only second-order Hermite expansion for the equilibrium which does not satisfy this requirement.
Similarly, $4^{th}$-order Hermite expansion ($s=4$) and at least $8^{th}$-order Gauss-Hermite quadrature ($r\geq8$) are needed at the Burnett level $O(\tau^2)$.
For higher-order approximations to the Boltzmann equation beyond the Navier-Stokes level (namely, $O(\tau^n)$ with $n\geq3$), both the higher-order Hermite expansion for the equilibrium distribution function and the Gauss-Hermite quadrature with sufficiently high degree of precision are required for accurate evaluation.

Generally, to simulate rarefied gas flows, an alternative approach is to use the full Maxwellian equilibrium without applying the Hermite expansion. Therefore, Gauss-Hermite quadratures with sufficient degree of precision are needed to approximate the integral of the following type, namely,
\begin{eqnarray}\label{add}
	\int f(\xi)d\xi=\int\omega(\xi)\frac{f(\xi)}{\omega(\xi)}d\xi \cong\sum_{\alpha=1}^{d}W_{\alpha}\frac{f(\xi_\alpha)}{\omega(\xi_{\alpha})}.
\end{eqnarray}
By gradually increasing the number of discrete particle velocities (correspondingly, degree of precision of Gauss-Hermite quadrature), the final convergent solution of the BGK-Boltzmann equation can be obtained until no prominent changes are observed in the simulation results. For convenience, we use D$p$Q$q$A$r$H$s$ and D$p$Q$q$A$r$F to represent a Gauss-Hermite quadrature in $p$ dimensions, with $q$ discrete particle velocities and $r$-th degree of precision, where H$s$ denotes the $s$-th order truncated Hermite expansion for the Maxwellian equilibrium and F denotes the full Maxwellian equilibrium without applying the Hermite expansion.

It is worth mentioning that the regularized 26 (R26) momentum approximation is accurate up to the order of $O(\tau^5)$~\citep{WuLGuXJ2020,Rana2021,Struchtrup2022}, which is basically equivalent to the $7^{th}$-order Hermite expansion for the equilibrium and $14^{th}$-order Gauss-Hermite quadrature (at least 8 points should be used in one direction). Therefore, the accuracy of the Gauss-Hermite quadrature used in this paper is far beyond that of R26.

\section{Numerical simulation and analysis}\label{Section4}

\subsection{Problem description}
\begin{figure}[t]
	\centering
	\includegraphics[width=0.4\columnwidth,trim={8cm 3.0cm 8cm 3.5cm},clip]{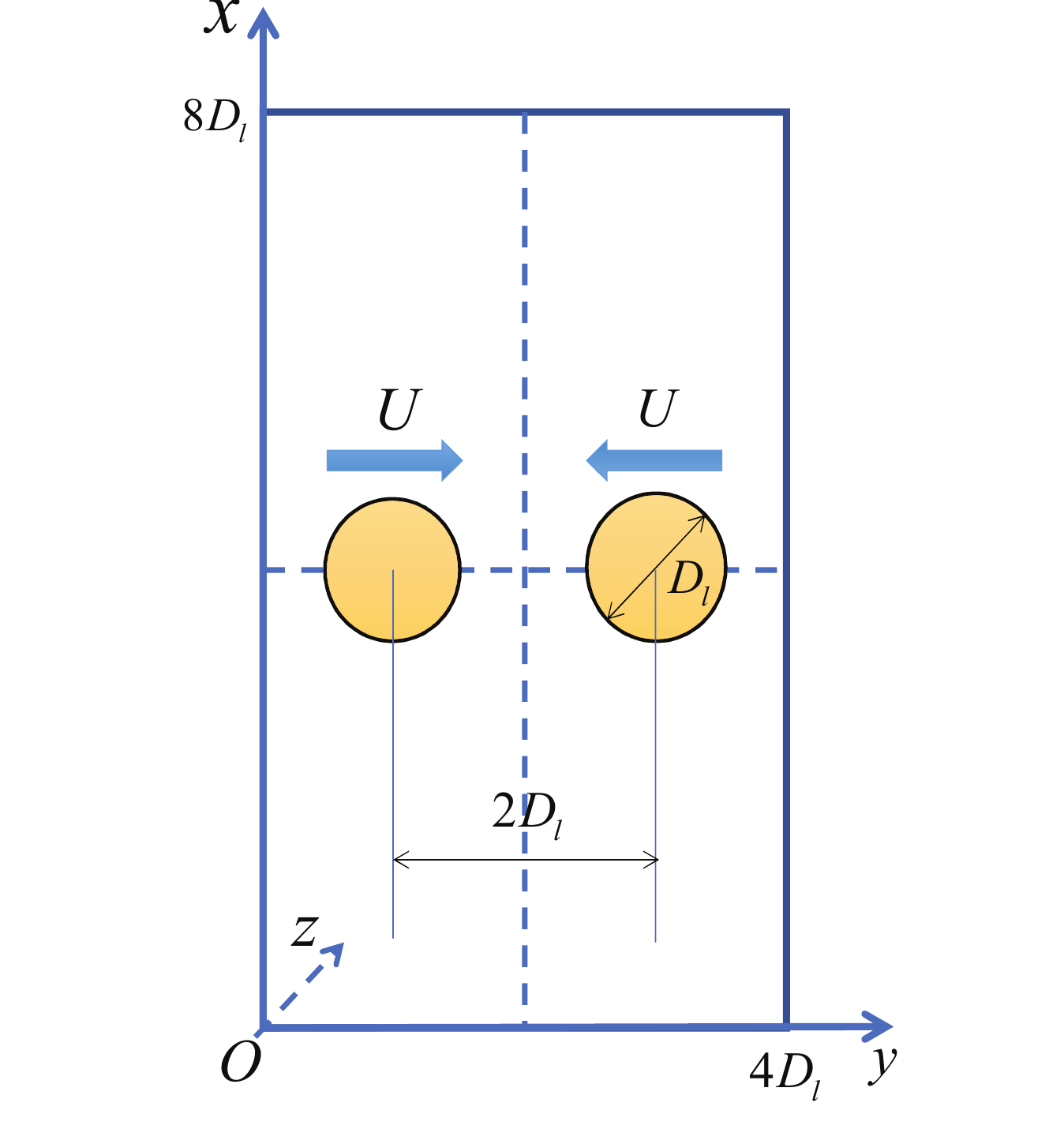}
	\caption{Geometrical configuration of head-on collision of two identical droplets. $D_l$ is the droplet diameter and $U$ is the initial droplet speed.} 
	\label{geometry}
\end{figure}

We perform two DUGKS simulations of head-on collision of two identical droplets by solving the BGK-Boltzmann equation. Compared to the Navier-Stokes-based simulations using the volume-of-fluid method~\citep{HXZ2019,HZ2020PRF,ChenXiaodong2020} or the front-tracking method~\citep{Nobari1996}, the interfacial fluid dynamics can be more physically described by the present kinetic model. Periodic boundary conditions are applied to the four boundaries with $x$ and $y$ being the vertical and horizontal directions, respectively. The dimensionless coordinates are normalized by $D_l/2$ so that $x^{*}\in[0,16]$ and $y^*\in[0,8]$, where $D_{l}$ is the droplet diameter.
As shown in figure~\ref{geometry}, two equal-sized droplets, whose centres are initially placed at ($x^*$,$y^*$)=(8,2) and ($x^*$,$y^*$)=(8,6), are moving with the speed $U$ towards the opposite directions along the $y$-axis (horizontal direction), respectively.
The uniform meshes $N_{x}\times N_{y}=800\times400$  are employed and $100$ grid points are used per drop diameter. 
The grid independence study shows that the resolution is sufficient for the simulated cases. Due to the restriction of the computational costs, only two-dimensional cases are considered in this paper. 

As a sensible first step to explore the rarefaction effects with a BGK-Boltzmann equation, the constant relaxation time is used in this research, which implies that
the kinematic viscosity ratio  $r_{\nu}\equiv\nu_l/\nu_g$ is always equal to the unity.
Obviously, the density ratio $r_{\rho}\equiv\rho_{l}/\rho_{g}$ is always equal to the shear viscosity ratio $r_\mu\equiv\mu_l/\mu_g$, because $\nu=\mu/(\rho RT)$ holds for the present model.
Since the simple double-well bulk free-energy density in~\eqref{bulk_free_energy_density} works as a good approximation only in the vicinity of the critical point of the equation of state~\citep{Jamet2000NED,LeeTaehun2006}, the range of the density ratio that can be handled by this approximation should not be arbitrarily exaggerated for different kinetic flow regimes. On the one hand, we have noticed that for continuum flows, a formulation for the order parameter similar to~\eqref{bulk_free_energy_density} can be applied to successfully simulate large-density-ratio ($r_\rho\geq1000$) immiscible two-phase flows by coupling with the Cahn-Hilliard (CH) or Allen-Cahn (AC) equation and by using some stability-enhanced numerical techniques~\citep{LiangHong2018,Kumar2019}, which drastically extends its theoretical limit. On the other hand, no results for its applicable density ratio are reported for non-continuum flows. Therefore, the small density ratio $r_{\rho}=2$ is considered in the present research without violating the theoretical limitation. Except for the viscosity and density ratios, the problem is also controlled by the conventional Weber number $We$ and the Reynolds number $Re$:
\begin{eqnarray}
We=\frac{4\rho_lD_lU^2}{\sigma_s},
\quad
Re=\frac{2\rho_l D_l U}{\mu_l},
\end{eqnarray}
as well as the Knudsen number $Kn$
\begin{eqnarray}\label{KnudsenNumber}
Kn=\frac{\sqrt{RT}\tau}{D_l}.
\end{eqnarray}
Usually, the continuum flow regime lies in the range $Kn\lesssim0.001$. The non-continuum effects gradually 
become prominent as the Knudsen number increases. It is noted that $D_l$ in the denominator of \eqref{KnudsenNumber} represents the maximum thickness of the interdroplet gas film when the droplets come to contact.

The interfacial thickness $W$ of each droplet occupies $4$ lattices, which satisfies the empirical condition $(W>3\Delta x)$ for numerically sustainable interface thickness~\citep{LeeTaehun2006}.
It is mentioned that the following simulations are performed without applying extra numerical techniques such as the van Leer limiter or the weighted essentially non-oscillatory (WENO) scheme~\citep{vanleer1977,JS1996}. All the spatial derivatives are evaluated using the second-order finite difference schemes at both the cell centres and the cell interfaces, which is consistent with the second-order spatial accuracy of the DUGKS. Therefore, with sufficient gird resolution, the artificial finite resolution effects inherent in these numerical schemes are believed to be constrained to such an extent that the critical physical information is not obviously contaminated.

Combining these careful numerical considerations, we believe that the present two-dimensional study could provide some new physical insights on higher-order non-continuum effects for head-on binary droplet collision problem.
\begin{figure}[h!]
	\centering
	\subfloat[$t^*=0.3$]{
		\begin{minipage}[t]{0.48\linewidth}
			\centering
			\includegraphics[width=1.0\columnwidth,trim={0.0cm 0.18cm 0.0cm 0.2cm},clip]{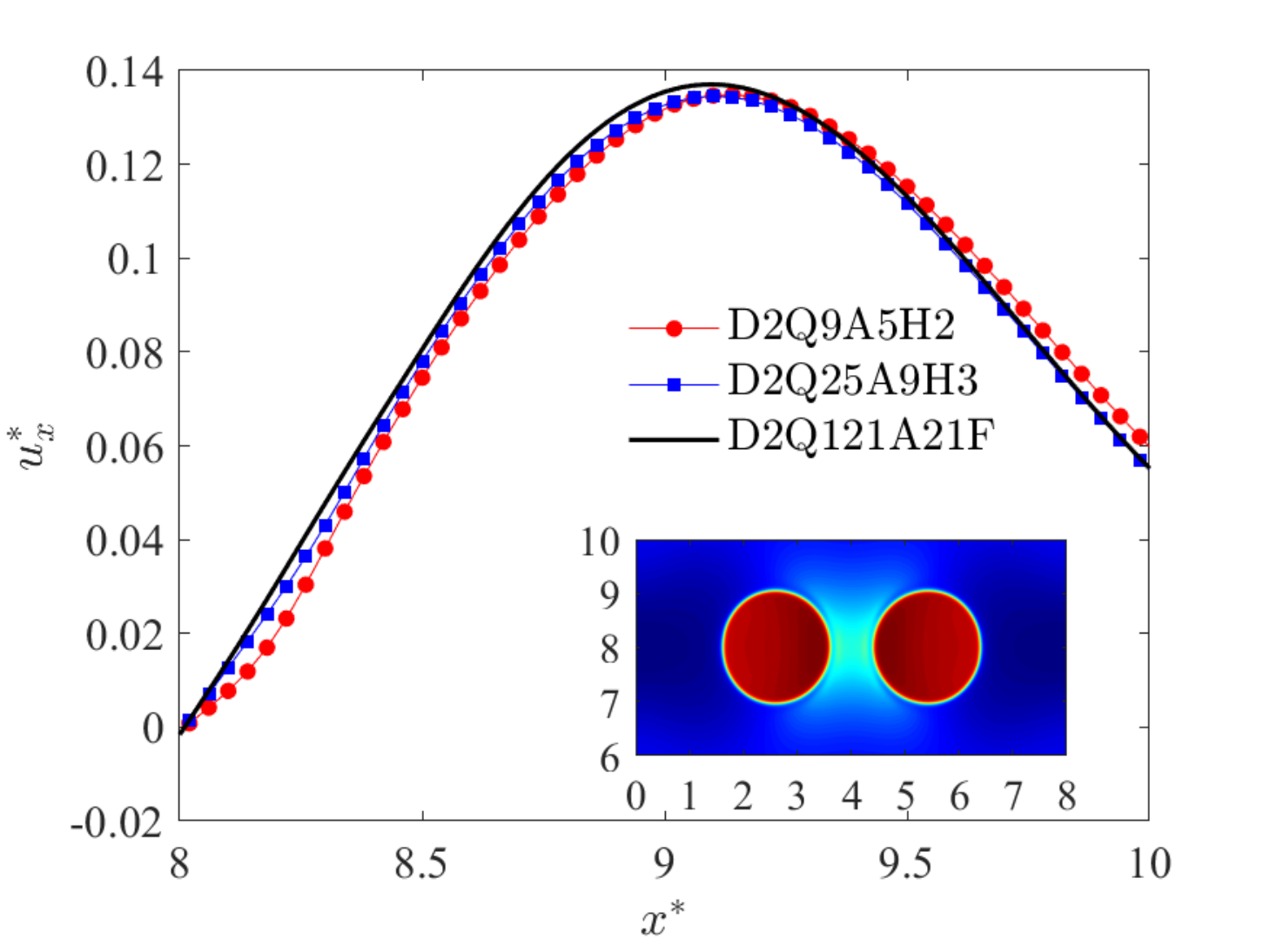}
			\label{ux_evol_1500}
		\end{minipage}%
	}%
	\subfloat[$t^*=0.4$]{
		\begin{minipage}[t]{0.48\linewidth}
			\centering
			\includegraphics[width=1.0\columnwidth,trim={0.0cm 0.18cm 0.0cm 0.2cm},clip]{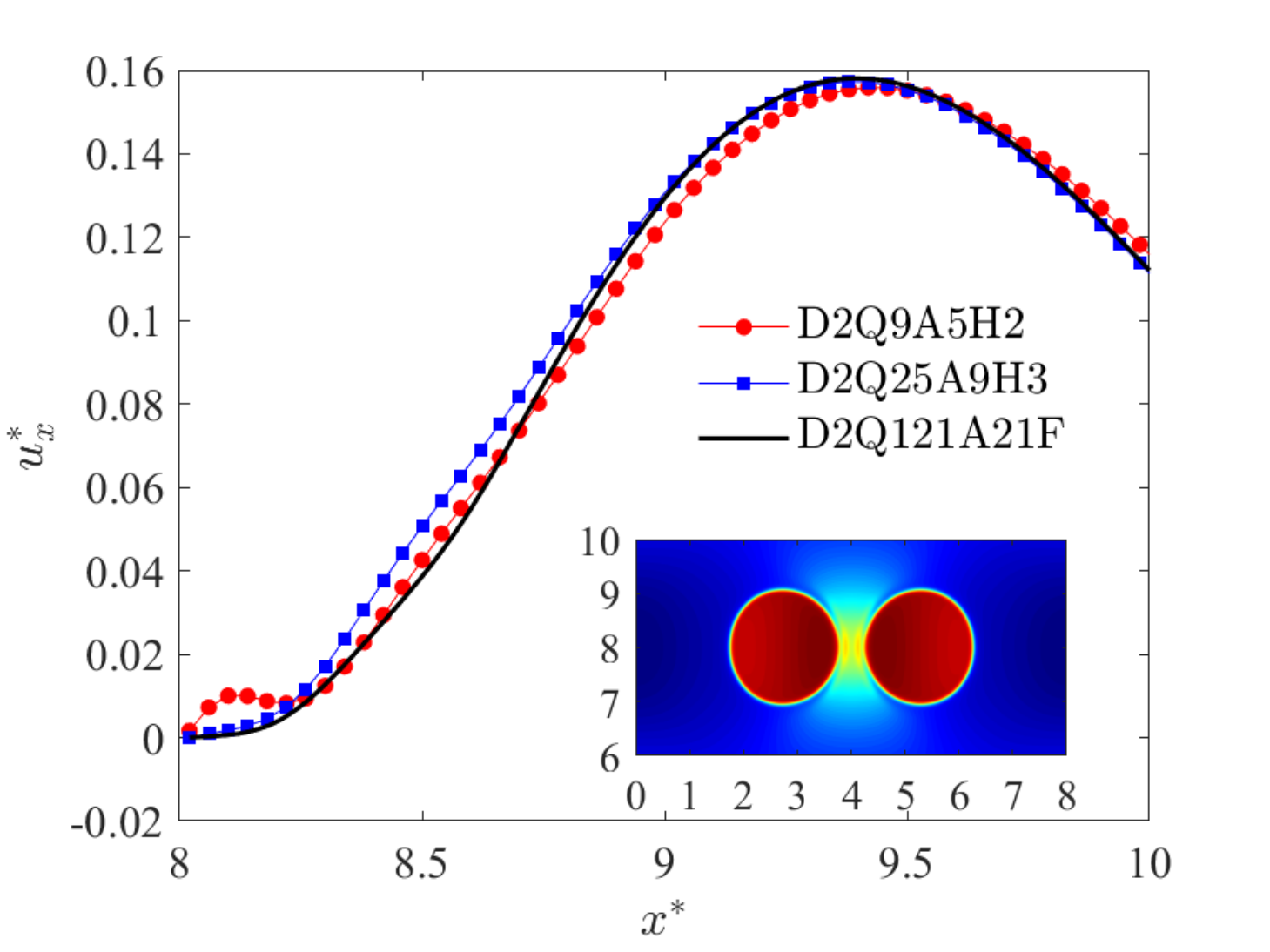}
			\label{ux_evol_2000}
		\end{minipage}%
	}%
	
	\subfloat[$t^*=0.5$]{
		\begin{minipage}[t]{0.48\linewidth}
			\centering
			\includegraphics[width=1.0\columnwidth,trim={0.0cm 0.18cm 0.0cm 0.2cm},clip]{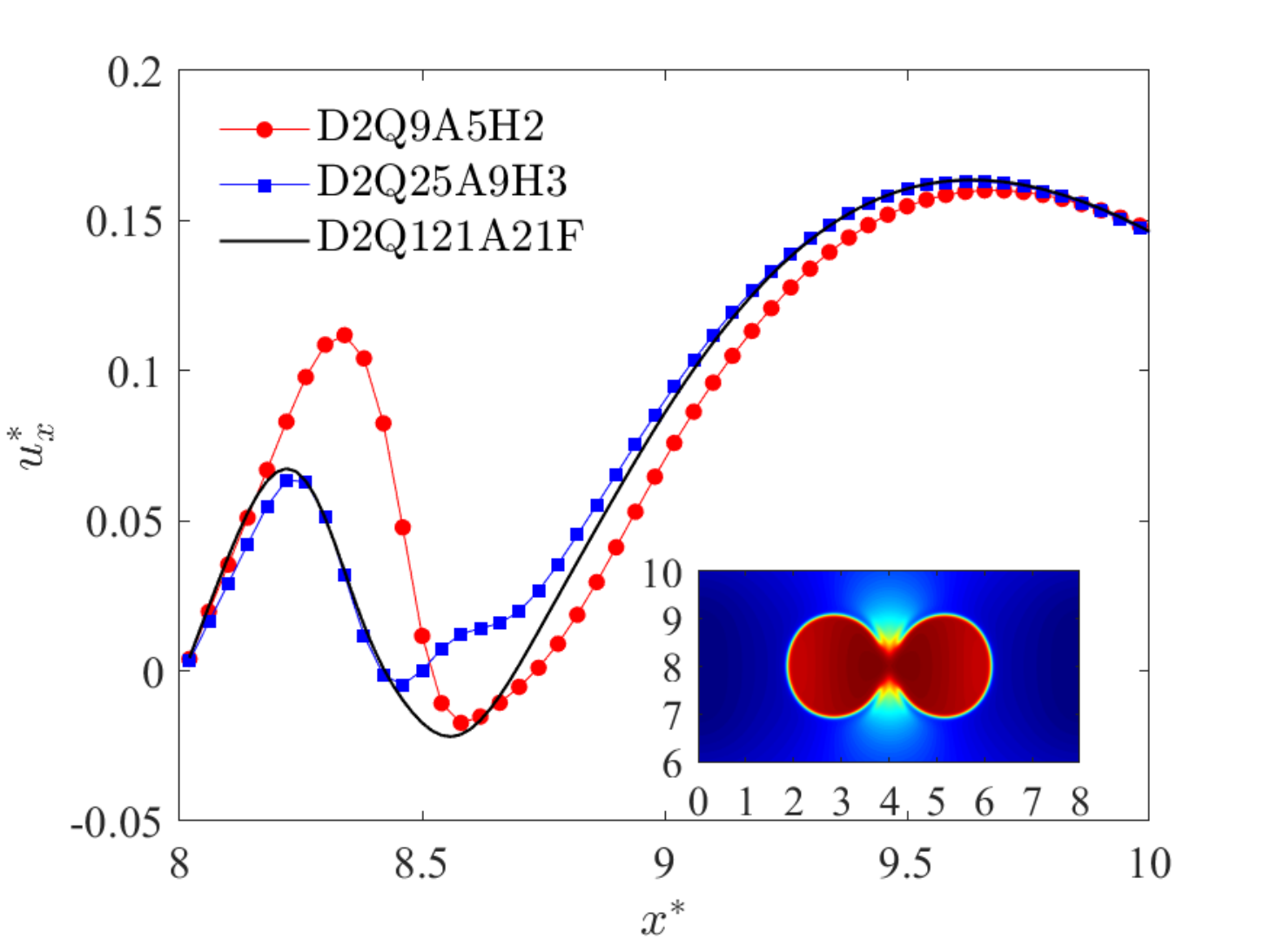}
			\label{ux_evol_2500}
		\end{minipage}%
	}%
	\subfloat[$t^*=0.6$]{
		\begin{minipage}[t]{0.48\linewidth}
			\centering
			\includegraphics[width=1.0\columnwidth,trim={0.0cm 0.18cm 0.0cm 0.2cm},clip]{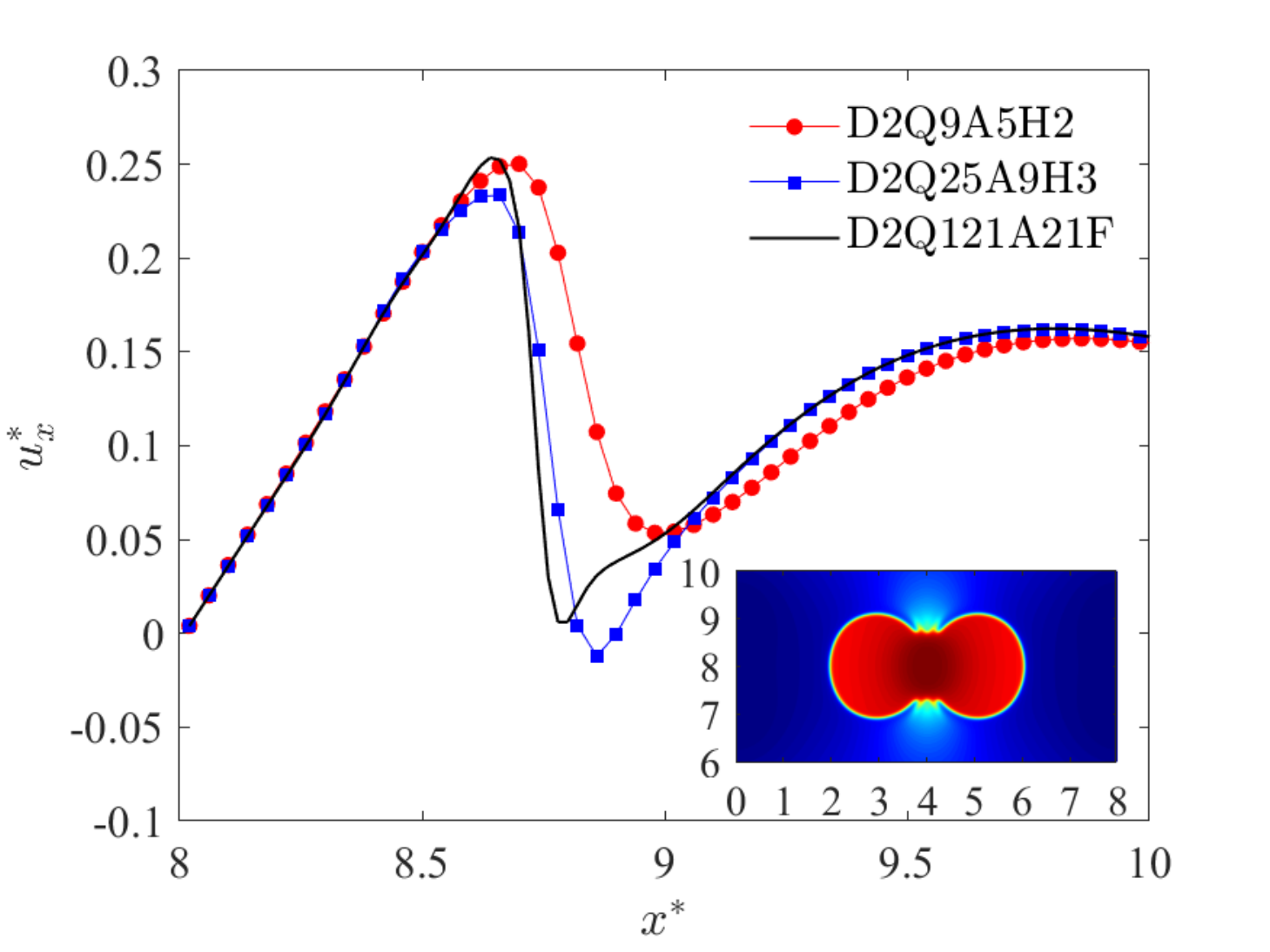}
			\label{ux_evol_3000}
		\end{minipage}%
	}%
	\caption{Evolution of the normalized velocity component $u_{x}^{*}$ in the $x$-direction. (a) $t^*=0.3$, (b) $t^*=0.4$, (c) $t^*=0.5$ and (d) $t^*=0.6$. $We=1600$, $Re=13.33$ and $Kn=0.052$.} 
	\label{case1_ux}
\end{figure}
\begin{figure}[h!]
	\centering
	\subfloat[$t^*=0.3$]{
		\begin{minipage}[t]{0.48\linewidth}
			\centering
			\includegraphics[width=1.0\columnwidth,trim={0.0cm 0.18cm 0.0cm 0.2cm},clip]{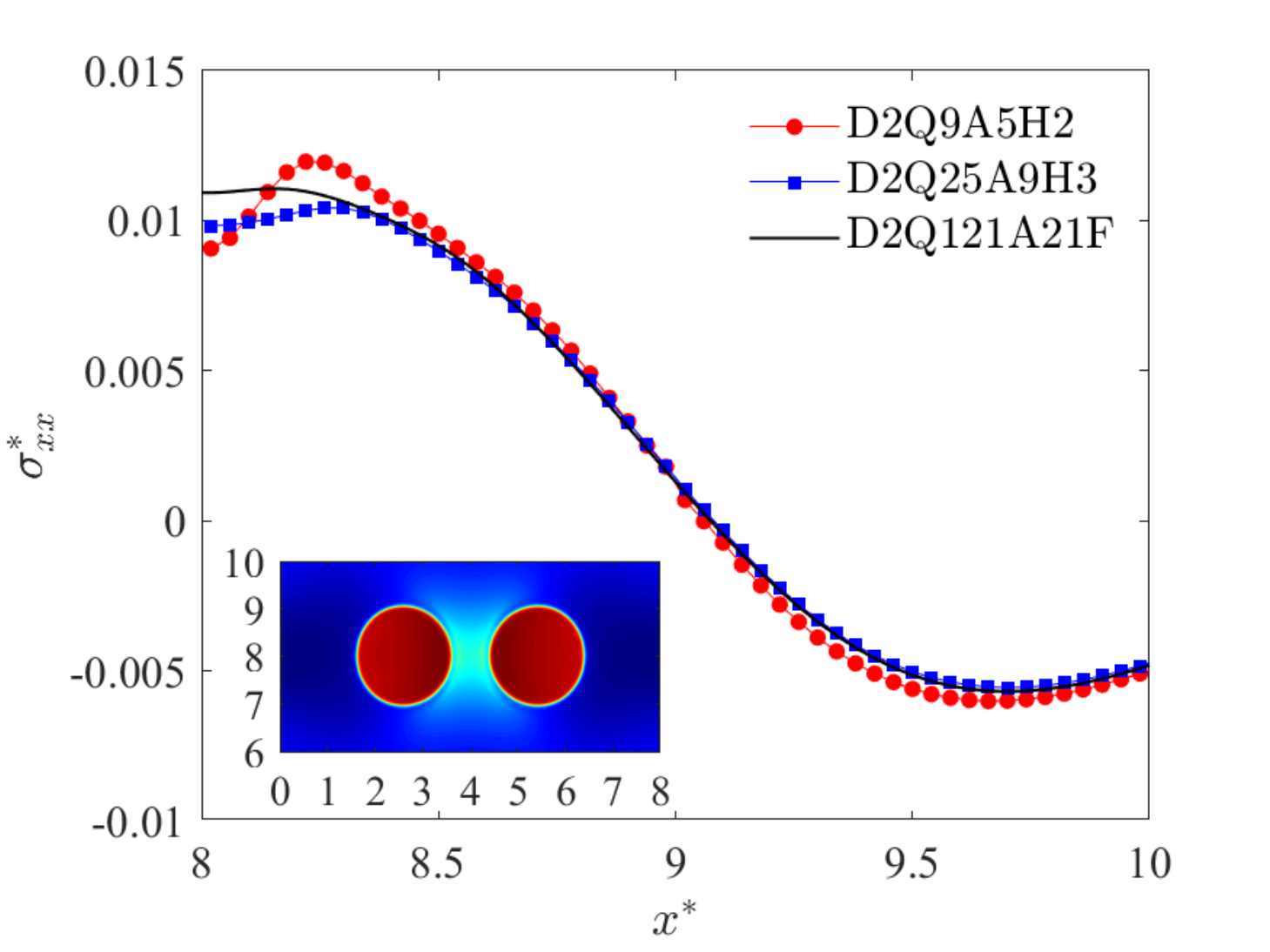}
			\label{sigmaxx_evol_1500}
		\end{minipage}%
	}%
	\subfloat[$t^*=0.4$]{
		\begin{minipage}[t]{0.48\linewidth}
			\centering
			\includegraphics[width=1.0\columnwidth,trim={0.0cm 0.18cm 0.0cm 0.2cm},clip]{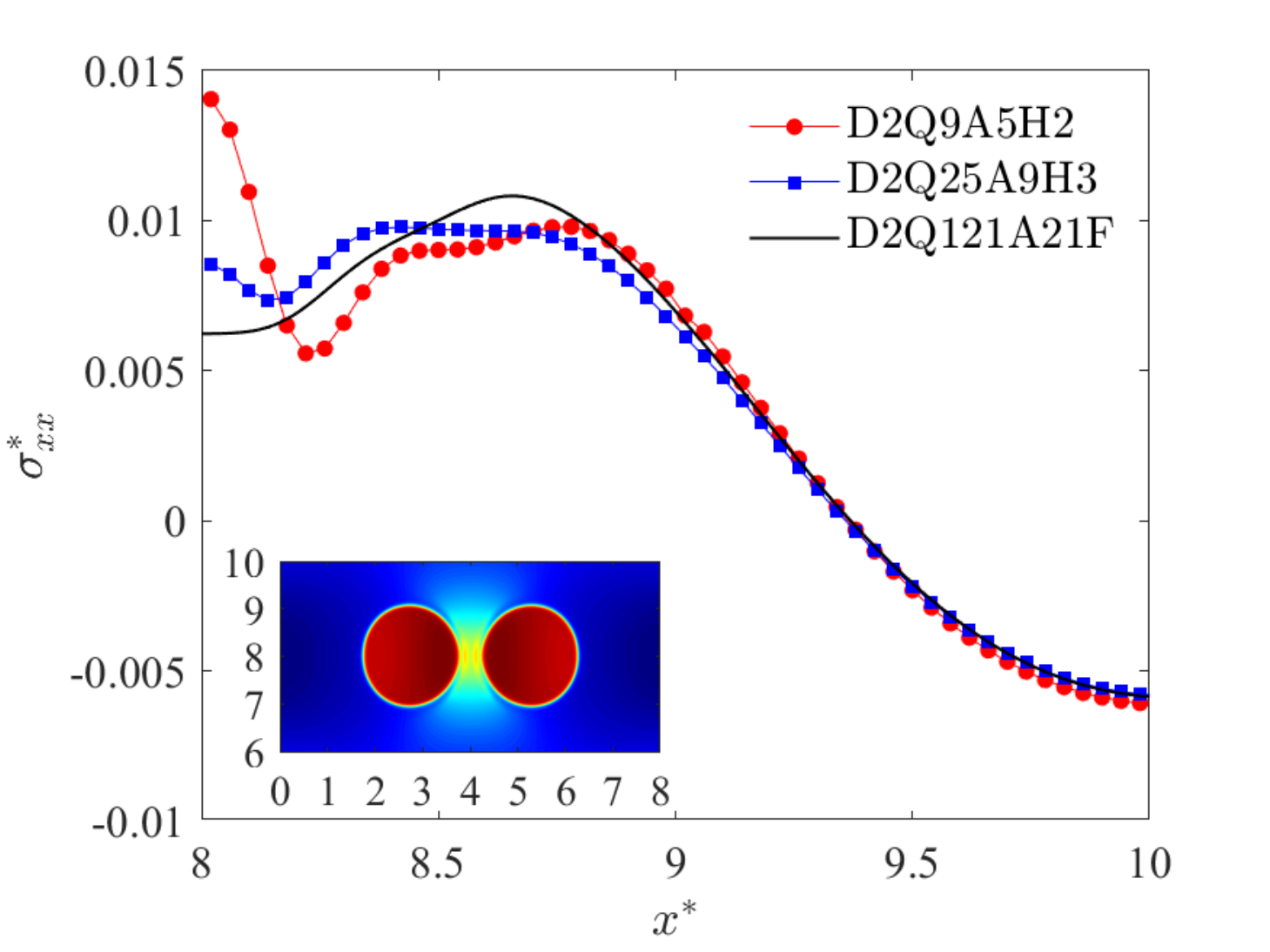}
			\label{sigmaxx_evol_2000}
		\end{minipage}%
	}%
	
	\subfloat[$t^*=0.5$]{
		\begin{minipage}[t]{0.48\linewidth}
			\centering
			\includegraphics[width=1.0\columnwidth,trim={0.0cm 0.18cm 0.0cm 0.2cm},clip]{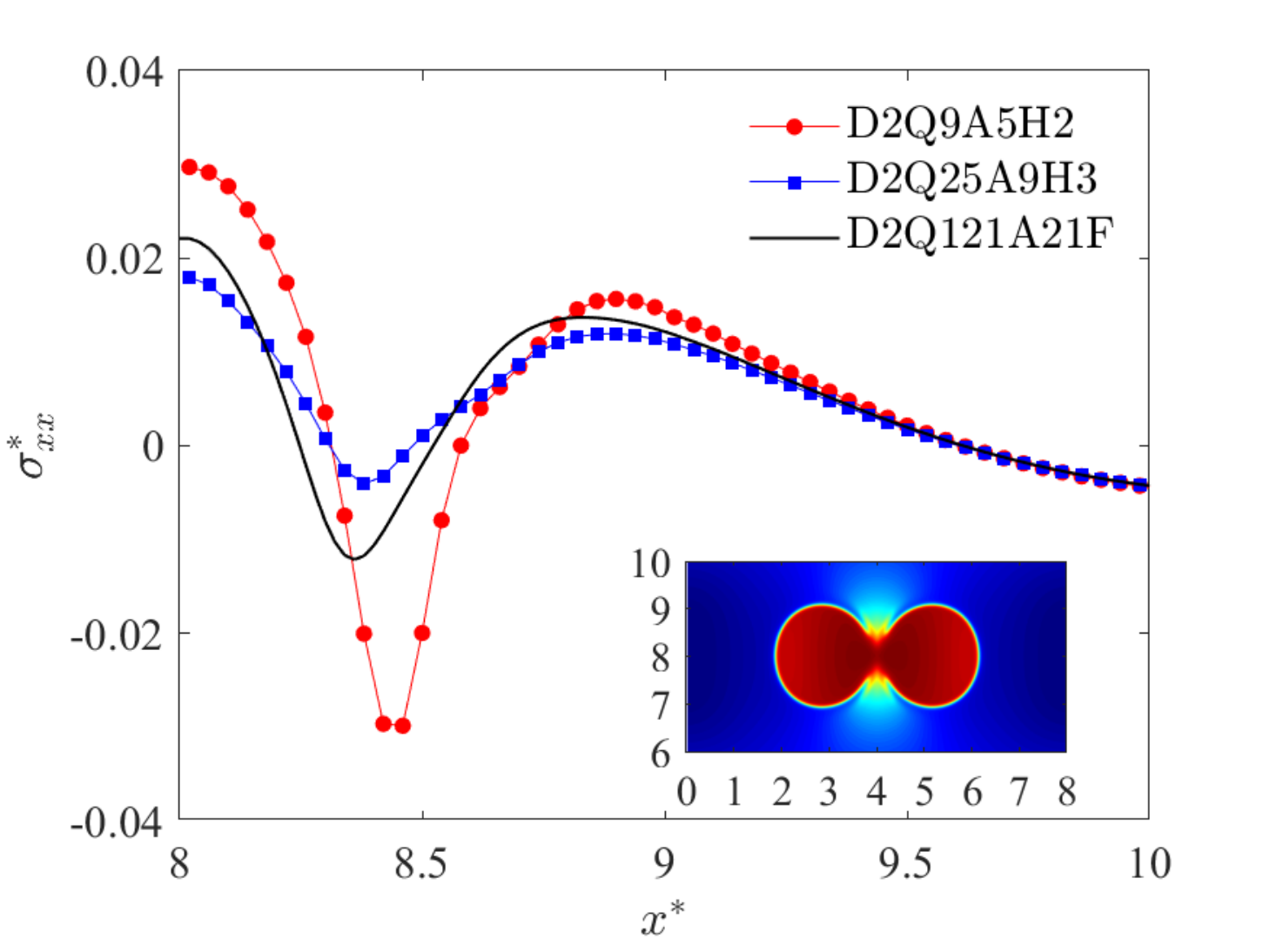}
			\label{sigmaxx_evol_2500}
		\end{minipage}%
	}%
	\subfloat[$t^*=0.6$]{
		\begin{minipage}[t]{0.48\linewidth}
			\centering
			\includegraphics[width=1.0\columnwidth,trim={0.0cm 0.18cm 0.0cm 0.2cm},clip]{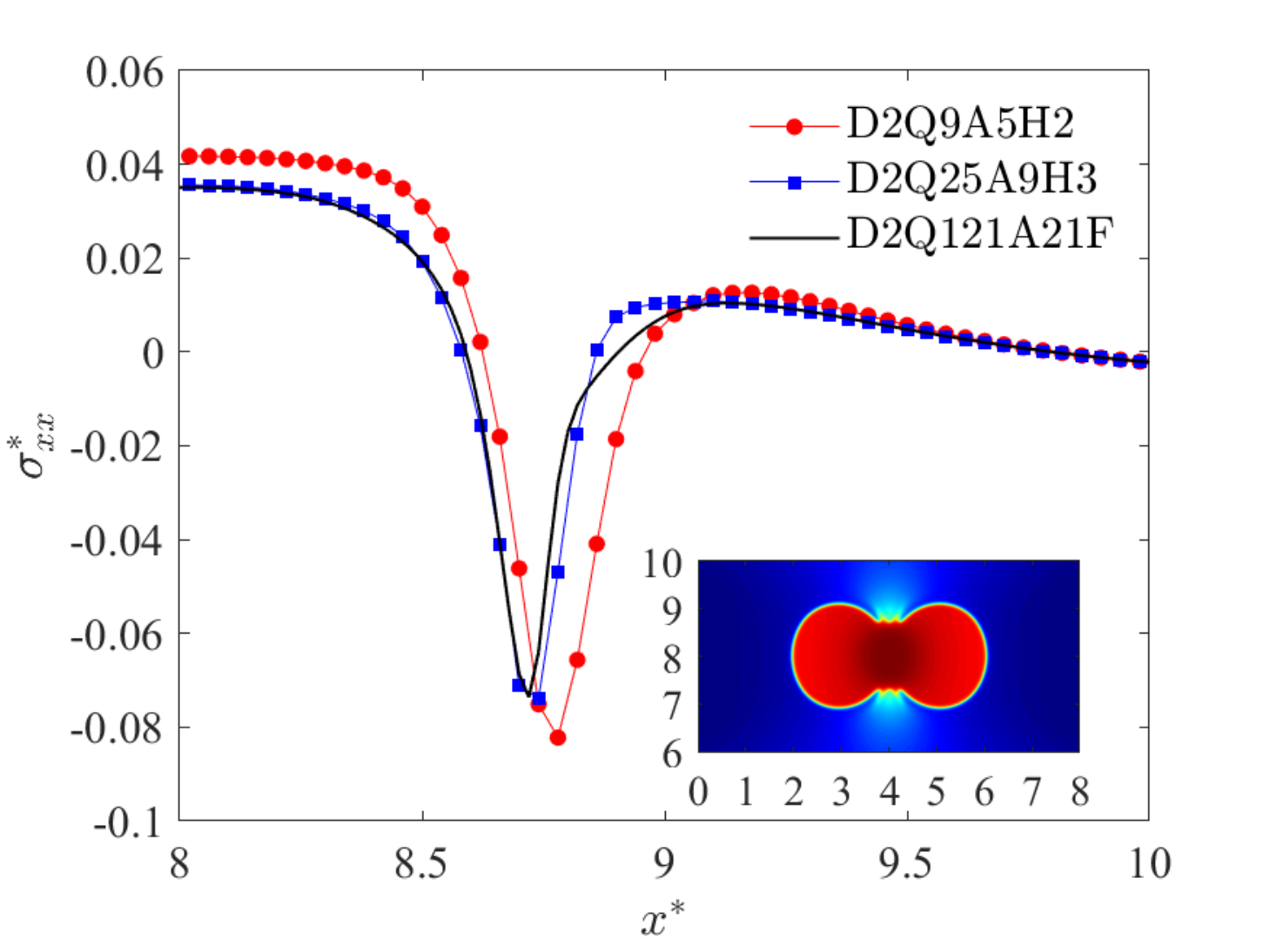}
			\label{sigmaxx_evol_3000}
		\end{minipage}%
	}%
	\caption{Evolution of the normalized viscous stress $\sigma_{xx}^{*}$ in the $x$-direction. (a) $t^*=0.3$, (b) $t^*=0.4$, (c) $t^*=0.5$ and (d) $t^*=0.6$. $We=1600$, $Re=13.33$ and $Kn=0.052$.} 
	\label{case1_sigmaxx}
\end{figure}
\subsection{Simulation results and discussions}
\subsubsection{$Kn=0.052$}
We simulate a case with the dimensionless parameters: $We=1600$, $Re=13.33$ and $Kn=0.052$. The ratio of the time step to the relaxation time is  $\Delta{t}/\tau\approx0.011$ and that of the grid spacing to the mean free path is $\Delta{x}/\lambda\approx0.19$, which implies that the kinetic scales have been fully resolved in the present simulation. The time in the following figures are normalized by $D_{l}/U$. Different Gauss-Hermite quadratures including D2Q9A5, D2Q25A9, D2Q81A17, D2Q121A21, D2Q169A25 and D2Q225A29 will be applied in the simulation. For neatness, not all the curves are presented in the following figures.

We mainly focus on the rarefaction effect during the formation of interdroplet gas film till the occurrence of coalescence. Coalescence happens if the interdroplet gas film can be squeezed out such that the contact point forms. The resistance with which the interdroplet gas film can be discharged depends on the droplet inertia as well as the dynamics of the film flow in particular to the pressure buildup within it~\citep{QianLaw1997}.
On the one hand, the high Weber number in the present simulation guarantees that the high drop inertia can overcome the viscous dissipation, surface tension work and the lubrication resistance inside the interdroplet gas film, rendering the successful coalescence. On the other hand, the droplets with high impact inertia greatly squeeze out the intervening gas film to some extent, causing more appreciable spatial variations of observable quantities in a neighbourhood of the contact point. 

\begin{figure}[h!]
	\centering
	\subfloat[D2Q25A9H3]{
		\begin{minipage}[t]{0.48\linewidth}
			\centering
			\includegraphics[width=1.0\columnwidth,trim={0.0cm 0.9cm 0.0cm 1.1cm},clip]{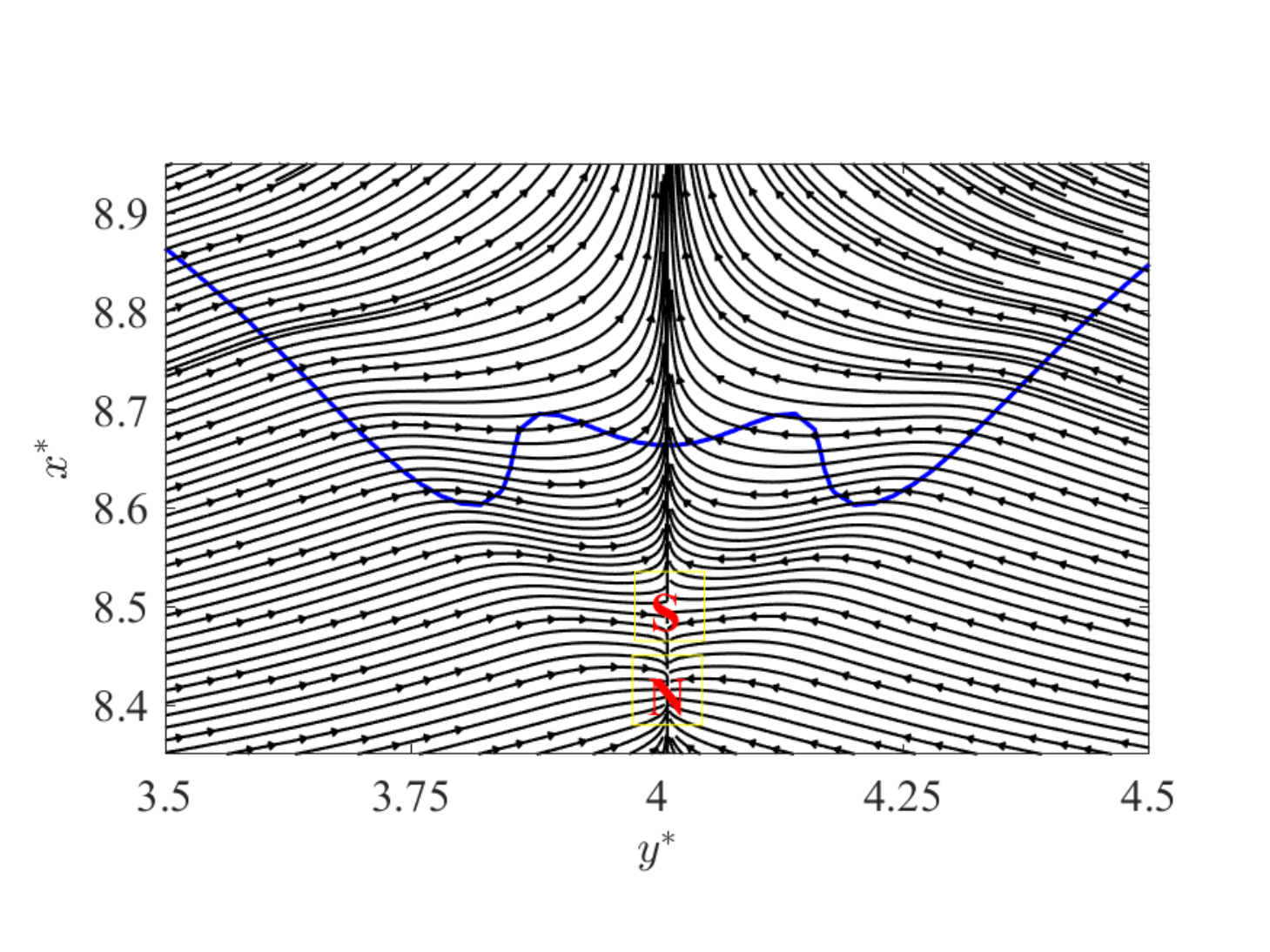}
			\label{streamlines_D2Q25H3}
		\end{minipage}%
	}%
	\subfloat[D2Q121A21F]{
		\begin{minipage}[t]{0.48\linewidth}
			\centering
			\includegraphics[width=1.0\columnwidth,trim={0.0cm 0.9cm 0.0cm 1.1cm},clip]{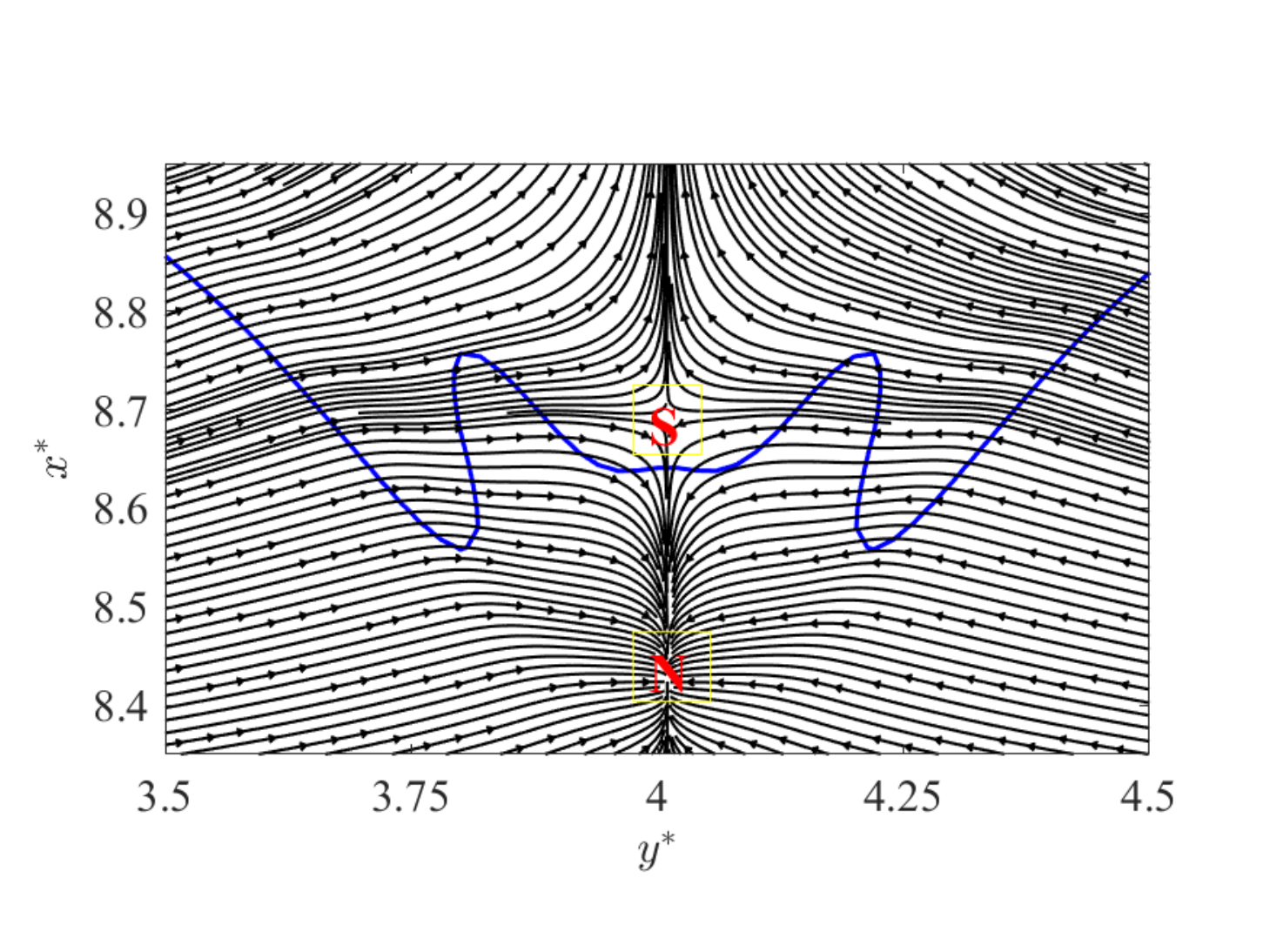}
			\label{streamlines_D2Q121F}
		\end{minipage}%
	}%
	
	\subfloat[D2Q169A25F]{
		\begin{minipage}[t]{0.48\linewidth}
			\centering
			\includegraphics[width=1.0\columnwidth,trim={0.0cm 0.9cm 0.0cm 1.1cm},clip]{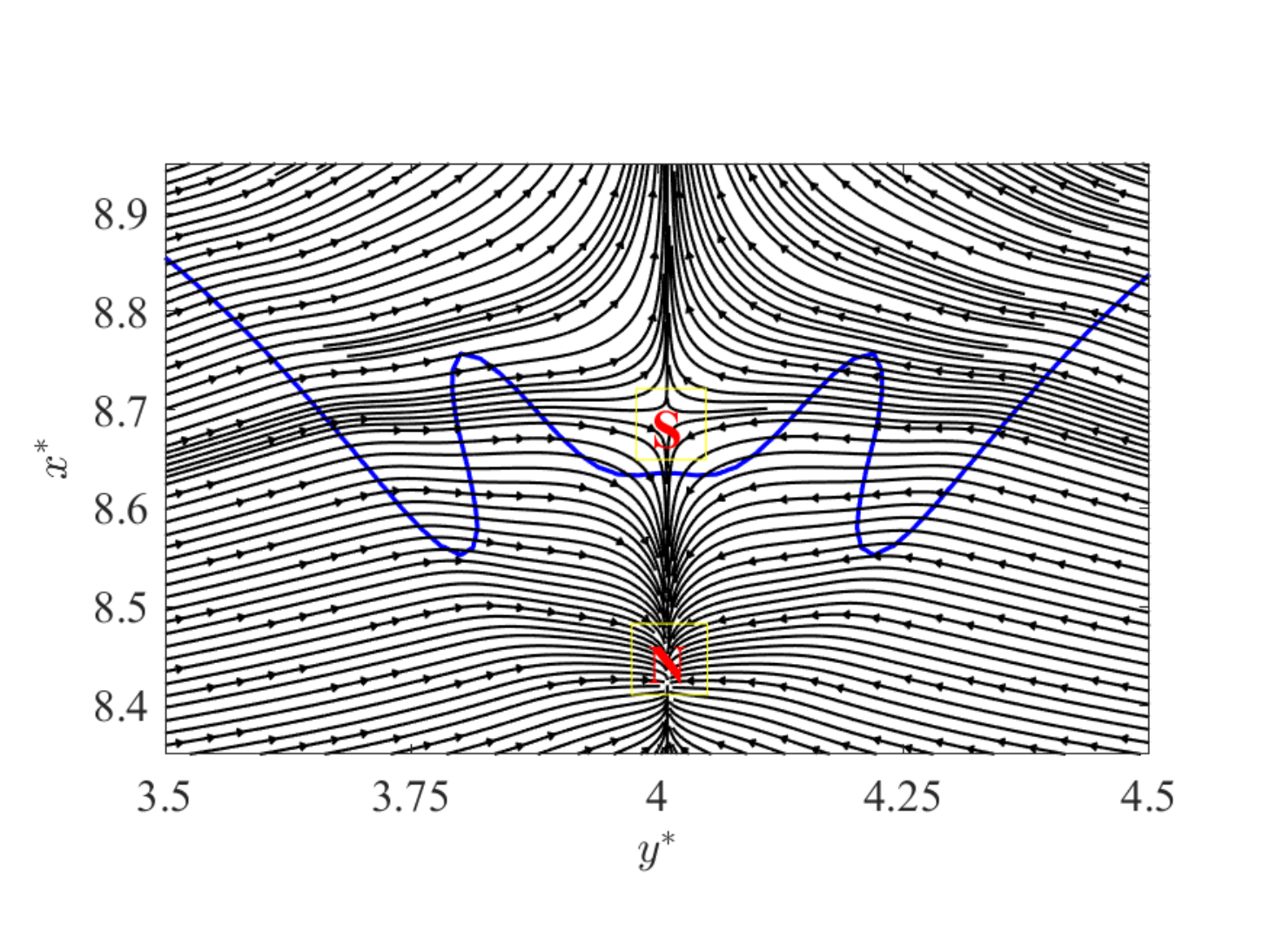}
			\label{streamlines_D2Q169F}
		\end{minipage}%
	}%
	\subfloat[D2Q225A29F]{
		\begin{minipage}[t]{0.48\linewidth}
			\centering
			\includegraphics[width=1.0\columnwidth,trim={0.0cm 0.9cm 0.0cm 1.1cm},clip]{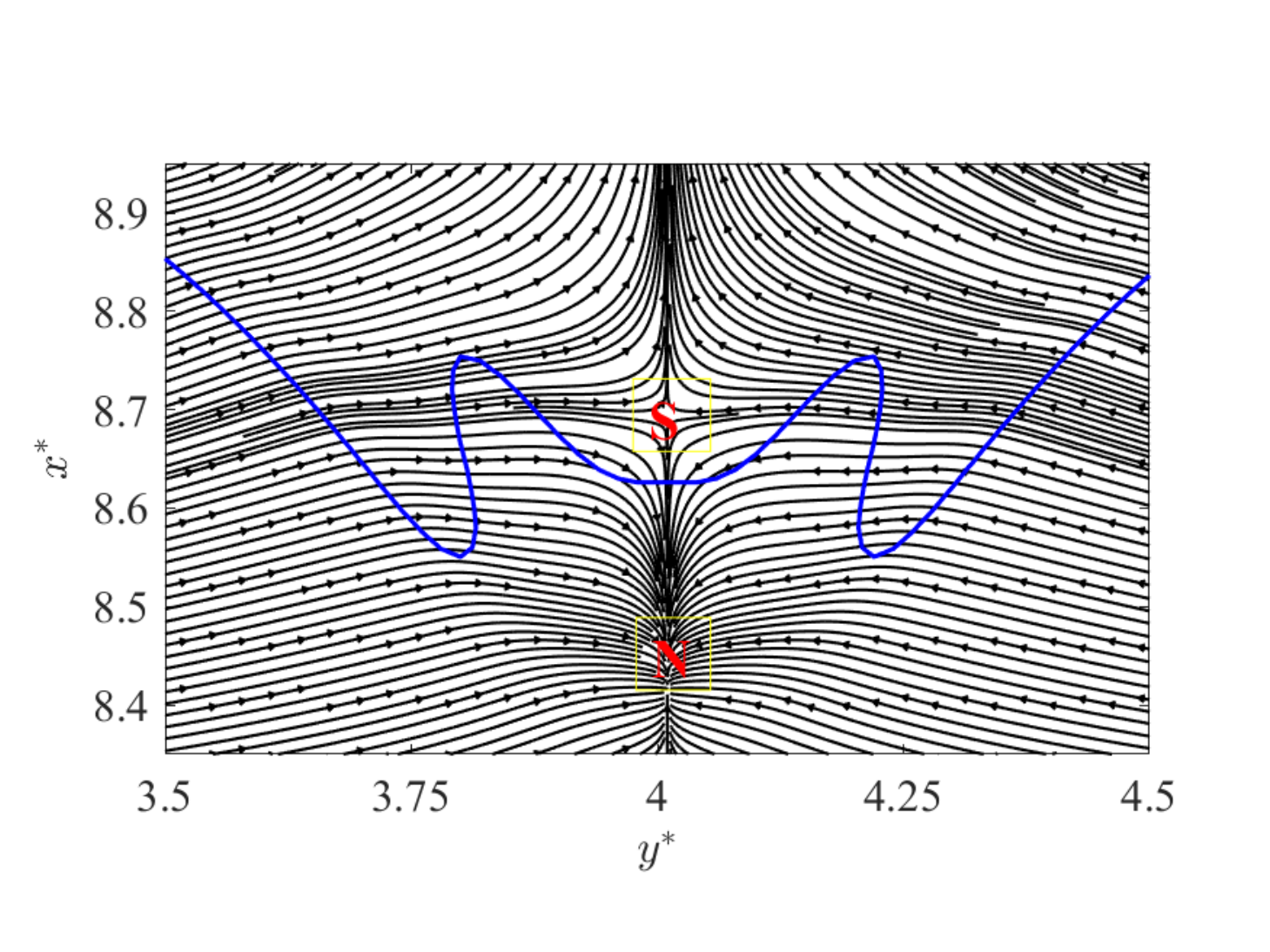}
			\label{streamlines_D2Q225F}
		\end{minipage}%
	}%
	\caption{Streamlines at $t^*=0.5$, where the red-marked $N$ and $S$ represent the node and saddle points, respectively. The blue contour line denotes the density $\rho=(\rho_{l}+\rho_{g})/2$. (a) D2Q25A9H3; (b) D2Q121A21F; (c) D2Q169A25F; (d) D2Q225A29F. $We=1600$, $Re=13.33$ and $Kn=0.052$.} 
	\label{xx3}
\end{figure}

\begin{figure}[h!]
	\centering
	\subfloat[D2Q25A9H3]{
		\begin{minipage}[t]{0.48\linewidth}
			\centering
			\includegraphics[width=1.0\columnwidth,trim={0.0cm 1.8cm 0.0cm 2.5cm},clip]{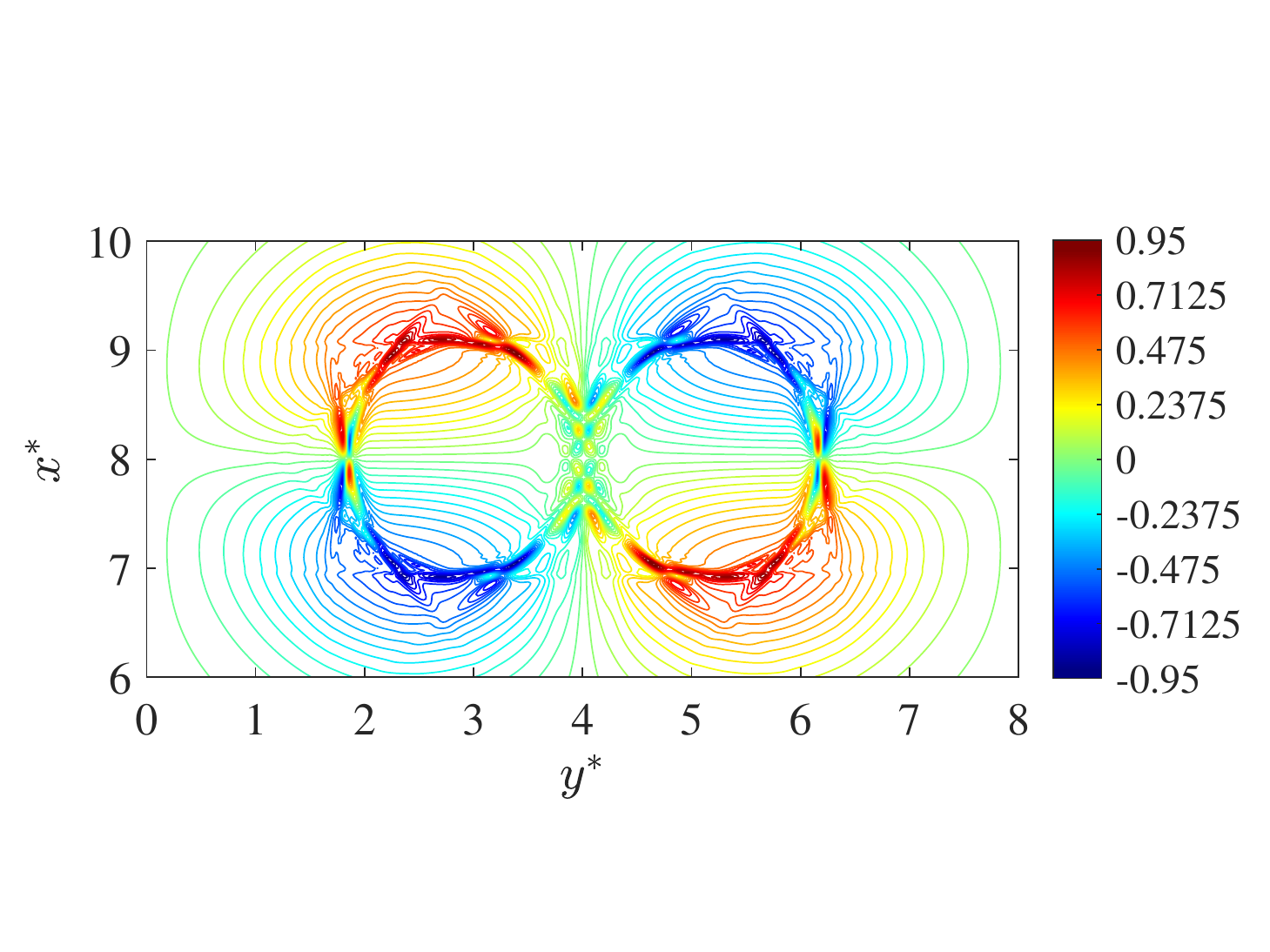}
			\label{contour_oz_D2Q25H3}
		\end{minipage}%
	}%
	\subfloat[D2Q121A21F]{
		\begin{minipage}[t]{0.48\linewidth}
			\centering
			\includegraphics[width=1.0\columnwidth,trim={0.0cm 1.8cm 0.0cm 2.5cm},clip]{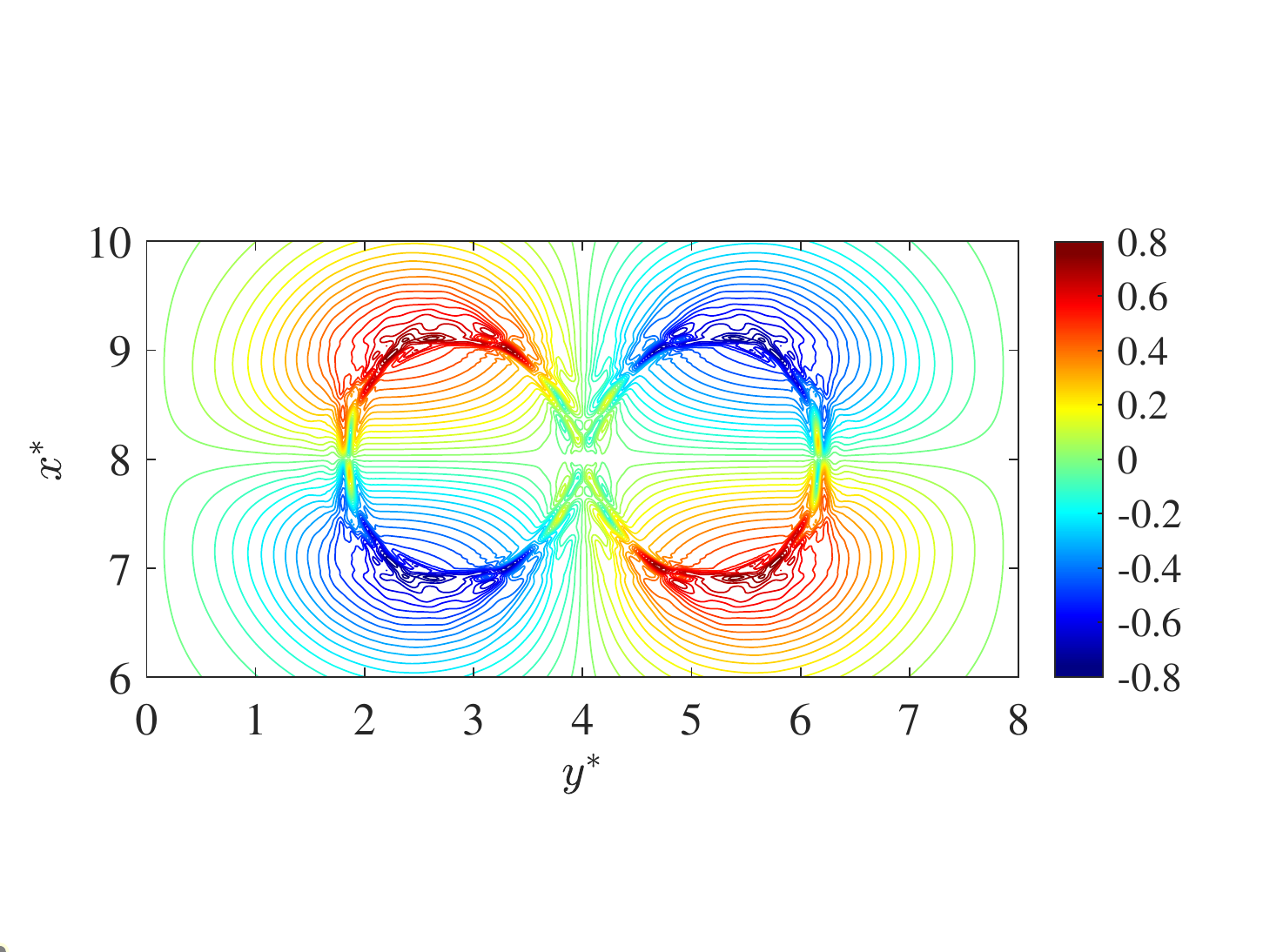}
			\label{contour_oz_D2Q121F}
		\end{minipage}%
	}%	
%	
%	\subfloat[D2Q169A25F]{
%		\begin{minipage}[t]{0.48\linewidth}
%			\centering
%			\includegraphics[width=1.0\columnwidth,trim={0.0cm 1.8cm 0.0cm 2.5cm},clip]{contour_oz_D2Q169F.eps}
%			\label{contour_oz_D2Q169F}
%		\end{minipage}%
%	}%
%	\subfloat[D2Q225A29F]{
%		\begin{minipage}[t]{0.48\linewidth}
%			\centering
%			\includegraphics[width=1.0\columnwidth,trim={0.0cm 1.8cm 0.0cm 2.5cm},clip]{contour_oz_D2Q225F.eps}
%			\label{contour_oz_D2Q225F}
%		\end{minipage}%
%	}%
	\caption{Contour of vorticity $\omega^*$ at $t^*=0.5$. (a) D2Q25A9H3 and (b) D2Q121A21F. $We=1600$, $Re=13.33$ and $Kn=0.052$.} 
	\label{xx4}
\end{figure}

\begin{figure}[h!]
	\centering
	\subfloat[$t^*=0.5$]{
		\begin{minipage}[t]{0.5\linewidth}
			\centering
			\includegraphics[width=1.0\columnwidth,trim={0.0cm 0.8cm 0.0cm 1.6cm},clip]{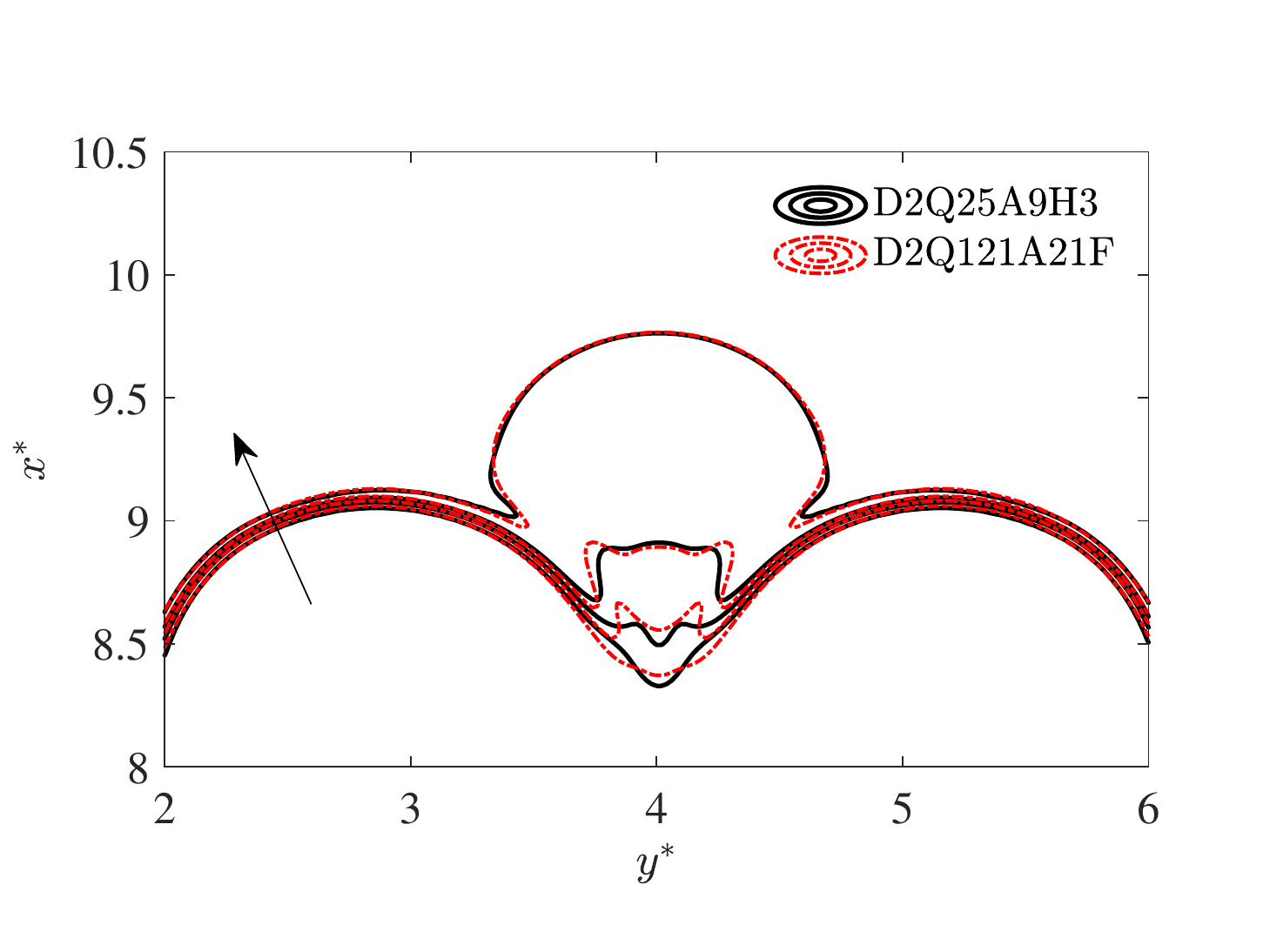}
			\label{droplet_case1_t2500}
		\end{minipage}%
	}%
	\subfloat[$t^*=0.6$]{
		\begin{minipage}[t]{0.5\linewidth}
			\centering
			\includegraphics[width=1.0\columnwidth,trim={0.0cm 0.8cm 0.0cm 1.6cm},clip]{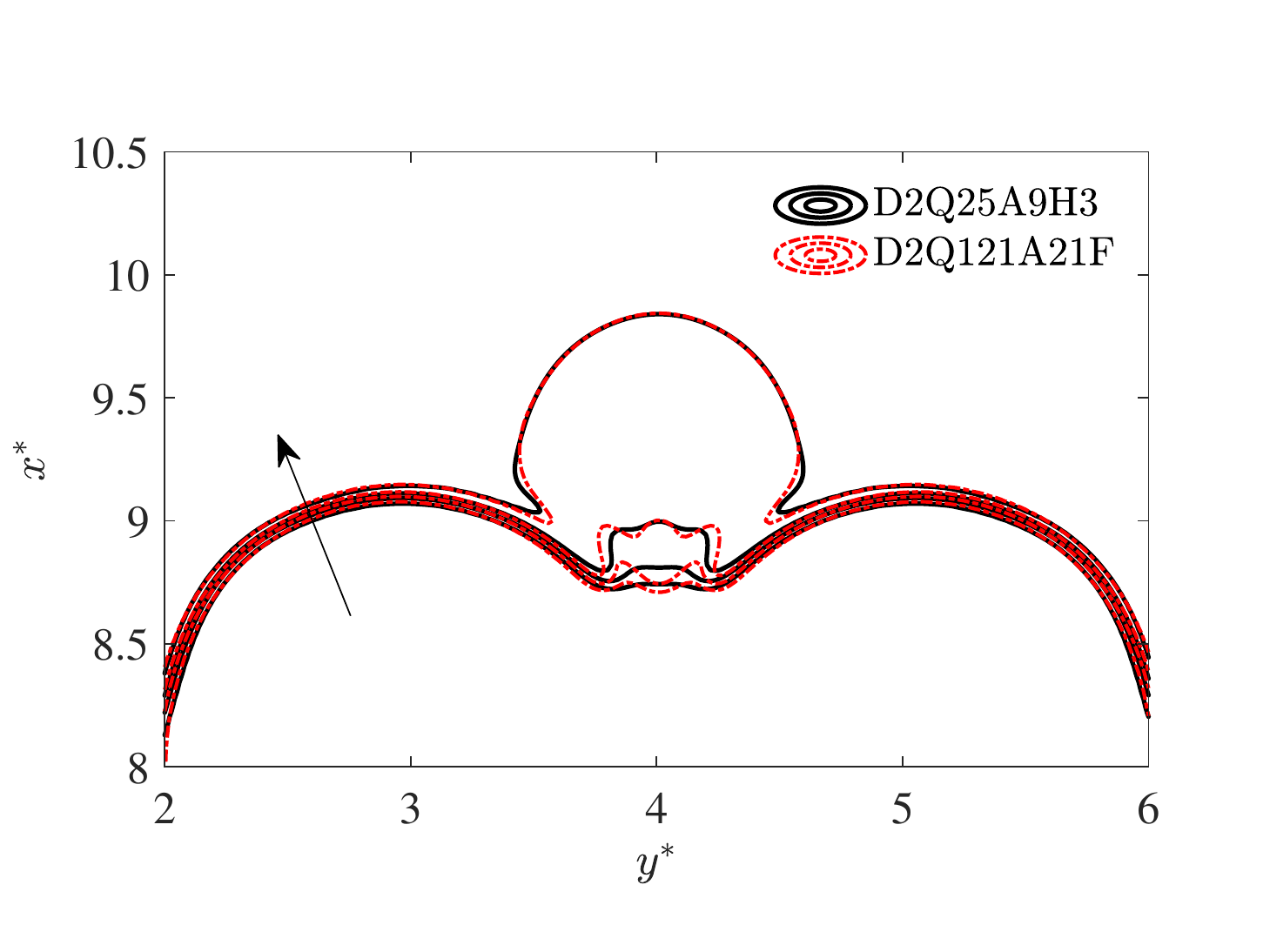}
			\label{droplet_case1_t3000}
		\end{minipage}%
	}%	
	\caption{Comparison of four dimensionless density profiles in the interfacial region of the droplets. Black: D2Q25A9H3, red: D2Q121A21F. The density decreases in the direction indicated by the arrow, where $\rho^*=0.75$, $0.90$, $1.05$ and $1.20$. (a) $t^*=0.5$ and (b) $t^*=0.6$. $We=1600$, $Re=13.33$ and $Kn=0.052$. } 
	\label{xx5}
\end{figure}

Figures~\ref{case1_ux} and~\ref{case1_sigmaxx} show the time evolutions of the normalized vertical velocity component $u_x^{*}\equiv u_{x}/U$ and the viscous stress component
 $\sigma_{xx}^{*}\equiv2\sigma_{xx}/(\rho_l+\rho_g)RT$ near the centreline of the interdroplet gas film (the left limiting position $y^*\rightarrow4$). At $t^*=0.3$, the droplets are moving closer towards each other but are separated by a relatively long centre-to-centre distance. The interdroplet gas region is being compressed such that a monotonic increase of $u_x^*$ is observed till around $x^*=9$, followed by a monotonic decrease of the velocity magnitude.
In figure~\ref{ux_evol_2000}, the maximum value of the velocity magnitude increases at $t^*=0.4$ compared to that shown at $t^*=0.3$, because the interdroplet gas film is further compressed as the centre-to-centre distance between the droplets becomes shorter.
In these two time instants, the rarefaction effect is not pronounced in the interdroplet gas region. Therefore, no obvious changes can be seen in the distributions of $u_x^*$ obtained from different Gauss-Hermite quadratures, except that an obvious variation of $\sigma_{xx}^{*}$ can be observed near $x^*=8$, as shown in figures~\ref{sigmaxx_evol_1500} and~\ref{sigmaxx_evol_2000}. 

As illustrated in figures~\ref{ux_evol_2500} and~\ref{sigmaxx_evol_2500}, the liquid bridge forms at the coalescence region and the droplets are merging due to the attractive intermolecular force at $t^*=0.5$.
Relatively large spatial variations of $u_x^{*}$ and $\sigma_{xx}^{*}$ are observed near the contact point in particular to the region around $x^*=8.5$. The differences between the results obtained using different Gauss-Hermite quadratures become pronounced compared to figures~\ref{sigmaxx_evol_1500} and~\ref{sigmaxx_evol_2000}. 
It is clear that both $u_x^{*}$ and $\sigma_{xx}^{*}$ converge with the increase of the number of discrete particle velocities.
According to the analysis in Section~\ref{GH33}, a Gauss-Hermite quadrature with at least $6^{th}$-order degree of precision is needed to compute the viscous stress tensor accurately at the Navier-Stokes order. Therefore, with the $3^{rd}$-order Hermite expansion for the equilibrium distribution function, D2Q25A9H3 has sufficient capability to capture the flow physics at the Navier-Stokes level, while incorporating the higher-order non-continuum effects is beyond its ability. 
Therefore, the differences between the results obtained by D2Q25A9H3 and higher-order Gauss-Hermite quadratures should be attributed to the rarefaction effect.
Although the second-order Hermite expansion cannot ensure accurate evaluation of the viscous stress tensor at the Navier-Stokes order, it is noted that D2Q9A5H2 is usually applied to simulate the continuum flows in the conventional LBM.
However, for this case with high impact inertia and approaching rate of two droplets, the viscous stress tensor cannot be accurately obtained using D2Q9A5H2, which will definitely influence the simulated droplet evolution process (see figures~\ref{case1_ux} and~\ref{case1_sigmaxx}).
Interestingly, D2Q9A5H2 still gives similar trends for the spatial variations of both $u_x^*$ and $\sigma_{xx}^{*}$ qualitatively. In contrast, higher-order Gauss-Hermite quadratures (D2Q81A17F, D2Q121A21F, D2Q169A25F and D2Q225A29F) can capture the non-continuum effects beyond the Navier-Stokes order to different levels. For this case, we find that D2Q81A17F can already give satisfying predictions with minor discrepancies which can be further reduced using D2Q121A21F. At the time $t^*=0.6$, the coalescence continues after the interdroplet gas film is forcibly discharged. The density in the merging region slightly arises and the rarefaction effect is attenuated compared to that observed at $t^*=0.5$. This can be further evidenced by figures~\ref{ux_evol_3000} and~\ref{sigmaxx_evol_3000}, where the differences between the results obtained from different Gauss-Hermite quadratures are not very prominent. The rarefaction effect mainly concentrates near the edge of the liquid bridge at $t^*=0.6$. 
\begin{figure}[p]
	\centering
	\subfloat[$t^*=0.3$]{
		\begin{minipage}[t]{0.48\linewidth}
			\centering
			\includegraphics[width=1.0\columnwidth,trim={0.0cm 0.18cm 0.0cm 0.2cm},clip]{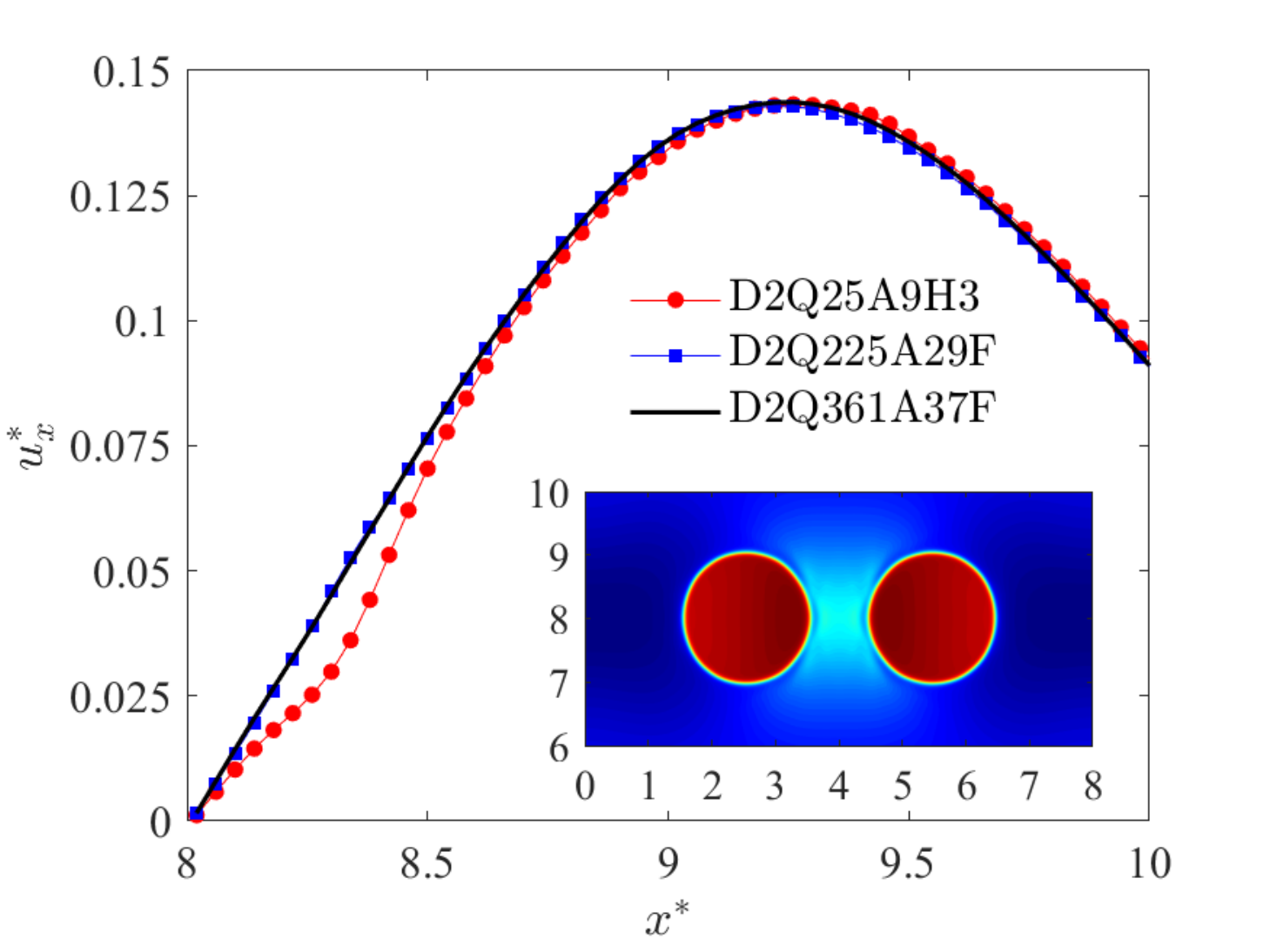}
			\label{ux_evol_1500_a}
		\end{minipage}%
	}%
	\subfloat[$t^*=0.4$]{
		\begin{minipage}[t]{0.48\linewidth}
			\centering
			\includegraphics[width=1.0\columnwidth,trim={0.0cm 0.18cm 0.0cm 0.2cm},clip]{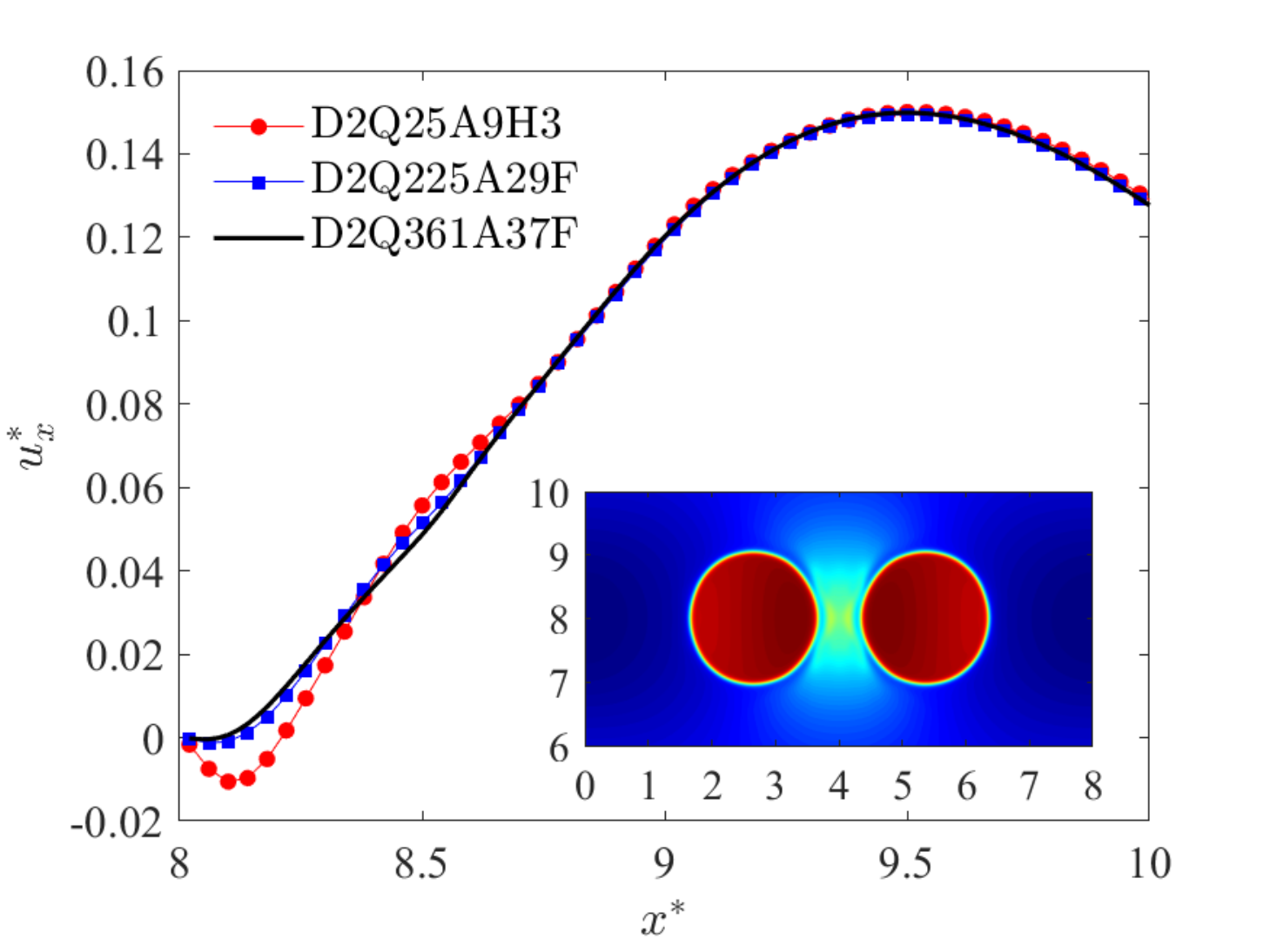}
			\label{ux_evol_2000_a}
		\end{minipage}%
	}\\
	\subfloat[$t^*=0.5$]{
		\begin{minipage}[t]{0.48\linewidth}
			\centering
			\includegraphics[width=1.0\columnwidth,trim={0.0cm 0.18cm 0.0cm 0.2cm},clip]{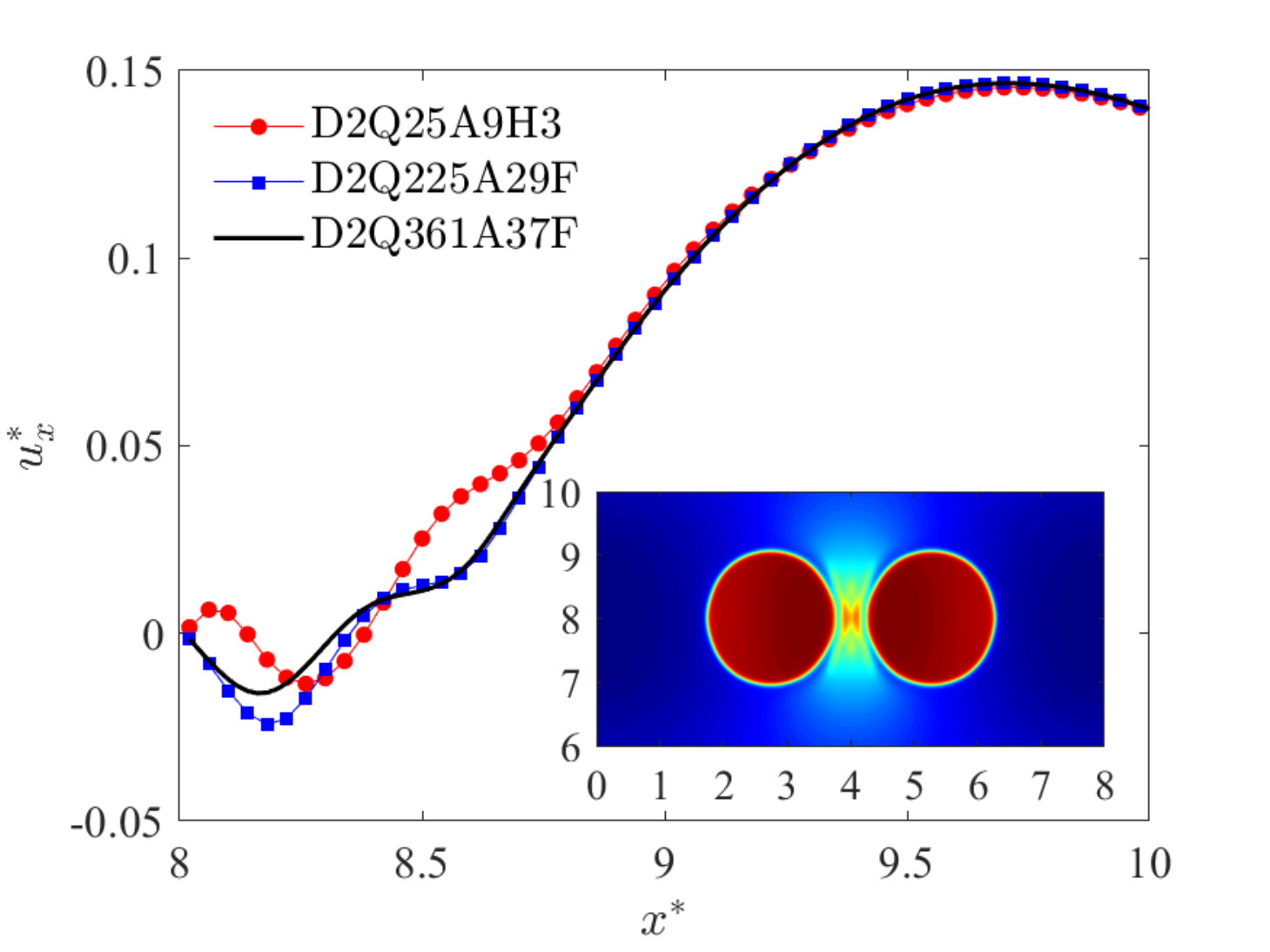}
			\label{ux_evol_2500_a}
		\end{minipage}
	}
	\subfloat[$t^*=0.6$]{
		\begin{minipage}[t]{0.48\linewidth}
			\centering
			\includegraphics[width=1.0\columnwidth,trim={0.0cm 0.18cm 0.0cm 0.2cm},clip]{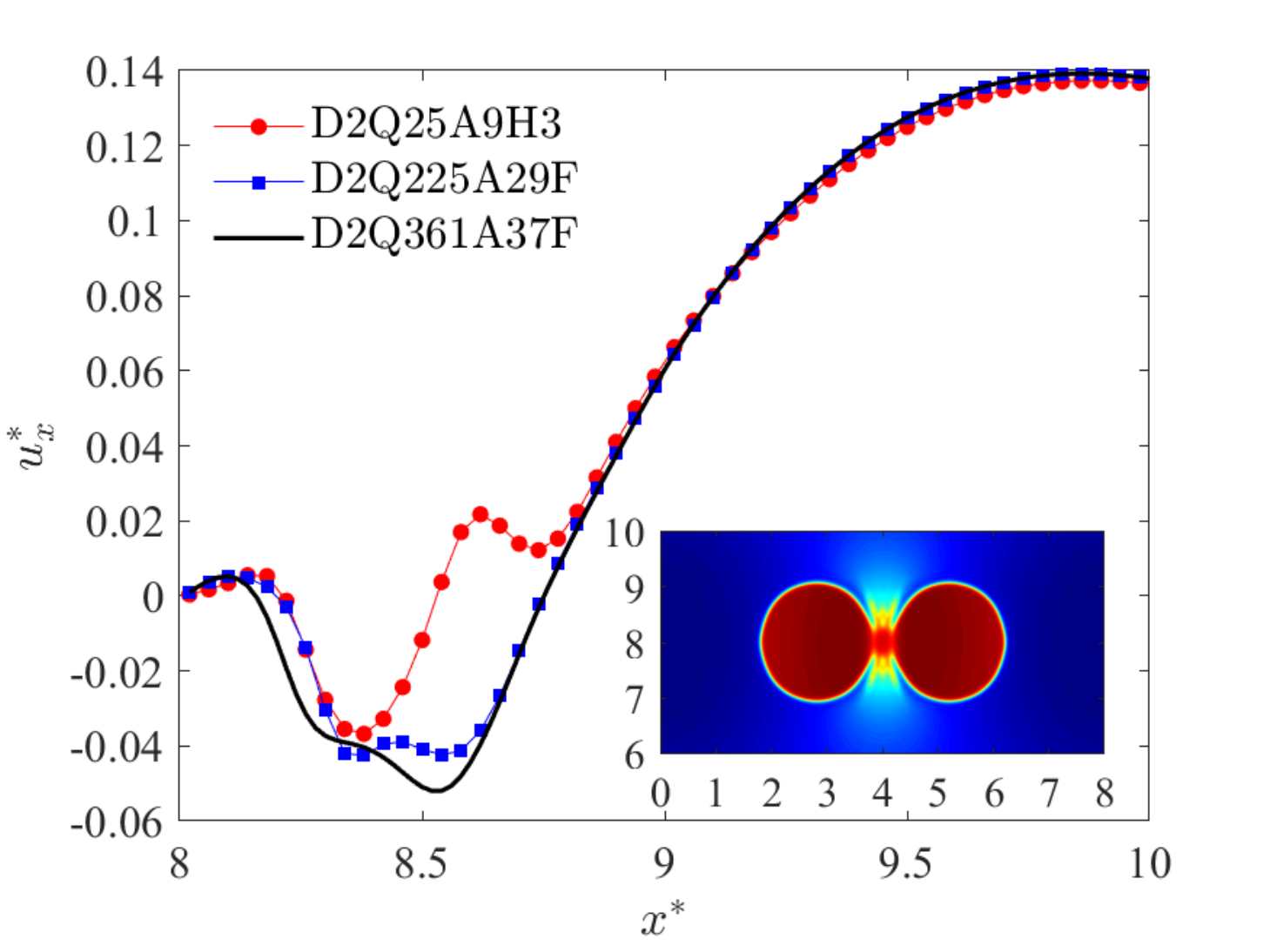}
			\label{ux_evol_3000_a}
		\end{minipage}%
	}\\
	\subfloat[$t^*=0.8$]{
		\begin{minipage}[t]{0.48\linewidth}
			\centering
			\includegraphics[width=1.0\columnwidth,trim={0.0cm 0.18cm 0.0cm 0.2cm},clip]{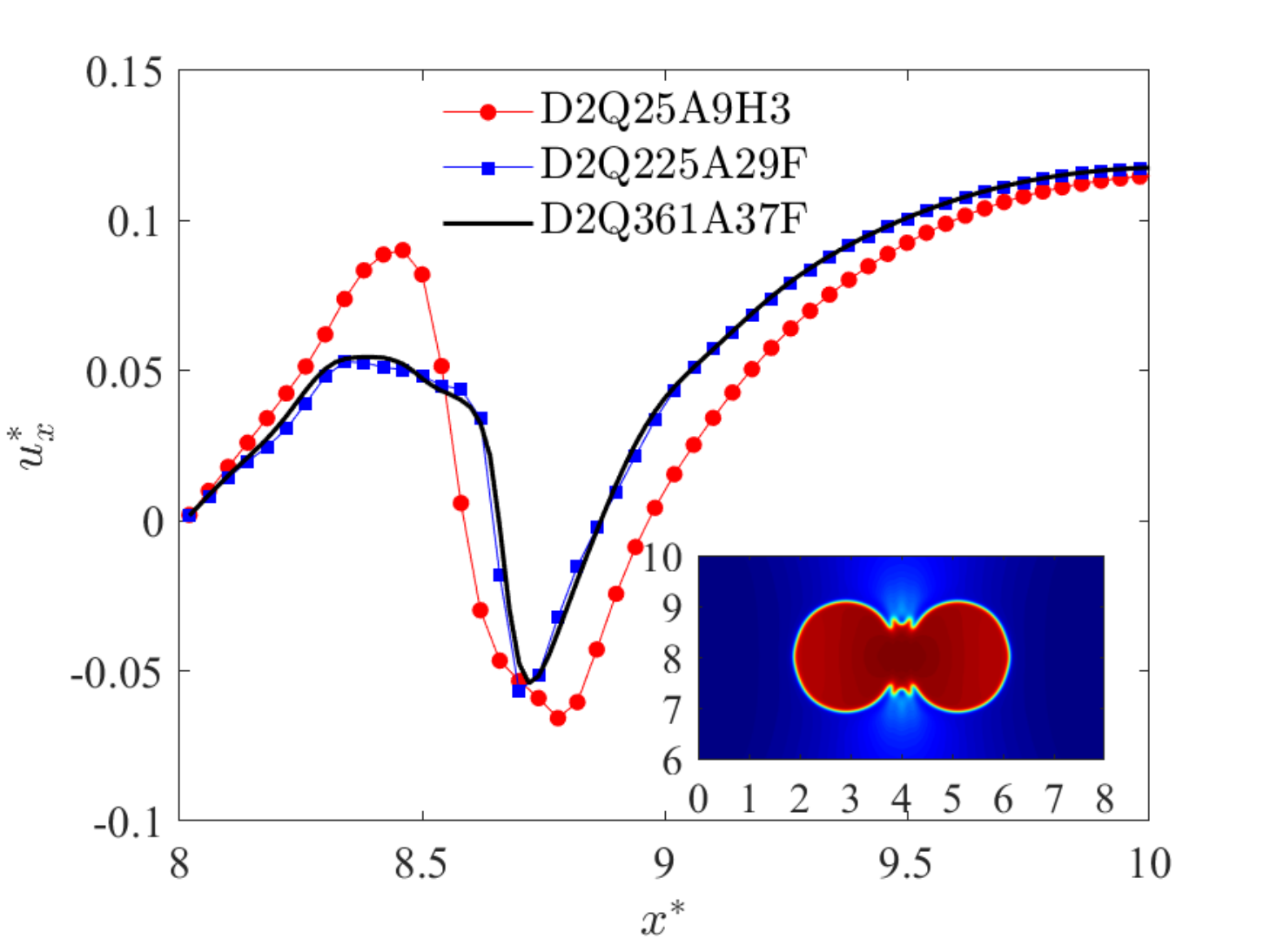}
			\label{ux_evol_4000_a}
		\end{minipage}%
	}%
	\subfloat[$t^*=1.0$]{
		\begin{minipage}[t]{0.48\linewidth}
			\centering
			\includegraphics[width=1.0\columnwidth,trim={0.0cm 0.18cm 0.0cm 0.2cm},clip]{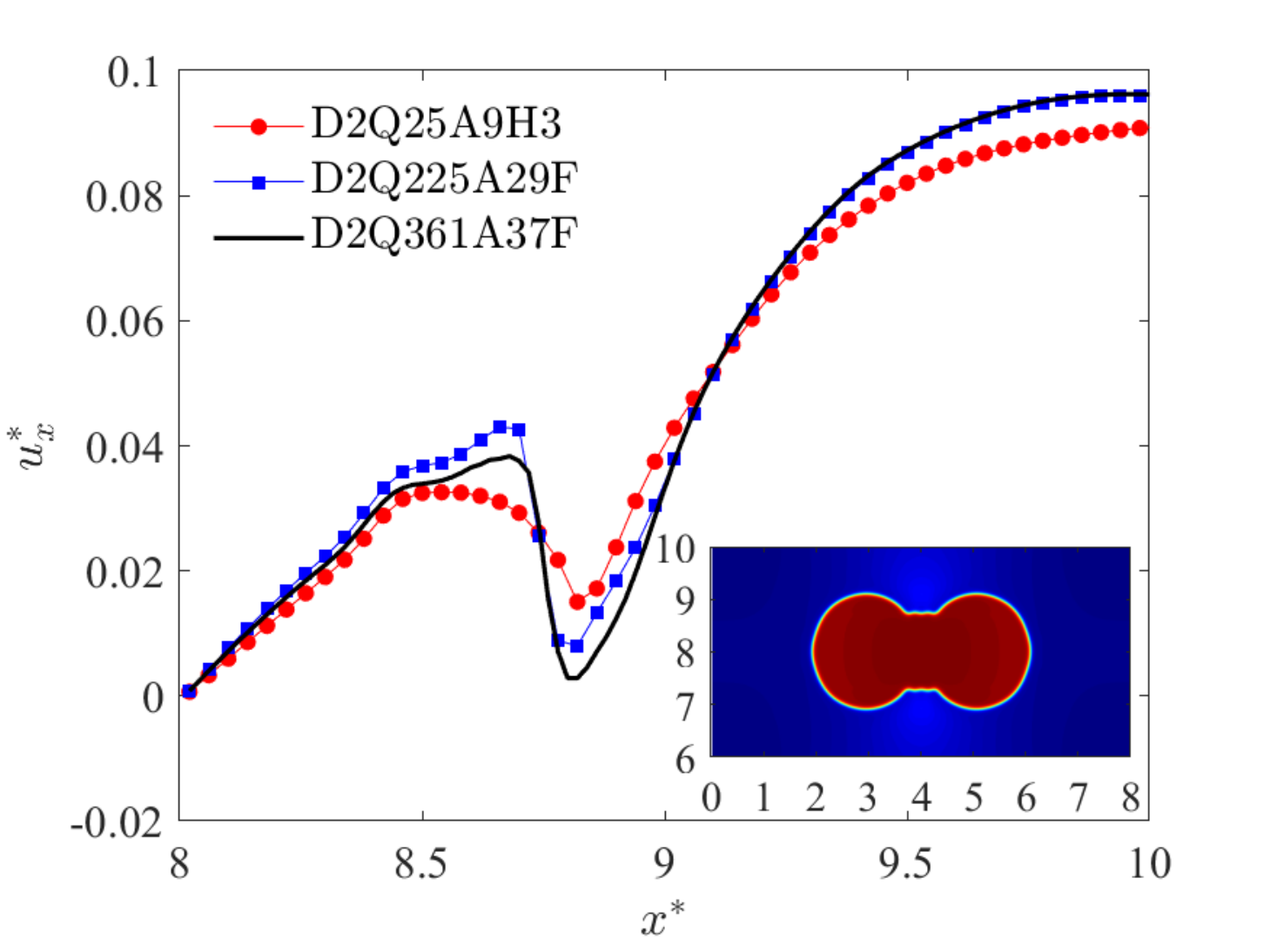}
			\label{ux_evol_5000_a}
		\end{minipage}%
	}%
	\caption{Evolution of the normalized velocity component $u_{x}^{*}$ in the $x$-direction. (a) $t^*=0.3$, (b) $t^*=0.4$, (c) $t^*=0.5$, (d) $t^*=0.6$, (e) $t^*=0.8$, (f) $t^*=1.0$. $We=1600$, $Re=6.67$ and $Kn=0.104$.} 
	\label{case2_ux}
\end{figure}

To further demonstrate the rarefaction effect, figures~\ref{xx3} and~\ref{xx4} show the snapshots of the streamlines and the vorticity contours at $t^*=0.5$, respectively.
In figure~\ref{xx3}, the profiles of the merged droplet are superposed on the streamlines for comparison. Overall, the directions of the streamlines are consistent with our physical intuition. Interestingly, a more distinct saddle-node pair can be clearly identified in figures~\ref{streamlines_D2Q121F},~\ref{streamlines_D2Q169F} and~\ref{streamlines_D2Q225F} when compared to that shown in figure~\ref{streamlines_D2Q25H3}.
On the one hand, from topological perspective, the resulting velocity field can be mapped onto a toroidal surface since the periodic boundary conditions are used in the simulations.
The Poincar\'e–Bendixson (P-B) index theorem~\citep{Flegg2001} says that the number of nodes must be equal to the number of saddles for any smooth vector field on a toroidal surface. A careful examination of the whole velocity field shows that there are six nodes and six saddles (including those in figure~\ref{xx3}) in the whole flow field, which implies the resulting streamline topology is reasonable (this also provides a self-consistent check for our calculation). On the other hand, D2Q121A21F, D2Q169A25F and D2Q225A29F can capture the higher-order non-continuum effect beyond the Navier-Stokes level, while D2Q25A9H3 can only pick up flow physics at the Navier-Stokes order. Therefore, the more distinct saddle-node pair and the relatively large crown structure are conjectured to be caused by the rarefaction effect. To some degree, since the present configuration is symmetrical with respect to the centreline $y^*=4$, the formation of these distinct flow structures could be comparable to those observed in droplet spreading and splash phenomena on a surface, which are already shown to be influenced by the rarefaction effect~\citep{Mandre2012,Sprittles2015,Sprittles2017}. The case shown here seems more complicated since the intervening liquid plane has more complex velocity and viscous stress variations compared to a stationary solid wall. 
In addition, due to the rarefaction effect, the vorticity concentration $\omega^*\equiv\omega/(2U/D_l)$ around outer interfaces of the droplets is attenuated and the number of the sharp shear layers also decreases, as illustrated in figure~\ref{xx4}. High-magnitude positive and negative vorticity centres in the interdroplet region are also suppressed. This could be related to the previous finding that the rarefaction effect can reduce the lubrication resistance force, facilitate the rupture of the interdroplet gas film and boost the droplet coalescence~\citep{Gopinath1997,LiJiePRL2016}.

Figure~\ref{xx5} compares the dimensionless density profiles $\rho^*\equiv2\rho/(\rho_{l}+\rho_{g})$ obtained from D2Q25A9H3 and D2Q121A21F in the interfacial region of the droplets, respectively. It is observed that the rarefaction effect mainly influences the density profiles between the two droplets while the outer density profiles overlap very well.
It is also seen that the region with obvious discrepancies in the density profiles has prominent variations of the vertical velocity component $u_{x}^{*}$ (figure~\ref{case1_ux}) and the viscous stress component $\sigma_{xx}^{*}$ (figure~\ref{case1_sigmaxx}).

\subsubsection{$Kn=0.104$}
\begin{figure}[p]
	\centering
	\subfloat[$t^*=0.3$]{
		\begin{minipage}[t]{0.48\linewidth}
			\centering
			\includegraphics[width=1.0\columnwidth,trim={0.0cm 0.18cm 0.0cm 0.2cm},clip]{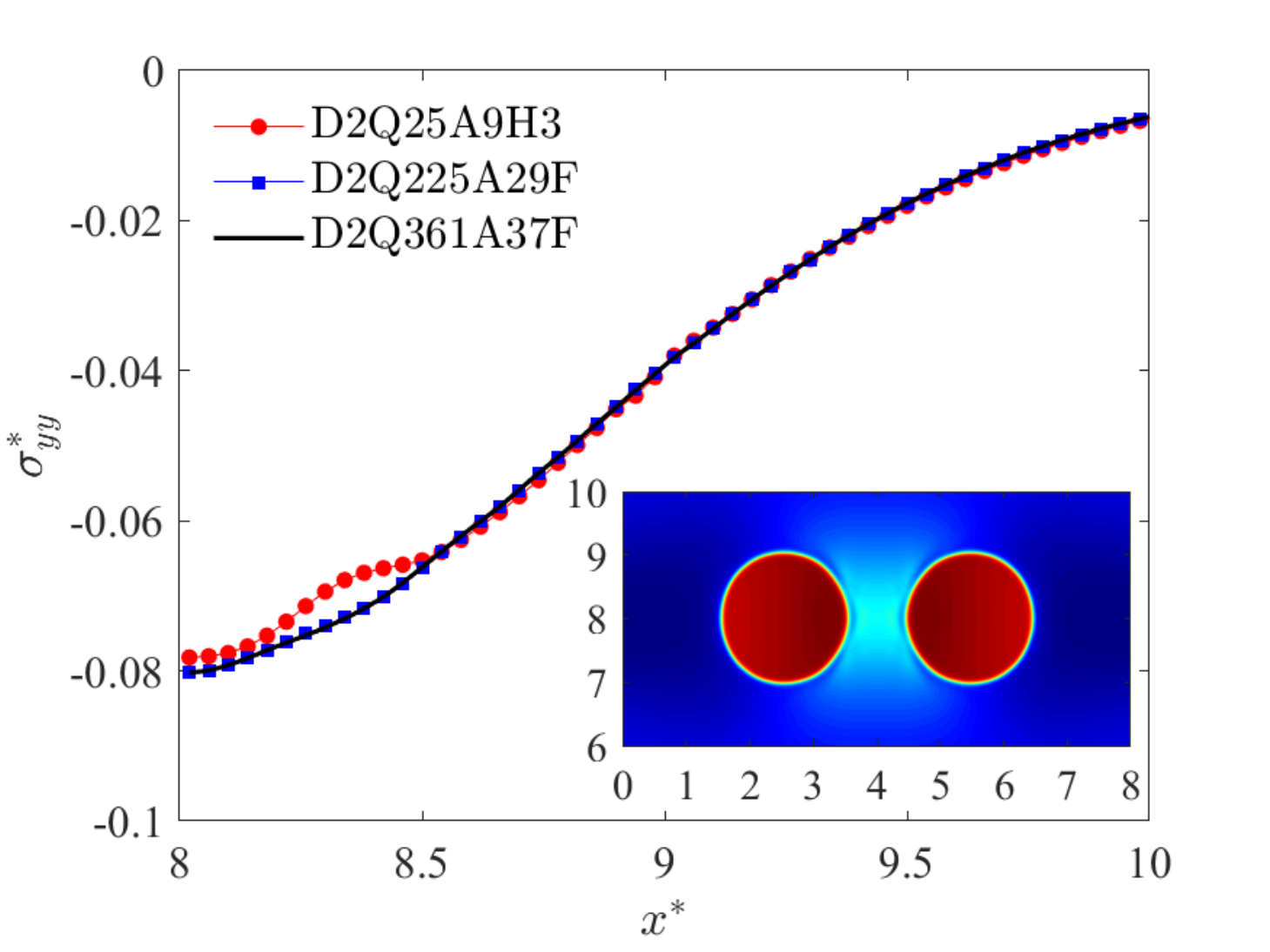}
			\label{sigmayy_evol_1500_a}
		\end{minipage}%
	}%
	\subfloat[$t^*=0.4$]{
		\begin{minipage}[t]{0.48\linewidth}
			\centering
			\includegraphics[width=1.0\columnwidth,trim={0.0cm 0.18cm 0.0cm 0.2cm},clip]{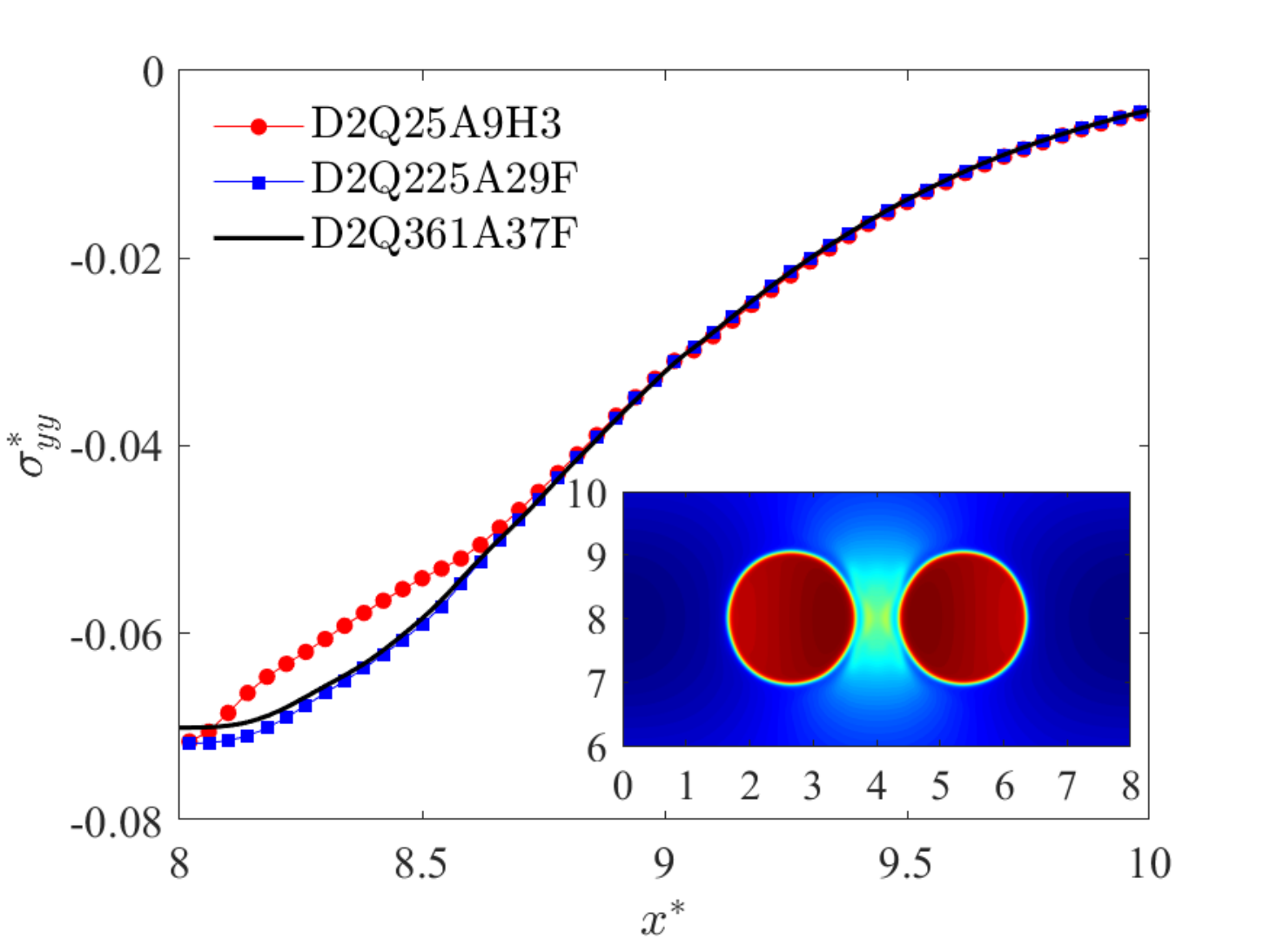}
			\label{sigmayy_evol_2000_a}
		\end{minipage}%
	}\\
	\subfloat[$t^*=0.5$]{
		\begin{minipage}[t]{0.48\linewidth}
			\centering
			\includegraphics[width=1.0\columnwidth,trim={0.0cm 0.18cm 0.0cm 0.2cm},clip]{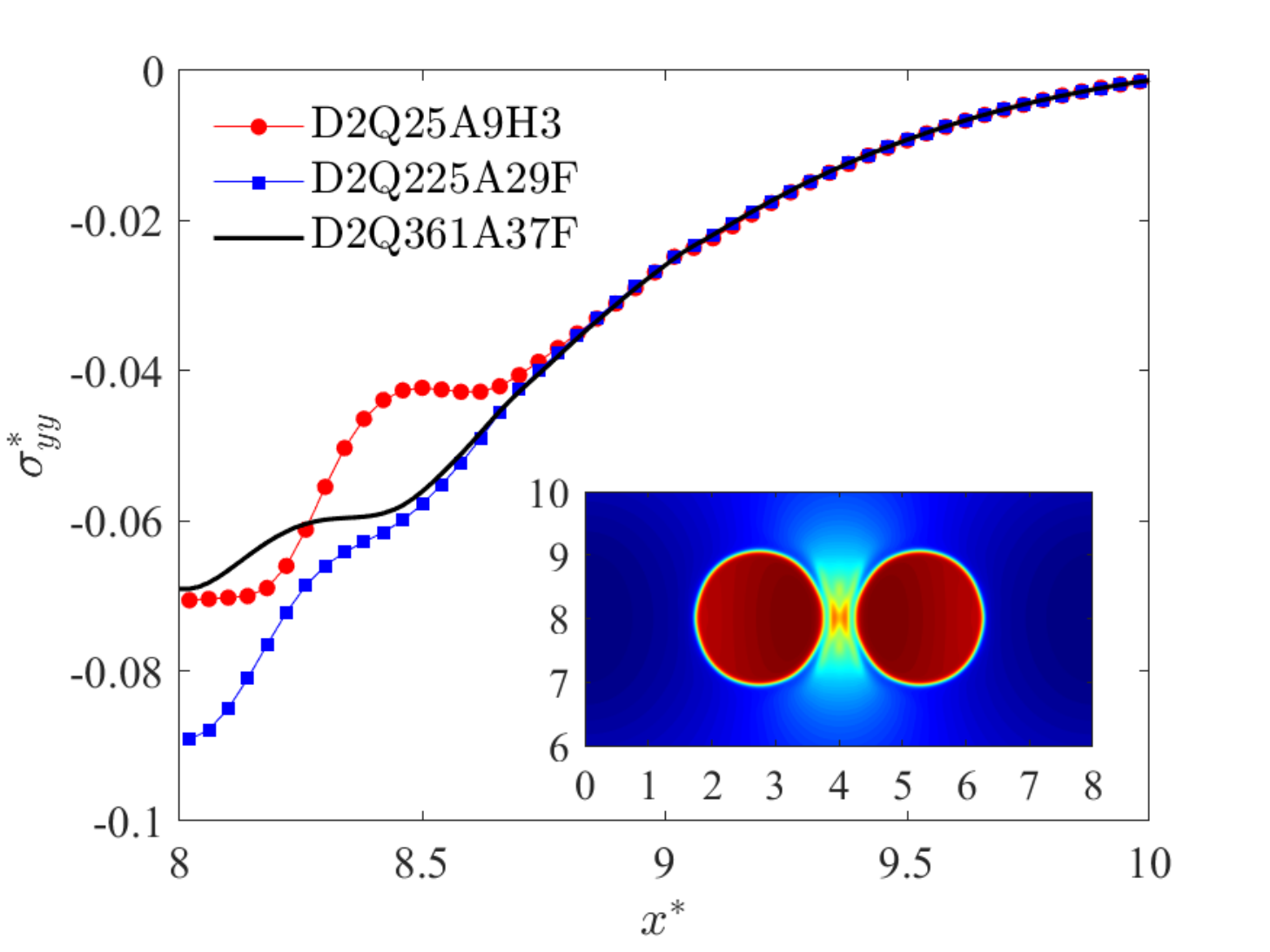}
			\label{sigmayy_evol_2500_a}
		\end{minipage}%
	}
	\subfloat[$t^*=0.6$]{
		\begin{minipage}[t]{0.48\linewidth}
			\centering
			\includegraphics[width=1.0\columnwidth,trim={0.0cm 0.18cm 0.0cm 0.2cm},clip]{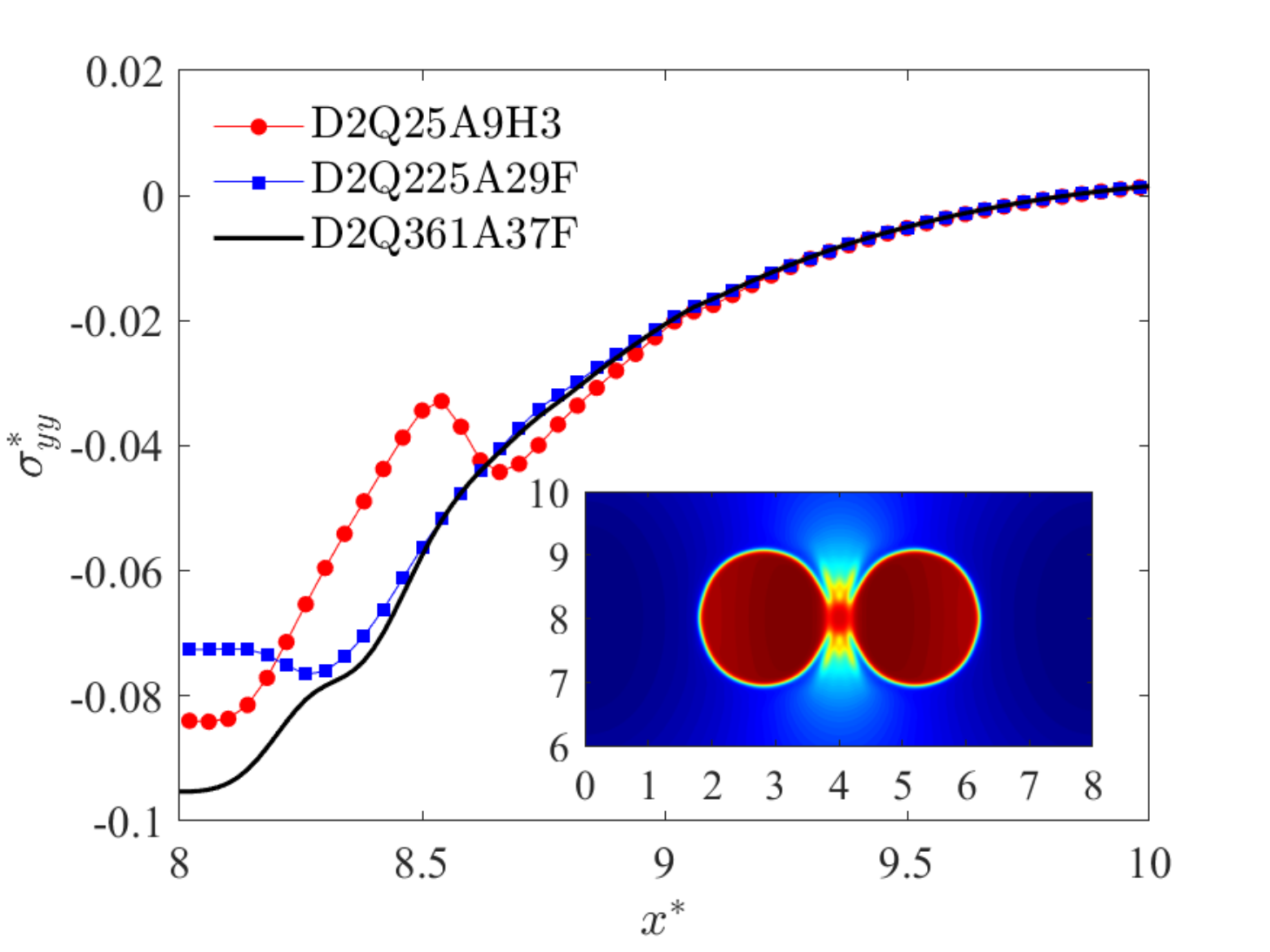}
			\label{sigmayy_evol_3000_a}
		\end{minipage}%
	}\\
	\subfloat[$t^*=0.8$]{
		\begin{minipage}[t]{0.48\linewidth}
			\centering
			\includegraphics[width=1.0\columnwidth,trim={0.0cm 0.18cm 0.0cm 0.2cm},clip]{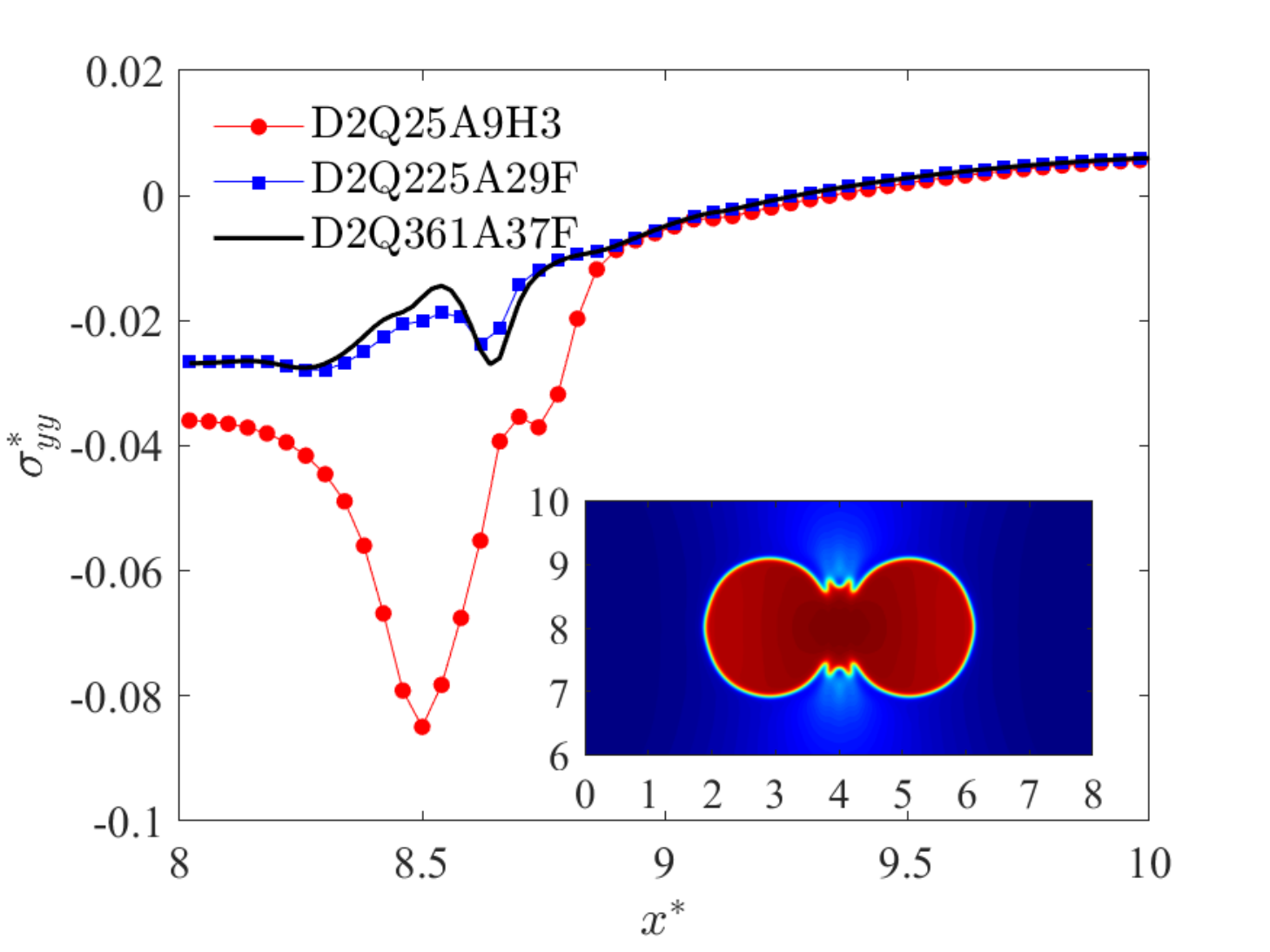}
			\label{sigmayy_evol_4000_a}
		\end{minipage}%
	}
	\subfloat[$t^*=1.0$]{
		\begin{minipage}[t]{0.48\linewidth}
			\centering
			\includegraphics[width=1.0\columnwidth,trim={0.0cm 0.18cm 0.0cm 0.2cm},clip]{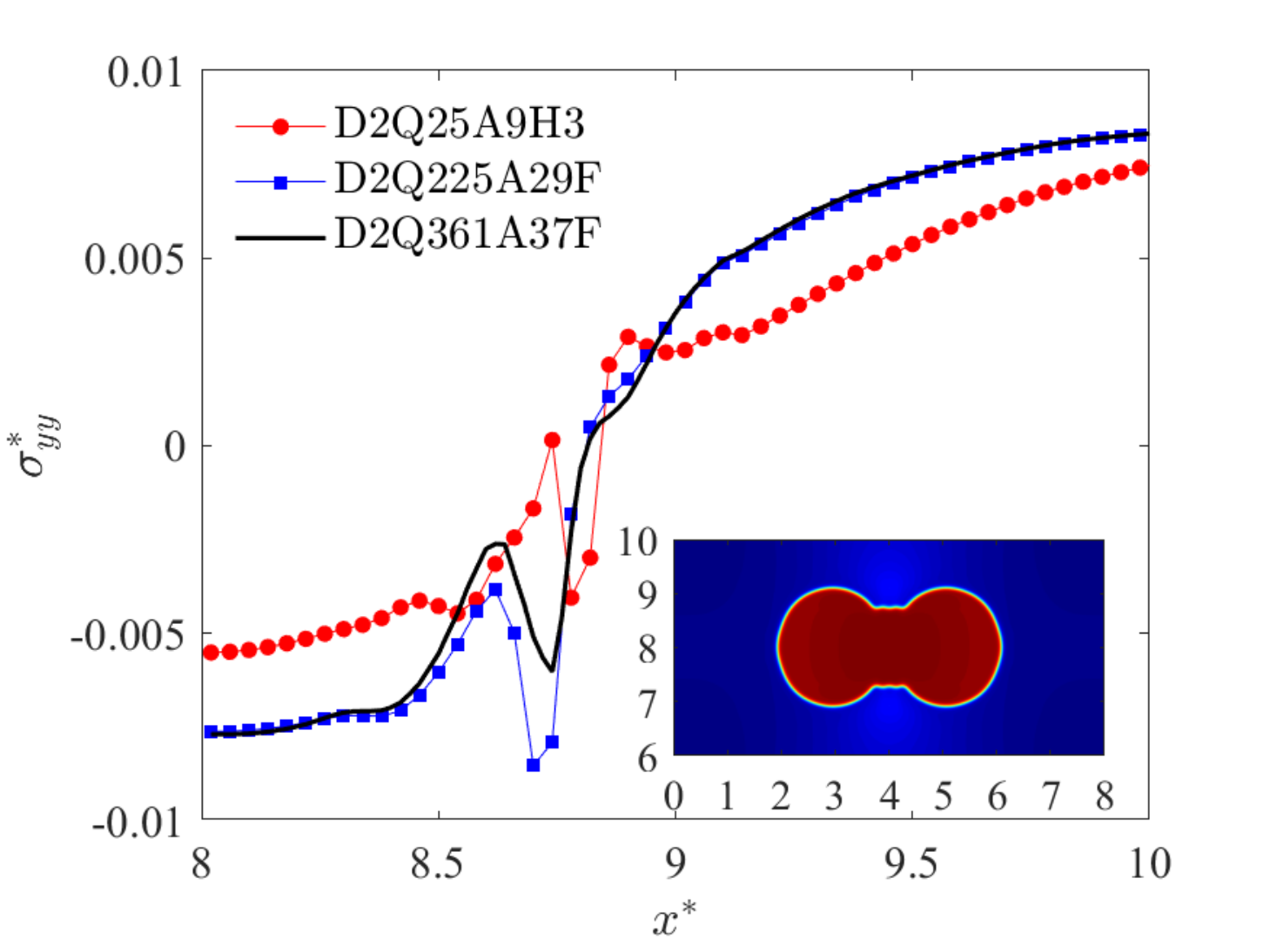}
			\label{sigmayy_evol_5000_a}
		\end{minipage}%
	}%
	\caption{Evolution of the normalized viscous stress $\sigma_{yy}^{*}$ in the $x$-direction. (a) $t^*=0.3$, (b) $t^*=0.4$, (c) $t^*=0.5$, (d) $t^*=0.6$, (e) $t^*=0.8$, (f) $t^*=1.0$. $We=1600$, $Re=6.67$ and $Kn=0.104$.} 
	\label{case2_sigmayy}
\end{figure}

Next, we simulate a case with higher Knudsen number: $We=1600$, $Re=6.67$ and $Kn=0.104$. The ratio of the time step to the relaxation time is  $\Delta{t}/\tau\approx0.0056$ and that of the grid spacing to the mean free path is $\Delta{x}/\lambda\approx0.096$. Figures~\ref{case2_ux} and~\ref{case2_sigmayy} show the time evolutions of the normalized vertical velocity component $u_x^{*}\equiv u_{x}/U$ and the viscous stress component $\sigma_{yy}^{*}\equiv2\sigma_{yy}/(\rho_l+\rho_g)RT$ near the centreline of the interdroplet gas film (the left limiting position $y^*\rightarrow4$).
Owing to the higher Knudsen number, more discrete particle velocities are required to obtain the final convergent solution. It is observed that even the simulations with D2Q225A29F cannot give the final convergent solution during the evolution process. In contrast, D2Q361A37F and D2Q441A41F can capture almost all the higher-order non-continuum effects and therefore generate the satisfying convergent solution. 

\begin{figure}[h!]
	\centering
	\subfloat[D2Q25A9H3]{
		\begin{minipage}[t]{0.48\linewidth}
			\centering
			\includegraphics[width=1.0\columnwidth,trim={0.0cm 1.8cm 0.0cm 2.5cm},clip]{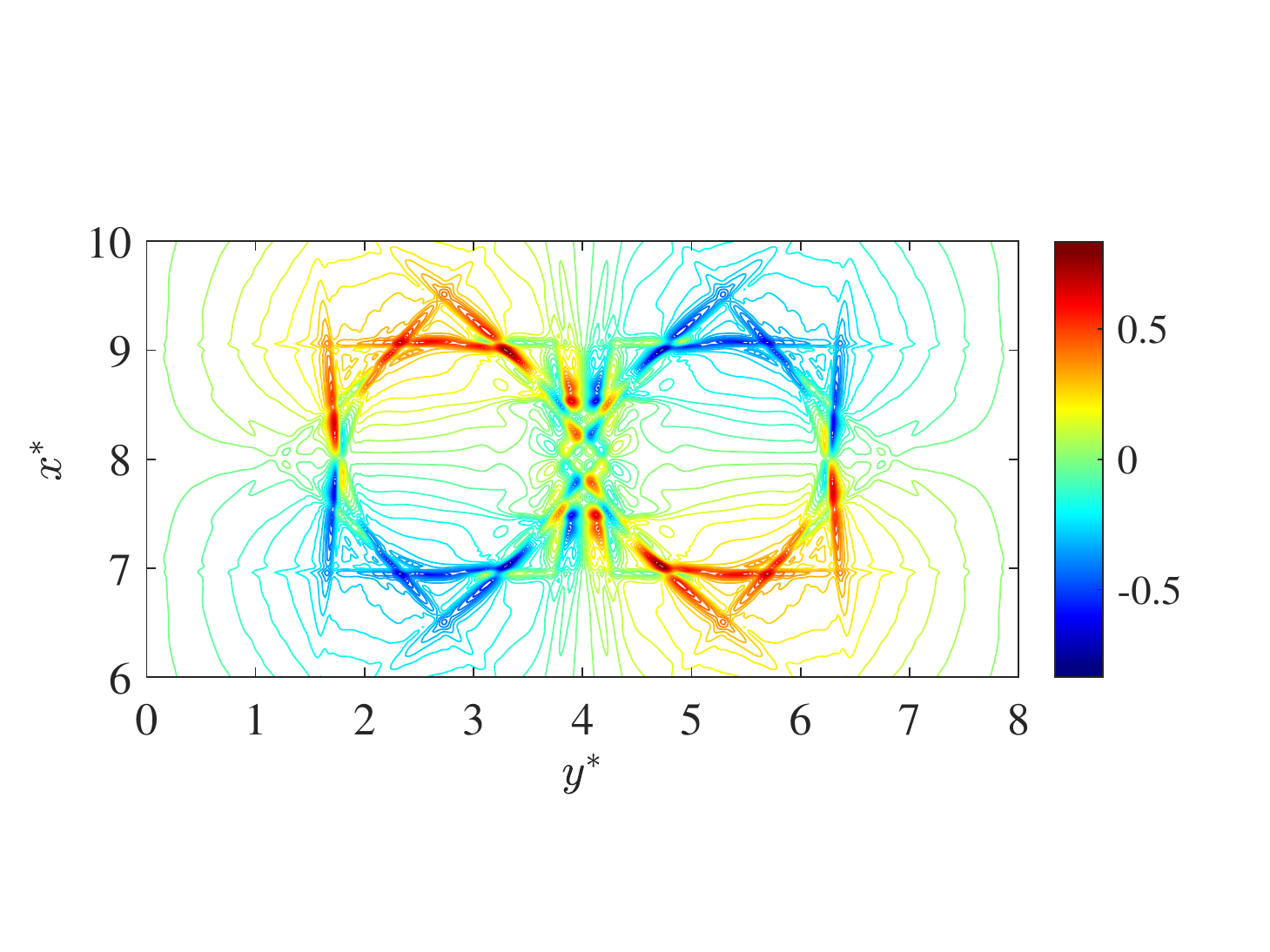}
			\label{contour_oz_D2Q25H3s}
		\end{minipage}%
	}%
%	\subfloat[D2Q289A33F]{
%		\begin{minipage}[t]{0.48\linewidth}
%			\centering
%			\includegraphics[width=1.0\columnwidth,trim={0.0cm 1.8cm 0.0cm 2.5cm},clip]{contour_oz_D2Q289Fs.eps}
%			\label{contour_oz_D2Q289Fs}
%		\end{minipage}%
%	}%	
	\subfloat[D2Q361A37F]{
		\begin{minipage}[t]{0.48\linewidth}
			\centering
			\includegraphics[width=1.0\columnwidth,trim={0.0cm 1.8cm 0.0cm 2.5cm},clip]{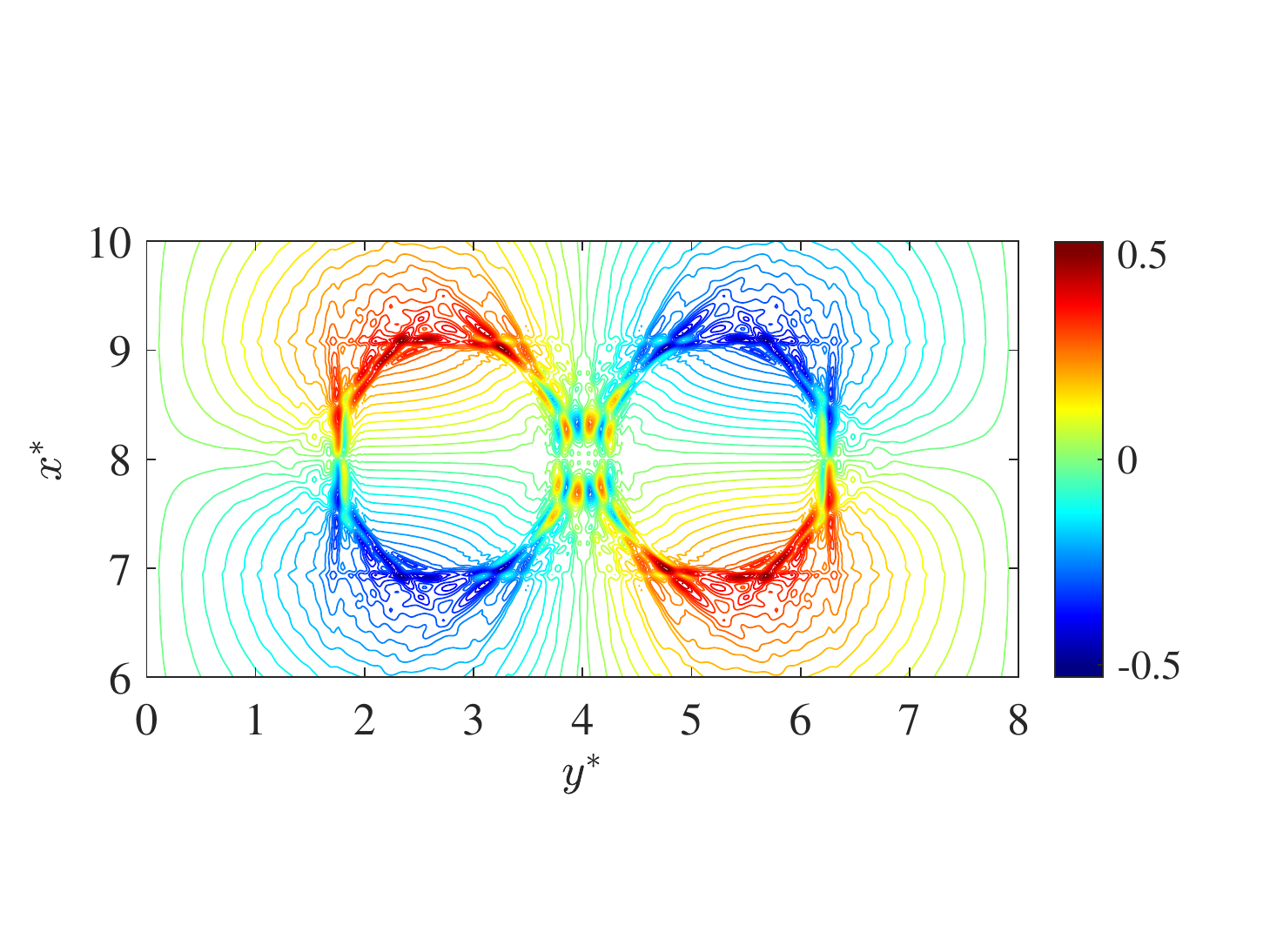}
			\label{contour_oz_D2Q361Fs}
		\end{minipage}%
	}%
%	\subfloat[D2Q441A41F]{
%		\begin{minipage}[t]{0.48\linewidth}
%			\centering
%			\includegraphics[width=1.0\columnwidth,trim={0.0cm 1.8cm 0.0cm 2.5cm},clip]{contour_oz_D2Q441Fs.eps}
%			\label{contour_oz_D2Q441Fs}
%		\end{minipage}%
%	}%
	\caption{Contour of vorticity $\omega^*$ at $t^*=0.6$. (a) D2Q25A9H3 and (b) D2Q361A37F. $We=1600$, $Re=6.67$ and $Kn=0.104$.} 
	\label{contour_omega_high_Kn}
\end{figure}
\begin{figure}[h!]
	\centering
	\subfloat[$t^*=0.5$]{
		\begin{minipage}[t]{0.5\linewidth}
			\centering
			\includegraphics[width=1.0\columnwidth,trim={0.0cm 0.8cm 0.0cm 1.6cm},clip]{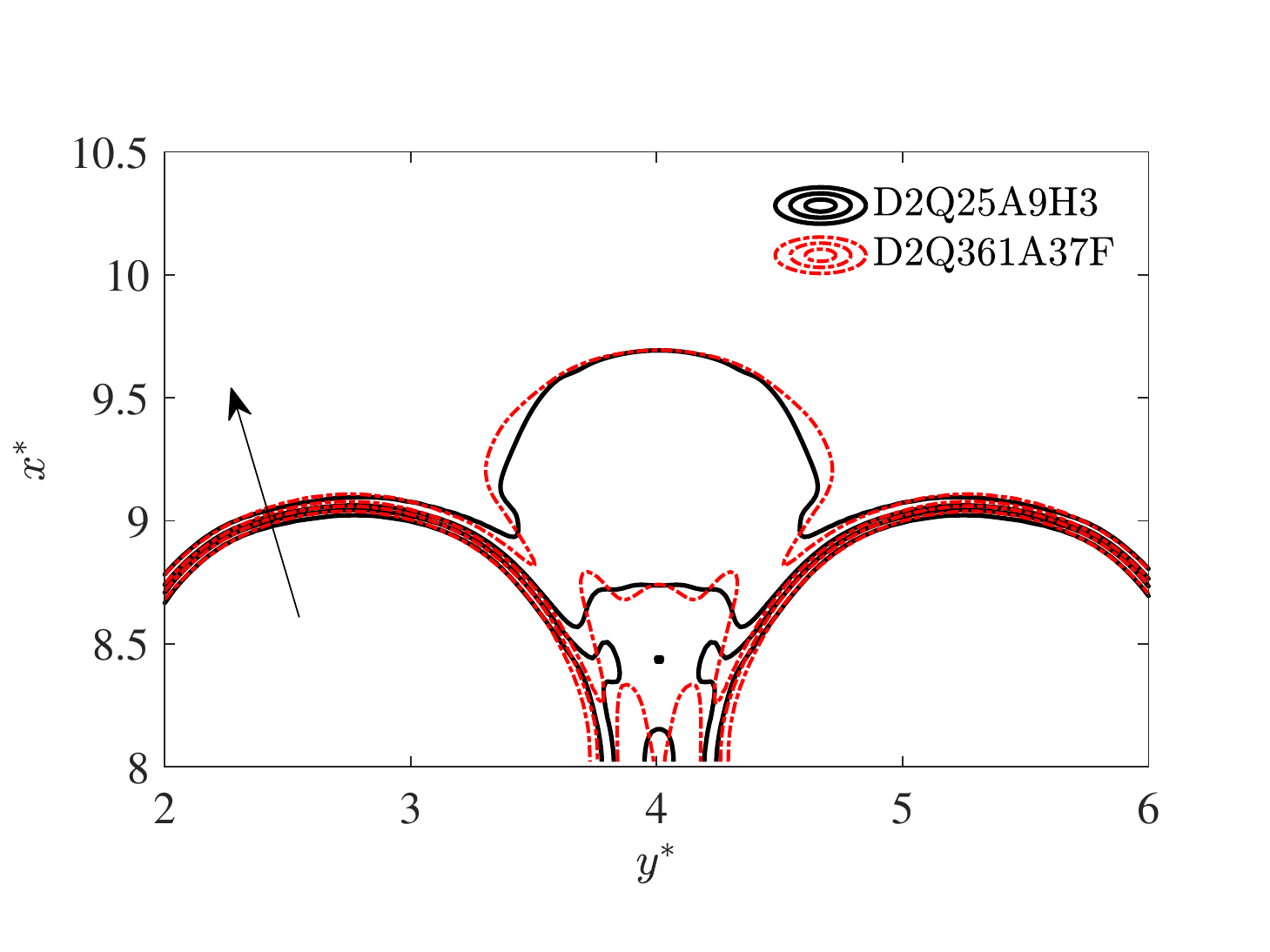}
			\label{droplet_case2_t2500}
		\end{minipage}%
	}%
	\subfloat[$t^*=0.6$]{
		\begin{minipage}[t]{0.5\linewidth}
			\centering
			\includegraphics[width=1.0\columnwidth,trim={0.0cm 0.8cm 0.0cm 1.6cm},clip]{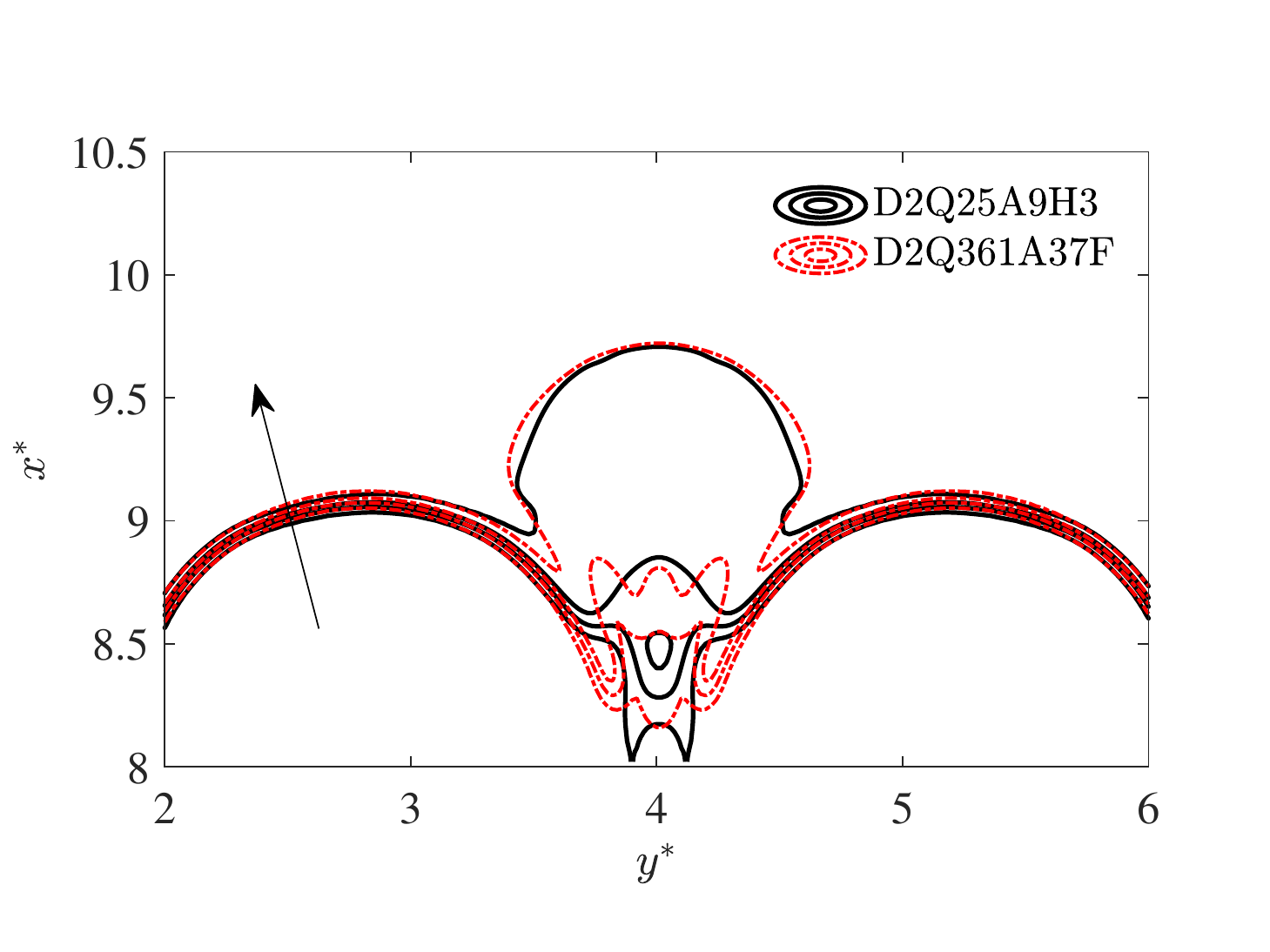}
			\label{droplet_case2_t3000}
		\end{minipage}%
	}%	
	\caption{Comparison of four dimensionless density profiles in the interfacial region of the droplets. Black: D2Q25A9H3, red: D2Q361A37F. The density decreases in the direction indicated by the arrow, where $\rho^*=0.75$, $0.90$, $1.05$ and $1.20$. (a) $t^*=0.5$ and (b) $t^*=0.6$. $We=1600$, $Re=6.67$ and $Kn=0.104$.} 
	\label{case2_xx5}
\end{figure}
\begin{figure}[h!]
	\centering
	\subfloat[$t^*=0.3$]{
		\begin{minipage}[t]{0.33\linewidth}
			\centering
			\includegraphics[width=1.0\columnwidth,trim={0.0cm 0.15cm 0.0cm 0.2cm},clip]{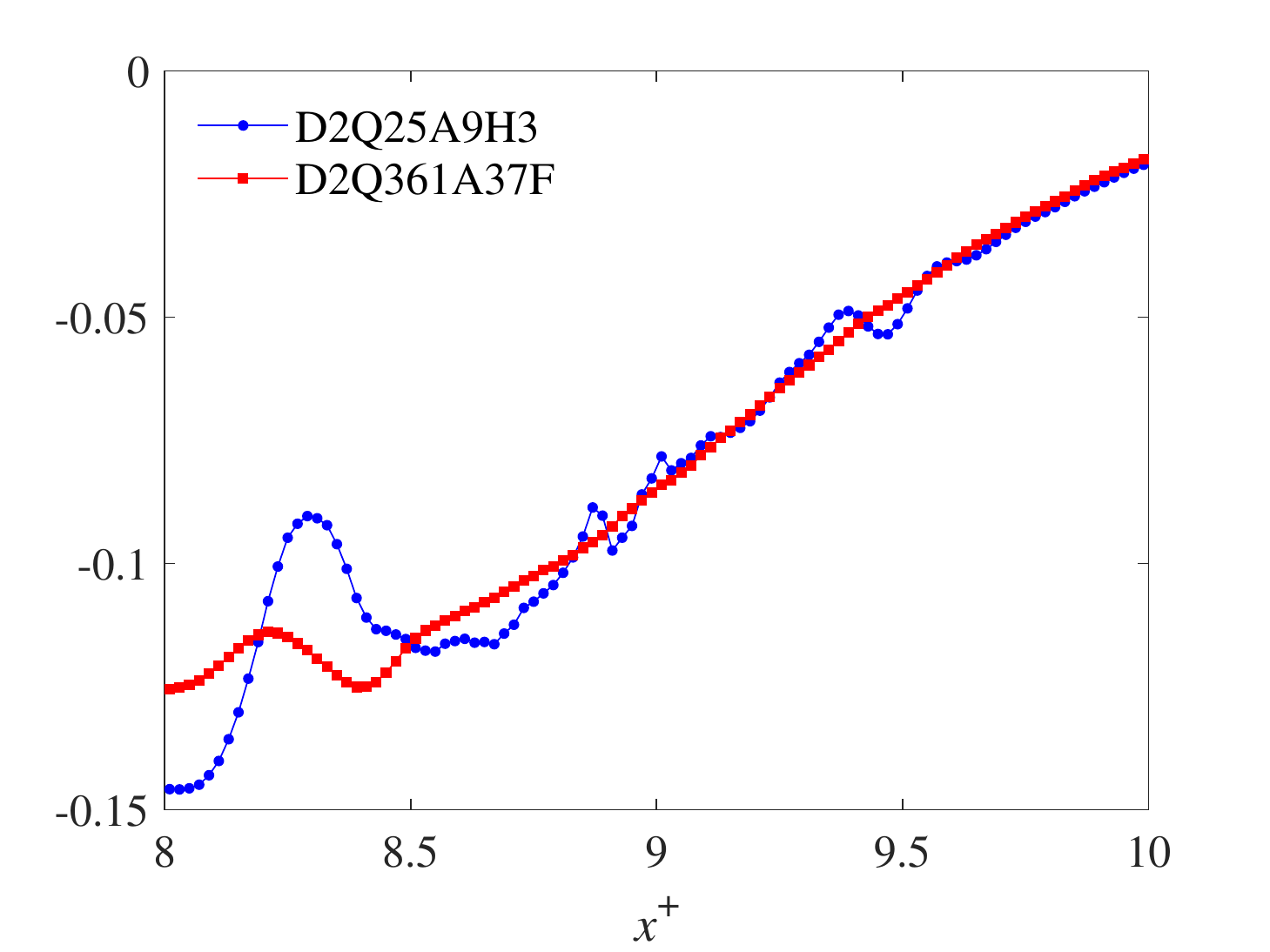}
			\label{ptheta1500}
		\end{minipage}%
	}%
	\subfloat[$t^*=0.4$]{
		\begin{minipage}[t]{0.33\linewidth}
			\centering
			\includegraphics[width=1.0\columnwidth,trim={0.0cm 0.15cm 0.0cm 0.2cm},clip]{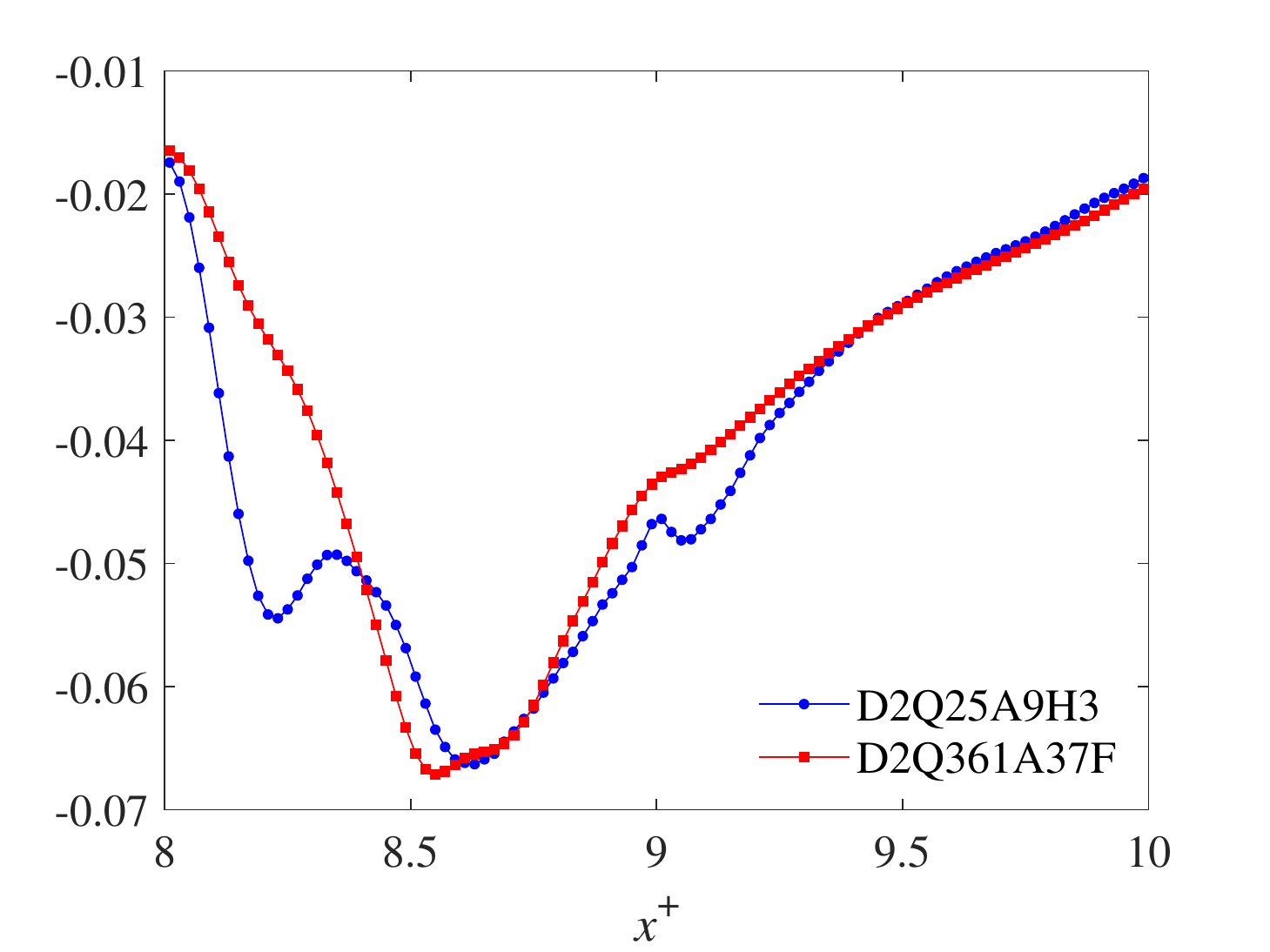}
			\label{ptheta2000}
		\end{minipage}%
	}%
	\subfloat[$t^*=0.5$]{
		\begin{minipage}[t]{0.33\linewidth}
			\centering
			\includegraphics[width=1.0\columnwidth,trim={0.0cm 0.15cm 0.0cm 0.2cm},clip]{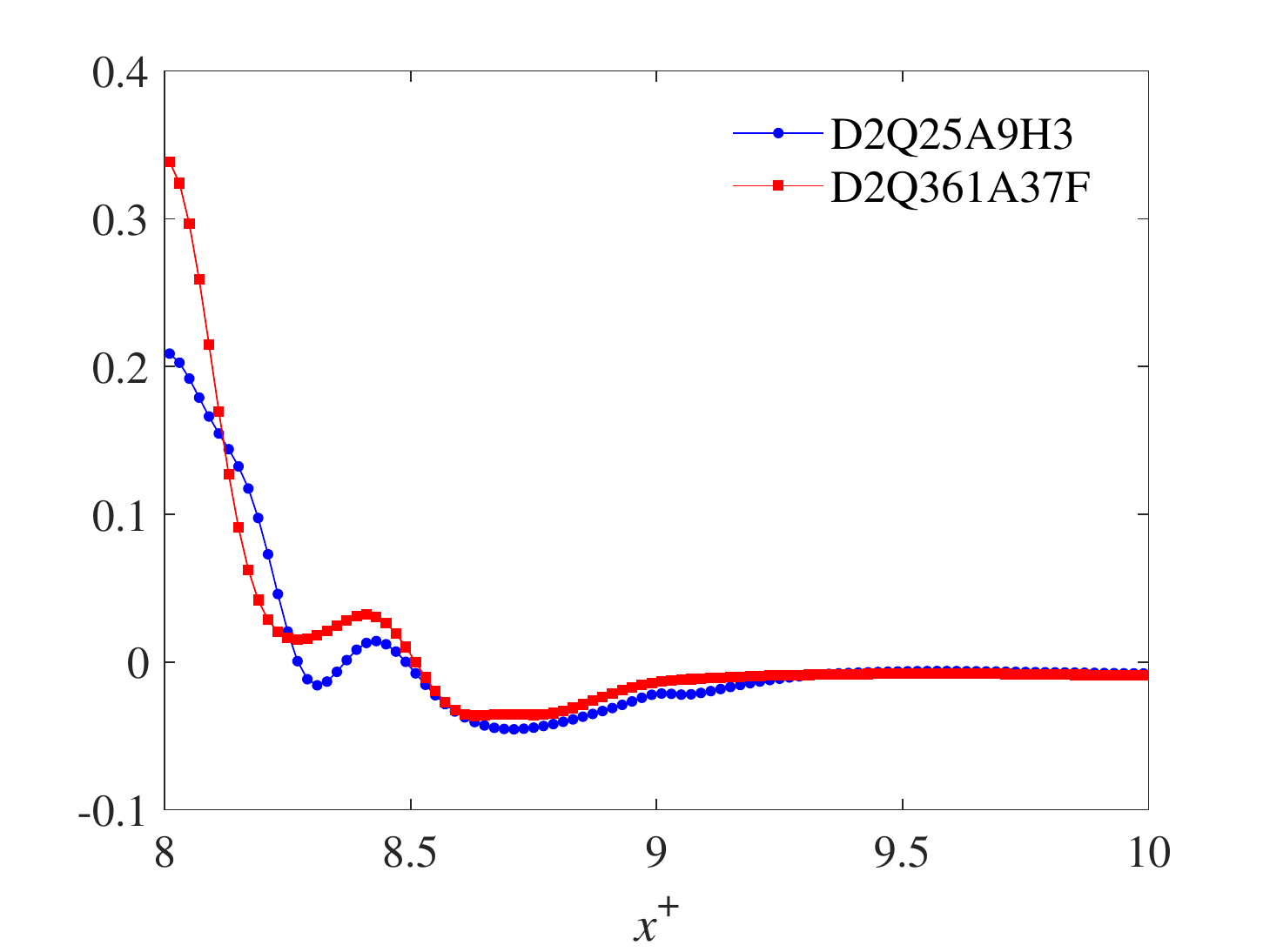}
			\label{ptheta2500}
		\end{minipage}%
	}%
	
	\subfloat[$t^*=0.6$]{
		\begin{minipage}[t]{0.33\linewidth}
			\centering
			\includegraphics[width=1.0\columnwidth,trim={0.0cm 0.15cm 0.0cm 0.2cm},clip]{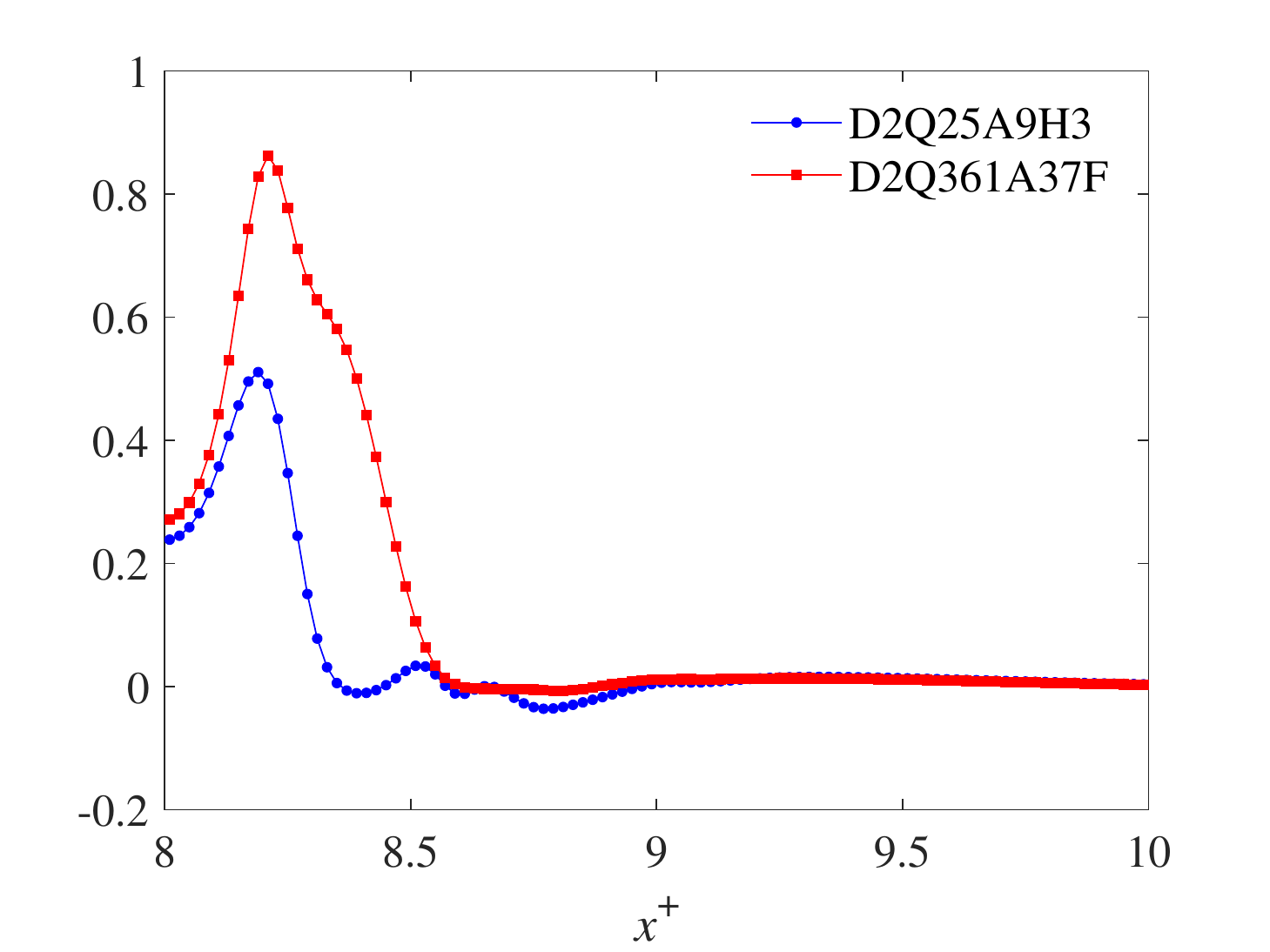}
			\label{ptheta3000}
		\end{minipage}%
	}%
	\subfloat[$t^*=0.8$]{
		\begin{minipage}[t]{0.33\linewidth}
			\centering
			\includegraphics[width=1.0\columnwidth,trim={0.0cm 0.15cm 0.0cm 0.2cm},clip]{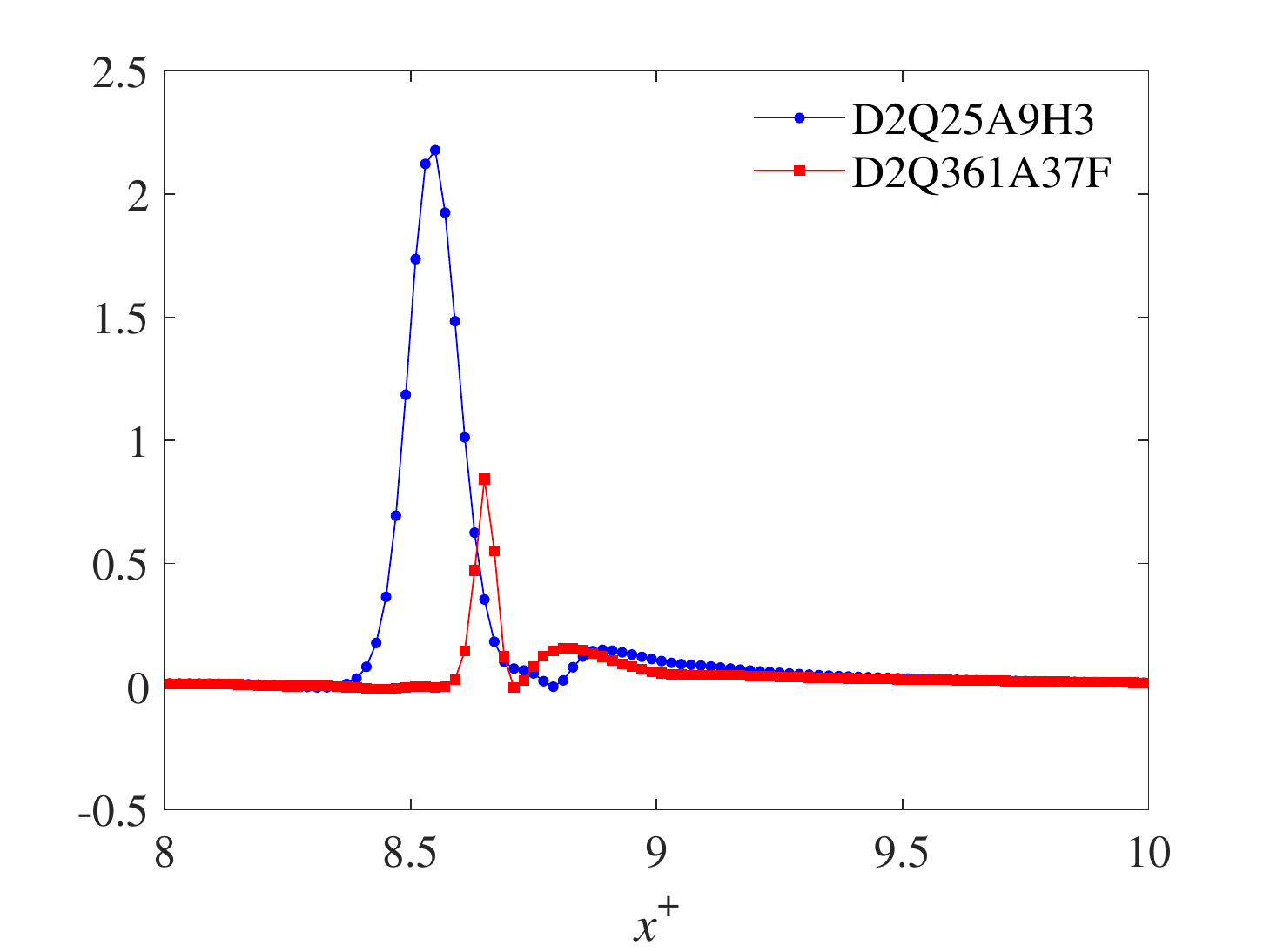}
			\label{ptheta4000}
		\end{minipage}%
	}%
	\subfloat[$t^*=1.0$]{
		\begin{minipage}[t]{0.33\linewidth}
			\centering
			\includegraphics[width=1.0\columnwidth,trim={0.0cm 0.15cm 0.0cm 0.2cm},clip]{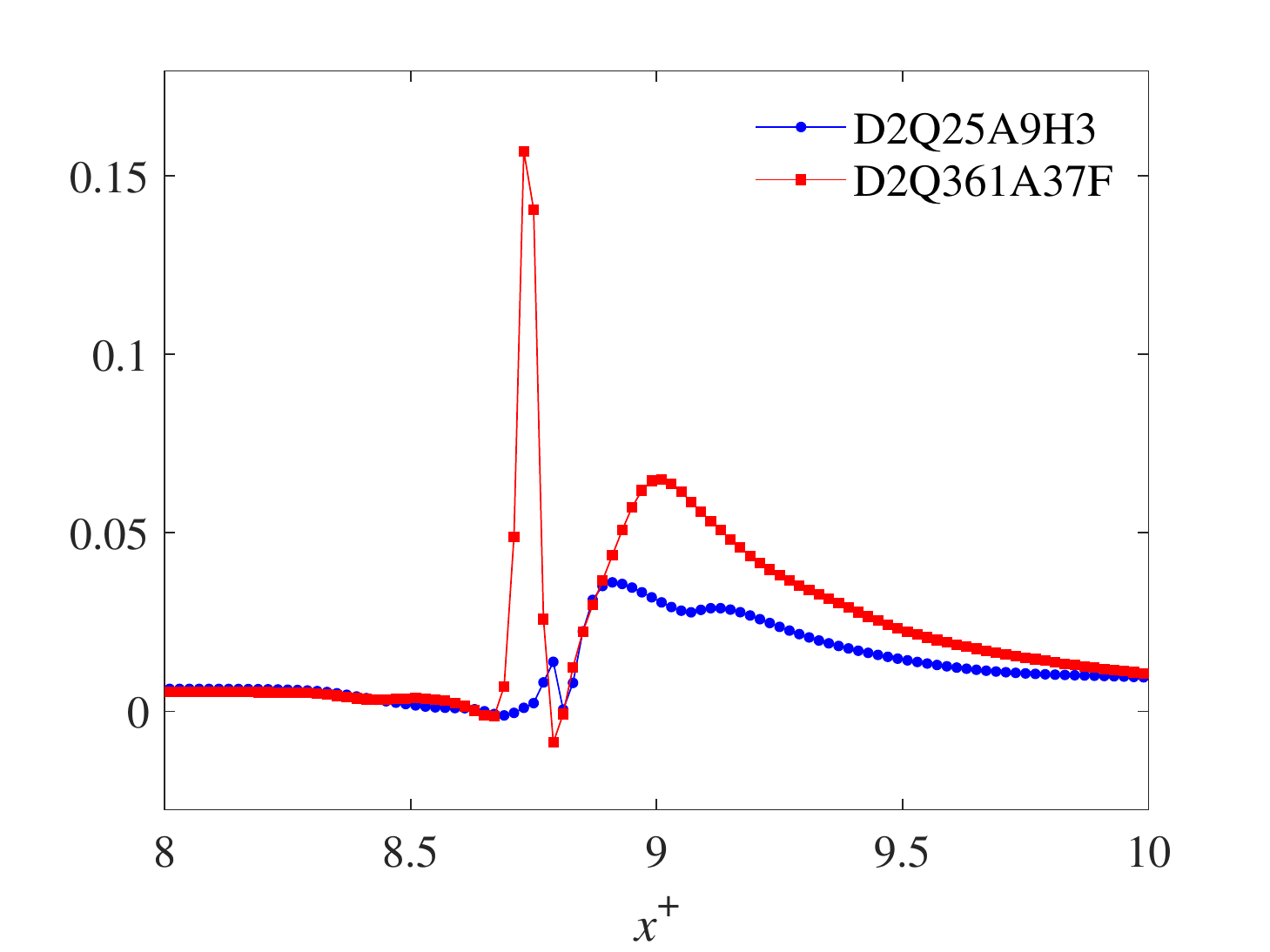}
			\label{ptheta5000}
		\end{minipage}%
	}%
	\caption{Evolution of $T_{1}^*$ at $y^*=4$. (a) $t^*=0.3$, (b) $t^*=0.4$, (c) $t^*=0.5$, (d) $t^*=0.6$, (e) $t^*=0.8$, (f) $t^*=1.0$. $We=1600$, $Re=6.67$ and $Kn=0.104$.} 
	\label{ptheta}
\end{figure}
\begin{figure}[h!]
	\centering
	\subfloat[$t^*=0.3$]{
		\begin{minipage}[t]{0.33\linewidth}
			\centering
			\includegraphics[width=1.0\columnwidth,trim={0.0cm 0.15cm 0.0cm 0.15cm},clip]{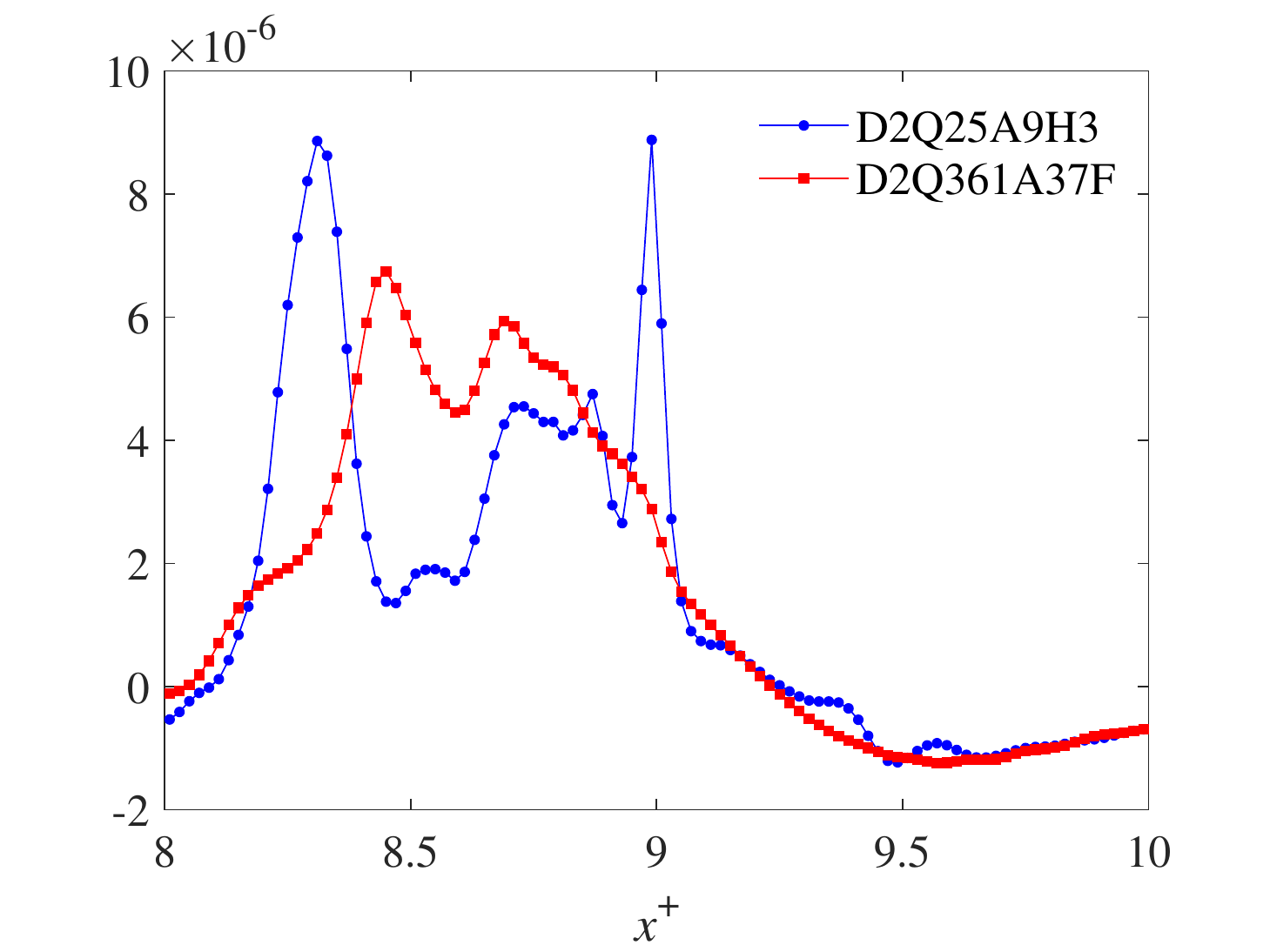}
			\label{KNNS1500}
		\end{minipage}%
	}%
	\subfloat[$t^*=0.4$]{
		\begin{minipage}[t]{0.33\linewidth}
			\centering
			\includegraphics[width=1.0\columnwidth,trim={0.0cm 0.15cm 0.0cm 0.15cm},clip]{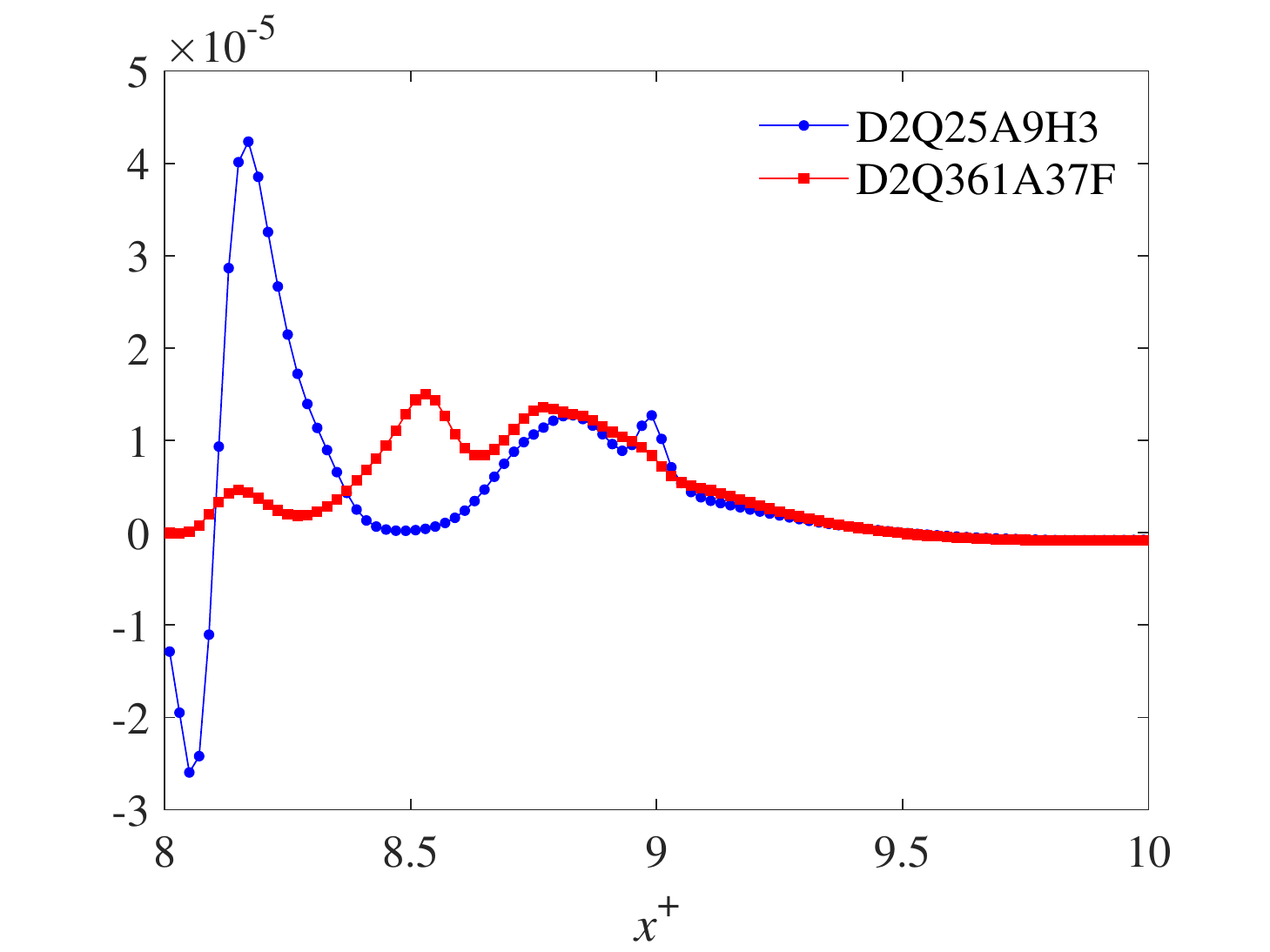}
			\label{KNNS2000}
		\end{minipage}%
	}%
	\subfloat[$t^*=0.5$]{
		\begin{minipage}[t]{0.33\linewidth}
			\centering
			\includegraphics[width=1.0\columnwidth,trim={0.0cm 0.15cm 0.0cm 0.15cm},clip]{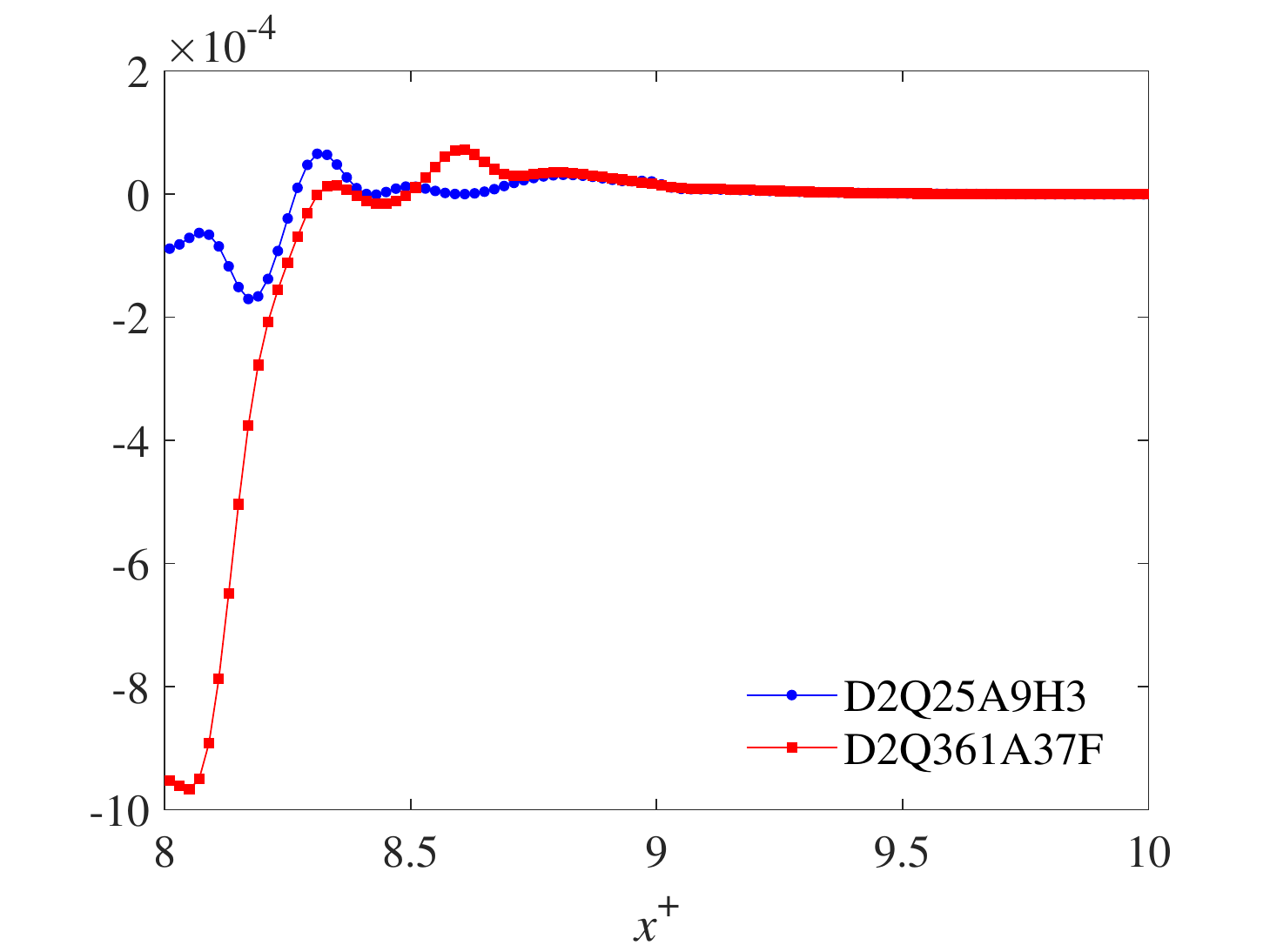}
			\label{KNNS2500}
		\end{minipage}%
	}%
	
	\subfloat[$t^*=0.6$]{
		\begin{minipage}[t]{0.33\linewidth}
			\centering
			\includegraphics[width=1.0\columnwidth,trim={0.0cm 0.15cm 0.0cm 0.15cm},clip]{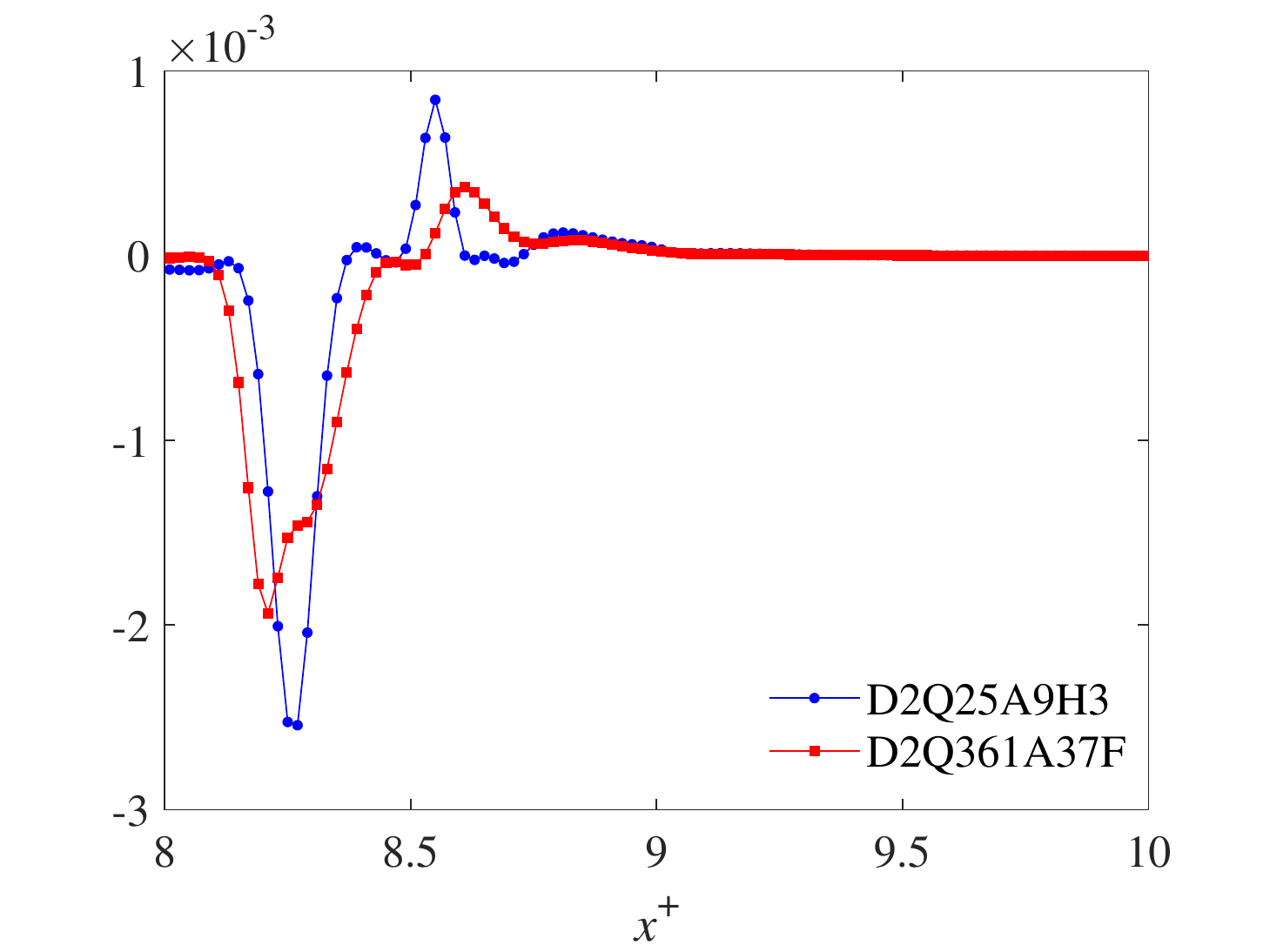}
			\label{KNNS3000}
		\end{minipage}%
	}%	
	\subfloat[$t^*=0.8$]{
		\begin{minipage}[t]{0.33\linewidth}
			\centering
			\includegraphics[width=1.0\columnwidth,trim={0.0cm 0.15cm 0.0cm 0.15cm},clip]{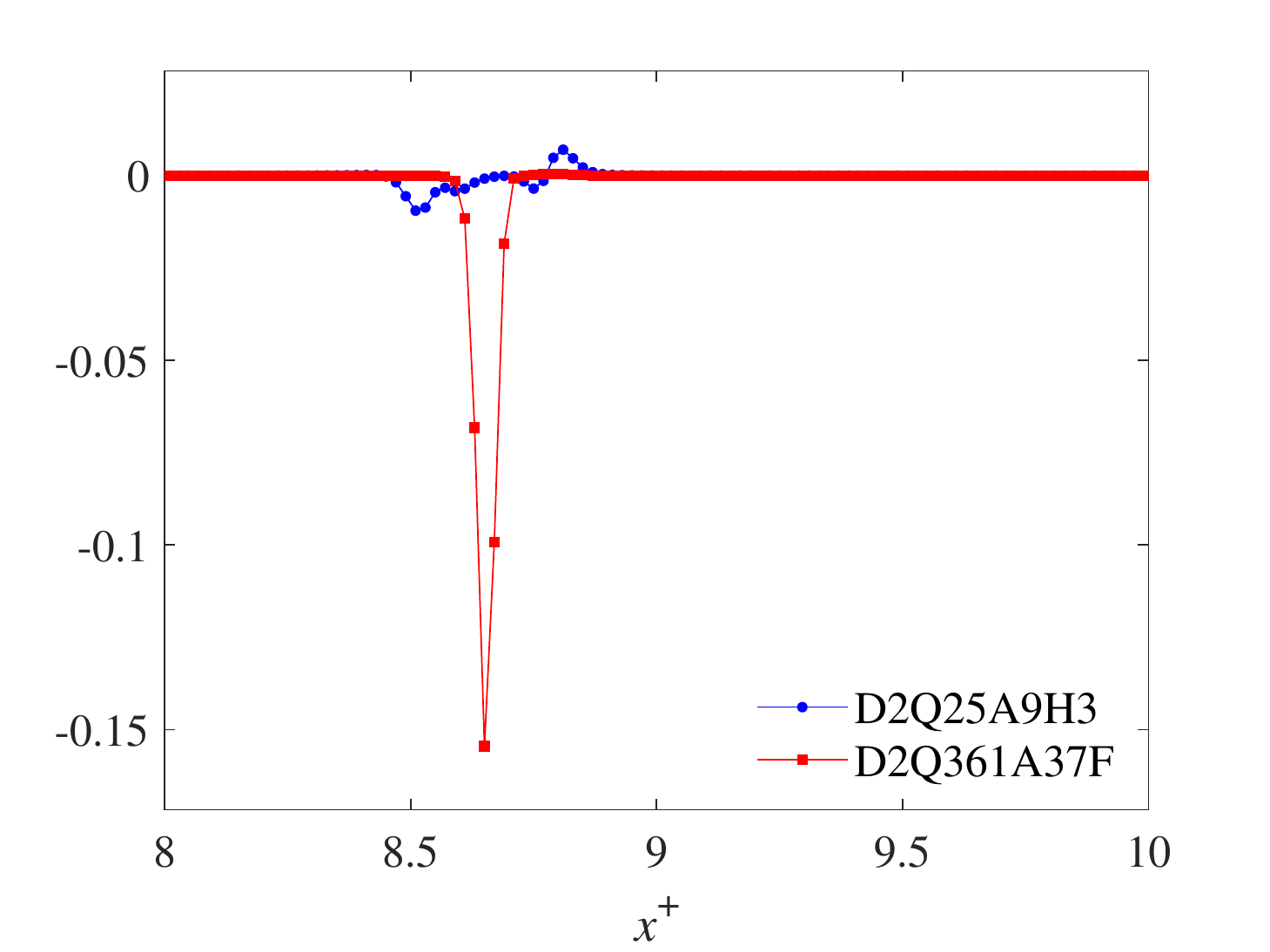}
			\label{KNNS4000}
		\end{minipage}%
	}%
	\subfloat[$t^*=1.0$]{
		\begin{minipage}[t]{0.33\linewidth}
			\centering
			\includegraphics[width=1.0\columnwidth,trim={0.0cm 0.15cm 0.0cm 0.15cm},clip]{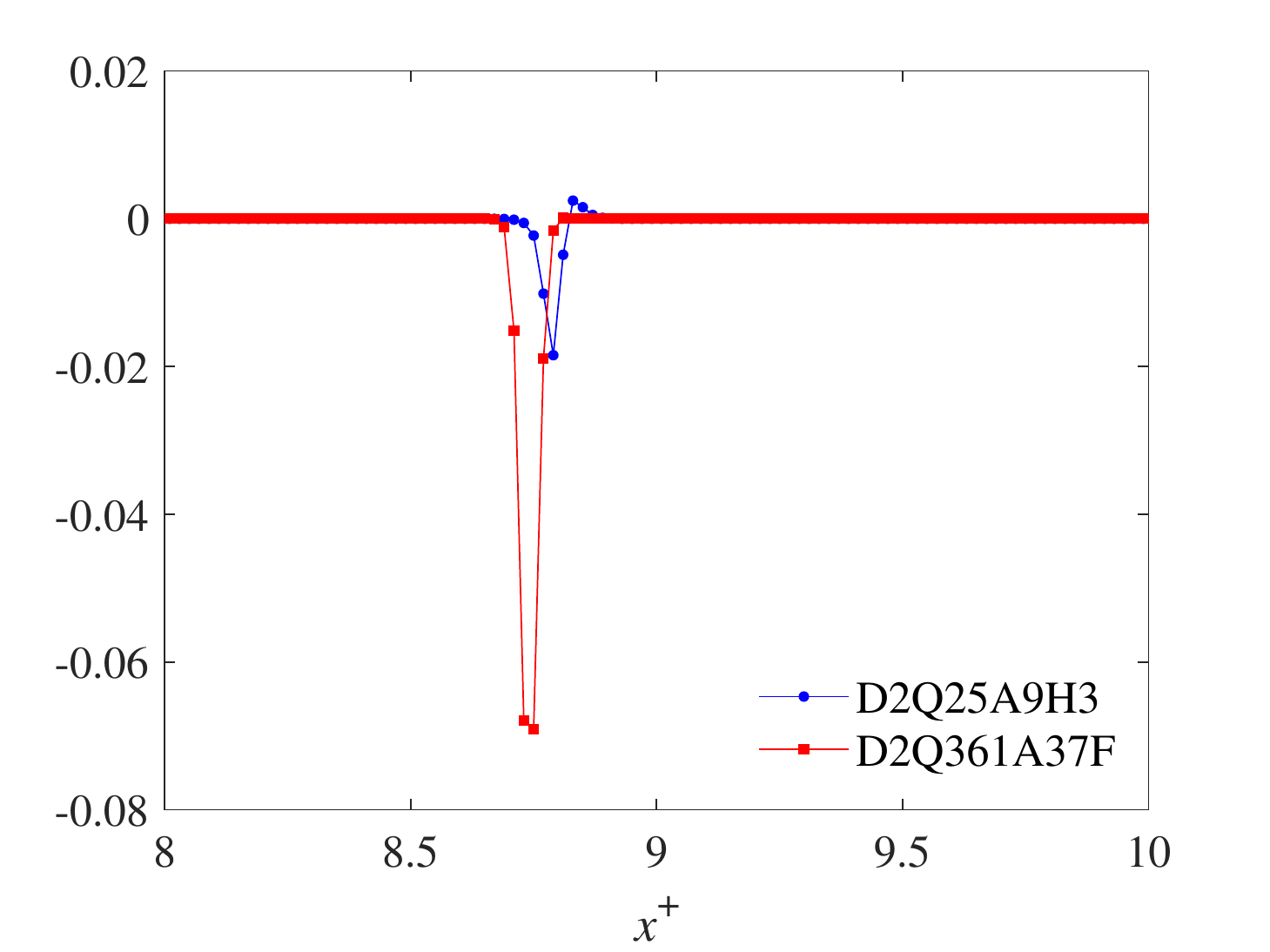}
			\label{KNNS5000}
		\end{minipage}%
	}%
	\caption{Evolution of $T_{2}^*$ at $y^*=4$. (a) $t^*=0.3$, (b) $t^*=0.4$, (c) $t^*=0.5$, (d) $t^*=0.6$, (e) $t^*=0.8$, (f) $t^*=1.0$. $We=1600$, $Re=6.67$ and $Kn=0.104$.} 
	\label{KNNS}
\end{figure}

\begin{figure}[h!]
	\centering
	\subfloat[$t^*=0.3$]{
		\begin{minipage}[t]{0.33\linewidth}
			\centering
			\includegraphics[width=1.0\columnwidth,trim={0.0cm 0.15cm 0.0cm 0.2cm},clip]{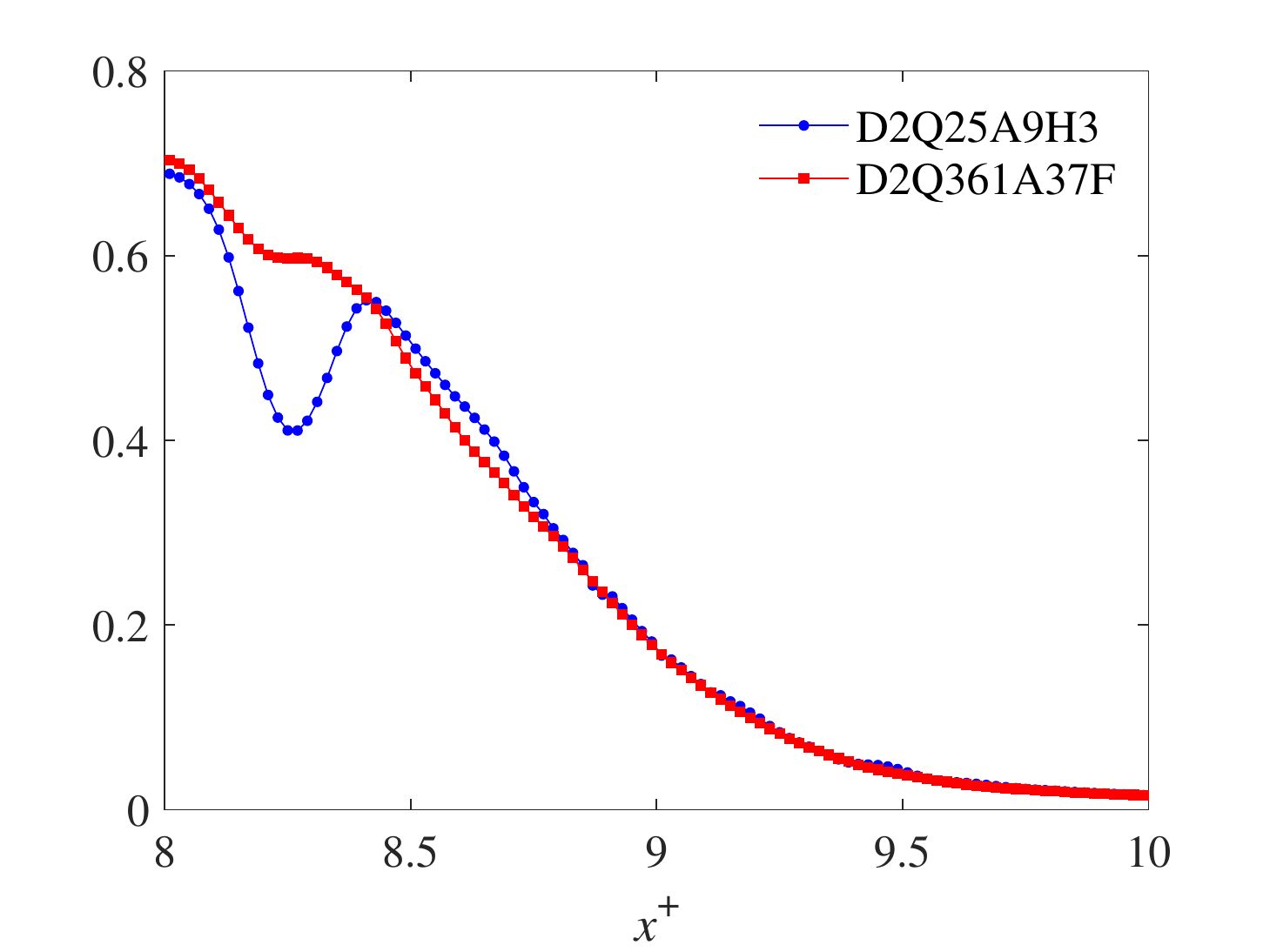}
			\label{disp1500}
		\end{minipage}%
	}%
	\subfloat[$t^*=0.4$]{
		\begin{minipage}[t]{0.33\linewidth}
			\centering
			\includegraphics[width=1.0\columnwidth,trim={0.0cm 0.15cm 0.0cm 0.2cm},clip]{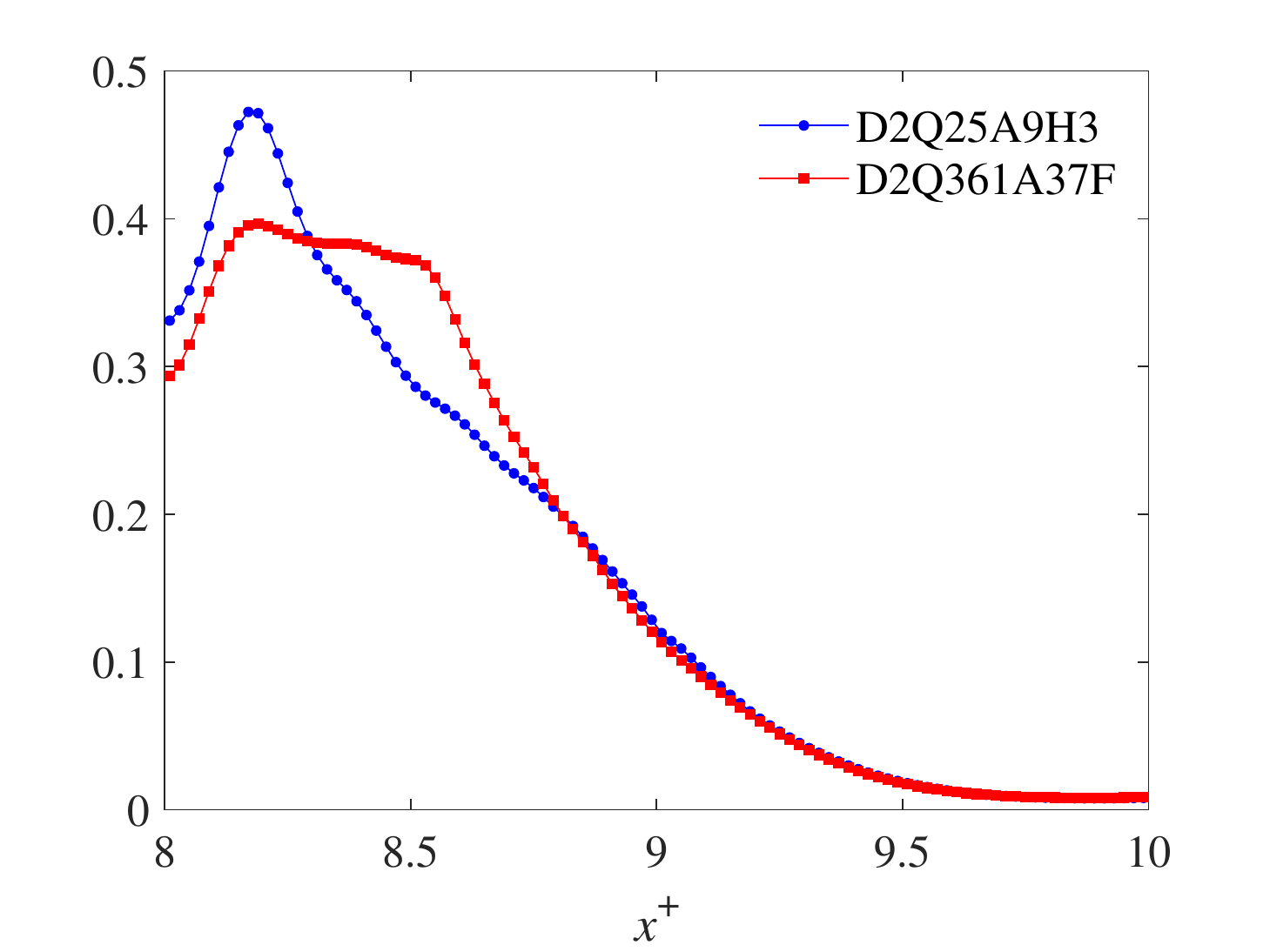}
			\label{disp2000}
		\end{minipage}%
	}%
	\subfloat[$t^*=0.5$]{
		\begin{minipage}[t]{0.33\linewidth}
			\centering
			\includegraphics[width=1.0\columnwidth,trim={0.0cm 0.15cm 0.0cm 0.2cm},clip]{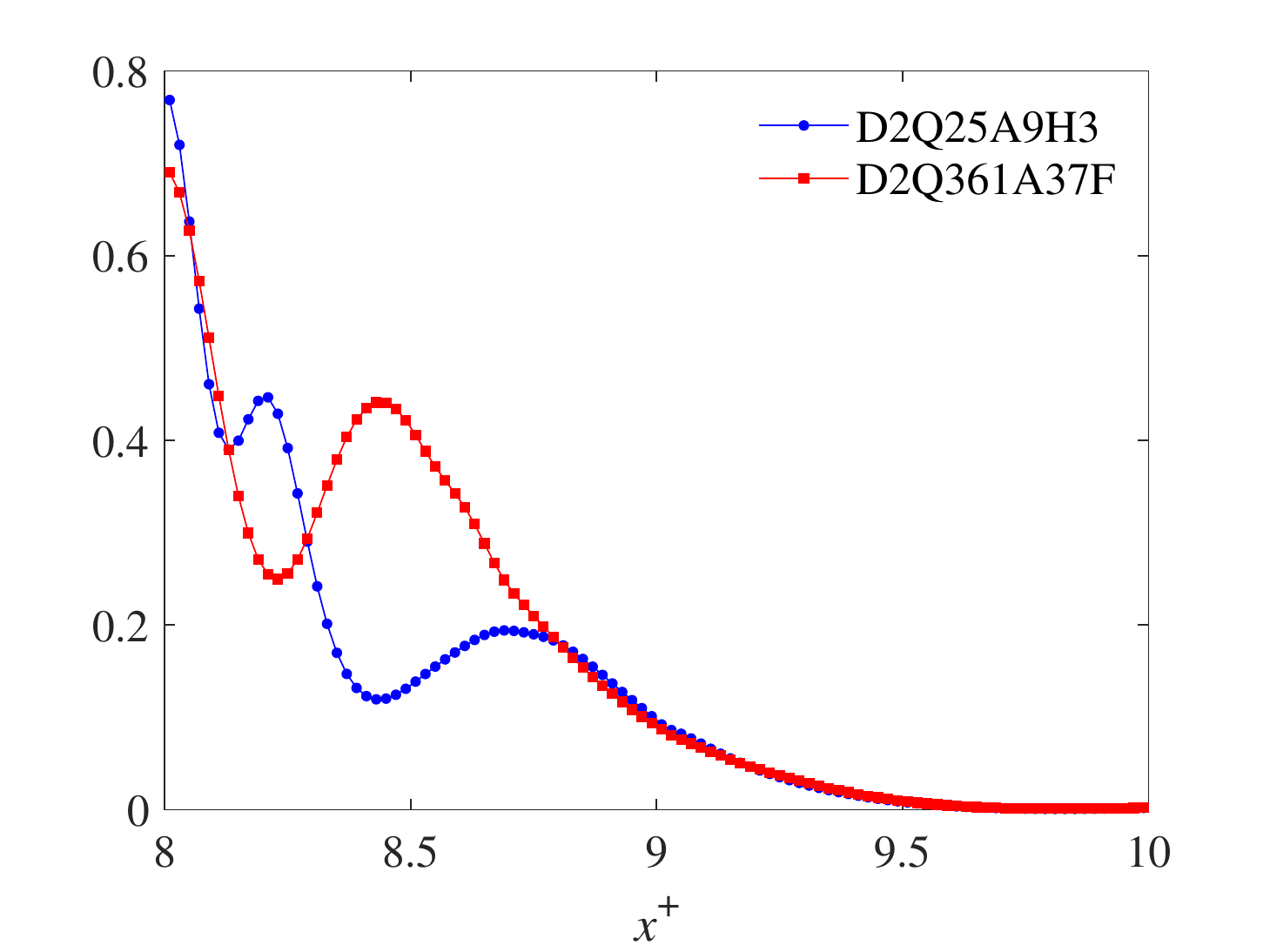}
			\label{disp2500}
		\end{minipage}%
	}%
	
	\subfloat[$t^*=0.6$]{
		\begin{minipage}[t]{0.33\linewidth}
			\centering
			\includegraphics[width=1.0\columnwidth,trim={0.0cm 0.15cm 0.0cm 0.2cm},clip]{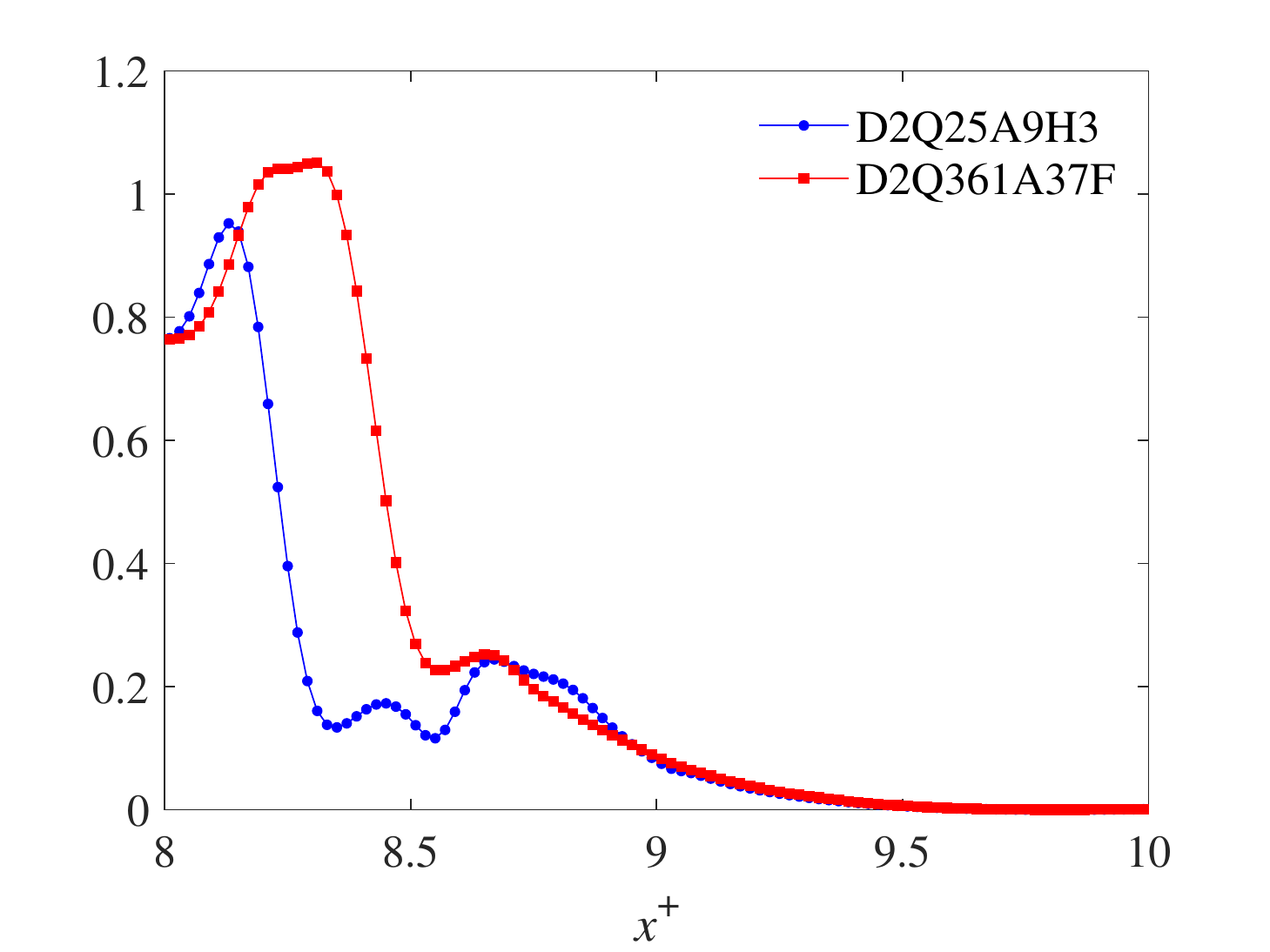}
			\label{disp3000}
		\end{minipage}%
	}%	
	\subfloat[$t^*=0.8$]{
		\begin{minipage}[t]{0.33\linewidth}
			\centering
			\includegraphics[width=1.0\columnwidth,trim={0.0cm 0.15cm 0.0cm 0.2cm},clip]{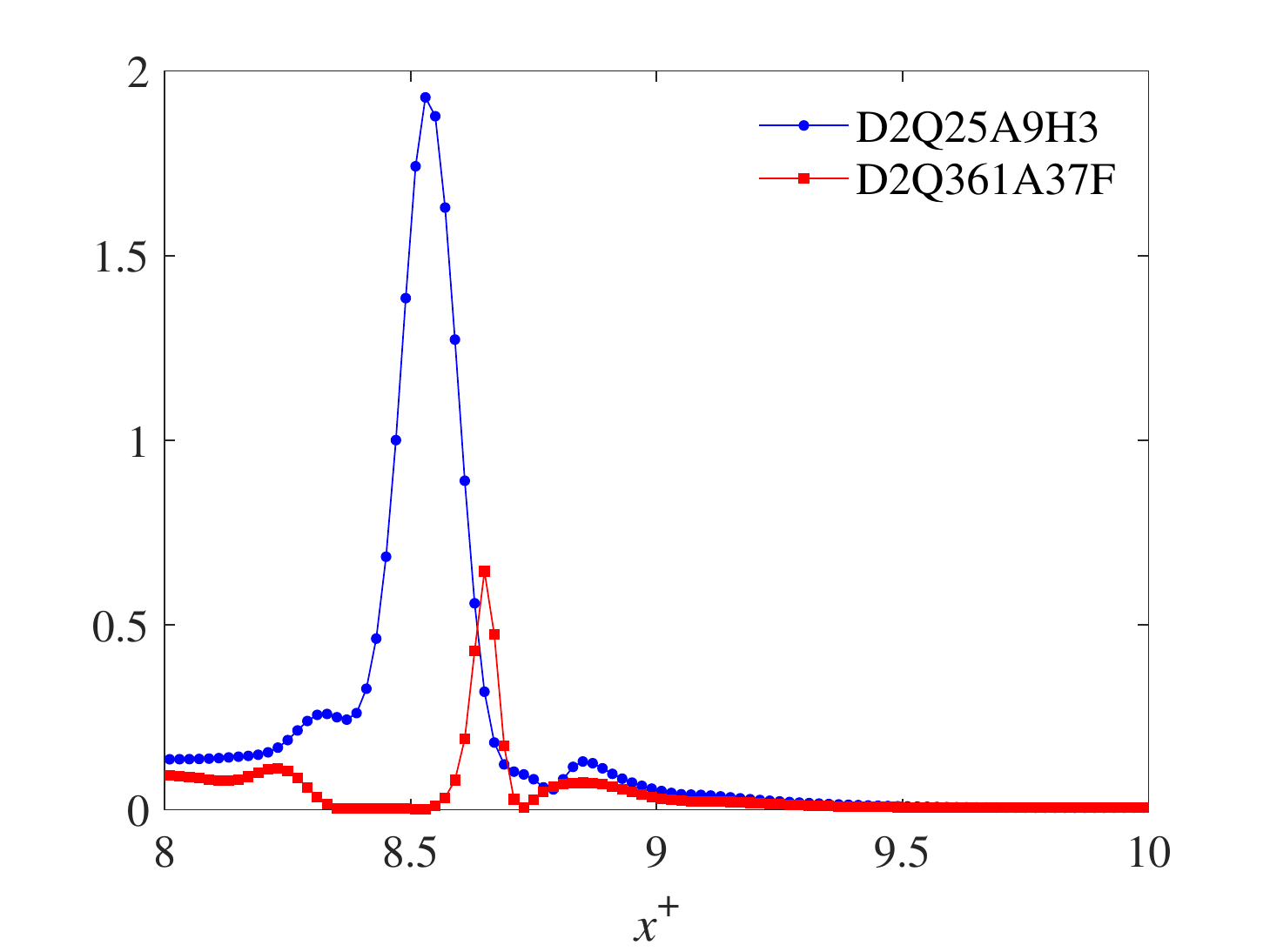}
			\label{disp4000}
		\end{minipage}%
	}%
	\subfloat[$t^*=1.0$]{
		\begin{minipage}[t]{0.33\linewidth}
			\centering
			\includegraphics[width=1.0\columnwidth,trim={0.0cm 0.15cm 0.0cm 0.2cm},clip]{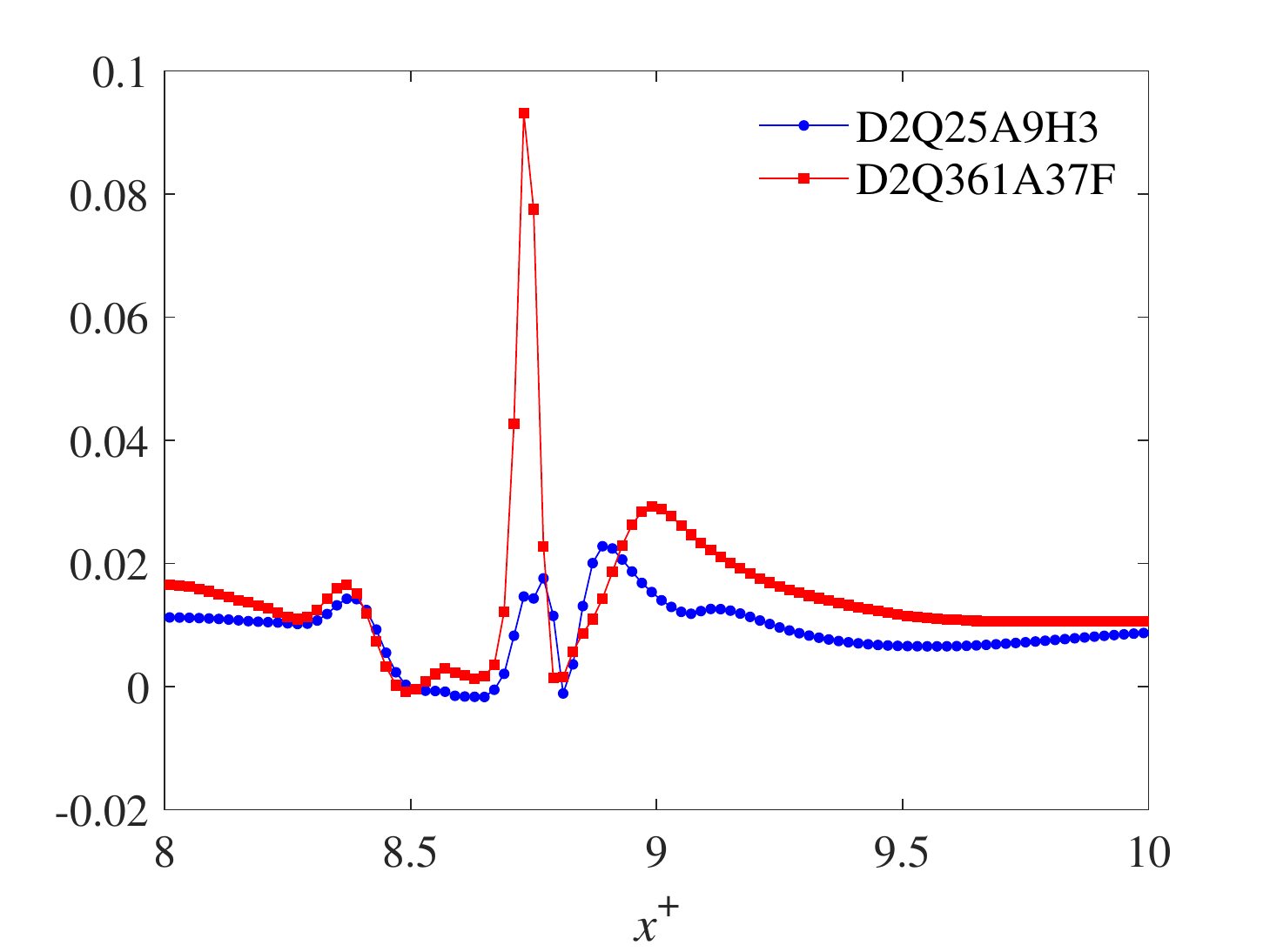}
			\label{disp5000}
		\end{minipage}%
	}%
	\caption{Evolution of $T_{3}^{*}$ at $y^*=4$. (a) $t^*=0.3$, (b) $t^*=0.4$, (c) $t^*=0.5$, (d) $t^*=0.6$, (e) $t^*=0.8$, (f) $t^*=1.0$. $We=1600$, $Re=6.67$ and $Kn=0.104$.} 
	\label{sigmaS}
\end{figure}
At $t^*=0.3$ and $t^*=0.4$, different Gauss-Hermite quadratures give similar results since the interdroplet gas film is only slightly compressed. As the droplets approach each other ($t^*=0.5$, $0.6$ and $0.8$), the lubrication layer between the droplets gradually forms until the emergence of the liquid bridge. It is clearly observed that the discrepancies between the results obtained from D2Q25A9H3 and those from other higher-order Gauss-Hermite quadratures become pronounced, particularly in the region from $x^*=8$ to $x^*=9$ (about one droplet radius). At $t^*=0.8$, a highly negative peak region is observed for D2Q25A9H3, which is suppressed due to the rarefaction effect. At $t^*=1.0$, a liquid bridge connecting the two droplets is formed. The discrepancies caused by the rarefaction effect can still be observed because the gas at the flank of the liquid bridge is still being squeezed out due to the movement of the droplets.

Figure~\ref{contour_omega_high_Kn} shows the normalized snapshots of the vorticity $\omega^*$ at $t^*=0.6$. Compared to the case where $Kn=0.052$, the vorticity distribution is significantly changed by the rarefaction effect. On the one hand, high-magnitude positive and negative vorticity centres around the interface are considerably diffused to form round regions, making the vorticity distribution more uniform there. One the other hand, the vorticity concentration inside the interdroplet region is also depleted as the coalescence occurs. 

Figure~\ref{case2_xx5} compares the dimensionless density profiles $\rho^*\equiv2\rho/(\rho_{l}+\rho_{g})$ obtained from D2Q25A9H3 and D2Q361A37F in the interfacial region of the droplets, respectively. Compared to the case with $Kn=0.052$, the density structures between the two droplets are significantly changed by the enhanced higher-order non-continuum effects.

\subsubsection{Rarefaction effects on energy conversion and viscous dissipation}
Although the rarefaction effects in the gas film have been noticed and described in several existing studies~\citep{ZhangPeng2011,LiJiePRL2016,Sprittles2015,Sprittles2017}, rarefaction effects on energy conversion and droplet coalescence are not explored yet.
For the present model, the total energy is the sum of the kinetic energy and the free energy.
Therefore, we first present a general derivation for the evolution equations of the free and kinetic energies in order to theoretically reveal the energy conversion mechanisms between these two energies. Then, rarefaction effects on relevant energy conversion terms are discussed using the simulation data.
It is mentioned that the following derivation is general and does not depend on the specific form of the bulk free energy density $\psi$, and hence the equation of state $p_0$.

The free energy density $f_{e}$ is a single-valued function of two variables $\rho$ and $\bm{\nabla}\rho$, namely,
\begin{eqnarray}\label{ep1}
	f_{e}(\rho,\bm{\nabla}\rho)=\psi(\rho)+\frac{1}{2}\kappa\lVert\bm{\nabla}\rho\rVert^2.
\end{eqnarray}
By applying the chain rule for the material derivative of $f_{e}$, we obtain
\begin{eqnarray}\label{ep2}
	\frac{Df_{e}}{Dt}=\frac{\partial f_{e}}{\partial\rho}\frac{D\rho}{Dt}
	+\frac{\partial f_{e}}{\partial\bm{\nabla}\rho}\bm{\cdot}\frac{D\bm{\nabla}\rho}{Dt},
\end{eqnarray}
The partial derivatives of $f_{e}$ with respect to $\rho$ and $\bm{\nabla}\rho$ are $\partial{f_e}/\partial\rho=\mu_0$ and ${\partial f_{e}}/{\partial\bm{\nabla}\rho}=\kappa\bm{\nabla}\rho$, respectively.
Using~\eqref{hy1}, the material derivatives of $\rho$ and $\bm{\nabla}\rho$ are respectively evaluated as
\begin{eqnarray}\label{ep4}
	\frac{D\rho}{Dt}=\frac{\partial\rho}{\partial t}+\bm{u}\bm{\cdot}\bm{\nabla}\rho=-\rho\vartheta,
\end{eqnarray}
and
\begin{eqnarray}\label{ep5}
	\frac{D\bm{\nabla}\rho}{Dt}=\bm{\nabla}\frac{\partial\rho}{\partial t}+\bm{u}\bm{\cdot}\bm{\nabla\nabla}\rho=-\bm{\nabla}(\rho\vartheta)-\bm{\nabla}\rho\bm{\cdot}\bm{\nabla}\bm{u},
\end{eqnarray}
where $\vartheta\equiv\bm{\nabla}\bm{\cdot}\bm{u}$ represents the fluid dilatation.
Substituting~\eqref{ep4} and~\eqref{ep5} into~\eqref{ep2} gives
\begin{eqnarray}\label{eq6}
	\frac{Df_{e}}{Dt}&=&-\rho\mu_{0}\vartheta-\kappa\lVert\bm{\nabla}\rho\rVert^2\vartheta
	-\kappa\rho\bm{\nabla}\rho\bm{\cdot}\bm{\nabla}\vartheta-{\kappa}\bm{\nabla}\rho\bm{\nabla}\rho\bm{:}\bm{S}\nonumber\\
	&=&-\rho\mu_{\rho}\vartheta-\kappa\lVert\bm{\nabla}\rho\rVert^2\vartheta-\kappa\rho\bm{\nabla}\bm{\cdot}\left(\vartheta\bm{\nabla}\rho\right)-{\kappa}\bm{\nabla}\rho\bm{\nabla}\rho\bm{:}\bm{S}\nonumber\\
	&=&-\rho\mu_{\rho}\vartheta-\bm{\nabla}\bm{\cdot}\left(\kappa\rho\vartheta\bm{\nabla}\rho\right)-{\kappa}\bm{\nabla}\rho\bm{\nabla}\rho\bm{:}\bm{S}.
\end{eqnarray}
Therefore, we have the evolution equation for $f_{e}$ as
\begin{eqnarray}\label{eq7}
	\frac{\partial f_{e}}{\partial t}+\bm{u}\bm{\cdot}\bm{\nabla}f_{e}=-\rho\mu_{\rho}\vartheta-\bm{\nabla}\bm{\cdot}\left(\kappa\rho\vartheta\bm{\nabla}\rho\right)-{\kappa}\bm{\nabla}\rho\bm{\nabla}\rho\bm{:}\bm{S}.
\end{eqnarray}
Using the following relation,
\begin{eqnarray}\label{ep8}
	\bm{u}\bm{\cdot}\bm{\nabla}f_{e}=\bm{\nabla}\bm{\cdot}(f_{e}\bm{u})-f_{e}\vartheta,
\end{eqnarray}
\eqref{eq7} can be rewritten as
\begin{eqnarray}\label{ep9}
	\frac{\partial f_{e}}{\partial t}=-(\rho\mu_{\rho}-f_{e})\vartheta-\bm{\nabla}\bm{\cdot}\left(\kappa\rho\vartheta\bm{\nabla}\rho+f_{e}\bm{u}\right)-{\kappa}\bm{\nabla}\rho\bm{\nabla}\rho:\bm{S}.
\end{eqnarray}
Interestingly, using~\eqref{chemical_potential},~\eqref{pth},~\eqref{Pressure_tensor} and~\eqref{ep1}, we have
\begin{eqnarray}\label{ep10}
	\rho\mu_{\rho}-f_{e}&=&\rho\left(\frac{\partial\psi}{\partial\rho}-\kappa\nabla^2\rho\right)-\left(\psi+\frac{1}{2}\kappa\lVert\bm{\nabla}\rho\rVert^2\right)\nonumber\\
	&=&\left(\rho\frac{\partial\psi}{\partial\rho}-\psi\right)-\kappa\rho\nabla^2\rho-\frac{1}{2}\kappa\lVert\bm{\nabla}\rho\rVert^2\nonumber\\
	&=&p_0-\kappa\rho\nabla^2\rho-\frac{1}{2}\kappa\lVert\bm{\nabla}\rho\rVert^2\nonumber\\
	&=&p,
\end{eqnarray}
which indicates that the difference between the chemical potential per unit volume and the free energy density is equal to the nonlocal total pressure, see~\eqref{Pressure_tensor}. Then, combining~\eqref{ep9} and~\eqref{ep10} yields
\begin{eqnarray}\label{ep11}
	\frac{\partial f_e}{\partial t}=-\bm{\nabla}\bm{\cdot}\left(\kappa\rho\vartheta\bm{\nabla}\rho+f_{e}\bm{u}\right)-T_{1}-T_{2},
\end{eqnarray}
where
\begin{eqnarray}\label{T1T2}
	T_{1}\equiv p\vartheta,~
	T_{2}\equiv{\kappa}\bm{\nabla}\rho\bm{\nabla}\rho\bm{:}\bm{S}.
\end{eqnarray}
In the right hand side of~\eqref{ep11}, the divergence term only contributes to the local variation of the free energy density but has no contribution to the energy conversion. $T_{1}$ represents the coupling between the total pressure and the dilatation. For the isosurface of the density field, $T_2$ can be rewritten as
\begin{eqnarray}
T_2=\kappa\lVert\bm{\nabla}\rho\rVert^2\left(\vartheta-\frac{1}{S_0}\frac{dS_0}{dt}\right),
\end{eqnarray}
which represents the contribution to the temporal evolution rate of the free energy density through the dilatational process and the stretching/compression of the small material surface element $S_0$ on the isosurface of the density field. 

Similarly, the evolution equation for the kinetic energy density $\rho u^2/2$ can be derived as 
\begin{eqnarray}\label{ep13}
	\frac{\partial}{\partial t}\left(\frac{1}{2}\rho u^2\right)=-\bm{\nabla}\bm{\cdot}\left(\bm{P}\bm{\cdot}\bm{u}-\bm{\sigma}\bm{\cdot}\bm{u}+\frac{1}{2}\rho u^2\bm{u}\right)+T_{1}+T_{2}-T_{3},
\end{eqnarray}
where
\begin{eqnarray}\label{T3}
	T_{3}\equiv\bm{\sigma}\bm{:}\bm{S}.
\end{eqnarray}

It follows from~\eqref{ep11} and~\eqref{ep13} that
the local energy conversion between the free energy and the kinetic energy can only be realized through two physical mechanisms, namely, the coupling between the total pressure and the dilatation ($T_{1}$), and the interaction between the density gradient and the strain rate tensor ($T_{2}$).
$T_{1}>0$ or $T_{2}>0$ represents the local conversion of the free energy into the kinetic energy. Conversely, the kinetic energy is locally converted to the free energy when $T_{1}<0$ or $T_{2}<0$. 
$T_{3}$ represents the viscous dissipation of the kinetic energy.
We focus on the case with $Kn=0.104$ in the following discussion.
$T_{1}$, $T_2$ and $T_3$ will be normalized by $\rho_{l}U^3/D_l$ for all the plots below.

Of particular interest are the rarefaction effects on local energy conversion and viscous dissipation along the vertical centreline $y^*=4$ where the gas film is squeezed out. Figure~\ref{ptheta} shows the time evolution of $T_1^{*}$ along $y^*=4$.
At $t^*=0.3$ and $t^*=0.4$ (figures~\ref{ptheta1500} and~\ref{ptheta2000}), the overall tendency of the results obtained from D2Q25A9H3 is similar to those from
D2Q361A37F. Compared to D2Q25A9H3, the oscillations in the results of D2Q361A37F are suppressed due to the high-order non-continuum effect, in particular to those around $x^*=8$. Negative $T_1^{*}$ contributes to the local conversion of the kinetic energy to the free energy. At $t^*=0.5$ and $t^*=0.6$ (figures~\ref{ptheta2500} and~\ref{ptheta3000}), $T_1^{*}$ basically shows positive values near $x^*=0.8$, which indicates that the free energy is locally converted to the kinetic energy. Comparison of the results from D2Q25A9H3 and D2Q361A37F shows that this local energy conversion is enhanced by the rarefaction effect.
At $t^*=0.8$ and $t^*=1.0$ (figures~\ref{ptheta4000} and~\ref{ptheta5000}), the liquid bridge is formed and the rarefaction effect mainly concentrates near its surface.

Figure~\ref{KNNS} shows the time evolution of $T_2^{*}$ along $y^*=4$. 
One common feature for the evolution of $T_{1}^{*}$ and $T_2^{*}$ is that the rarefaction effect suppresses the oscillation magnitude of $T_{1}^{*}$ and $T_{2}^{*}$ at $t^*=0.3$ and $t^*=0.4$, as shown in figures~\ref{KNNS1500} and~\ref{KNNS2000}.
At $t^*=0.5$ (figure~\ref{KNNS2500}), a highly negative peak region is observed for D2Q361A37F, which locally contributes to the conversion of the kinetic energy to the free energy. However, the final energy conversion rate between the kinetic energy and the free energy is determined by the competition between $T_{1}^{*}$ and $T_{2}^{*}$.
It is obvious that for the present case, the magnitude of $T_{2}^{*}$ is much lower than $T_{1}^{*}$, which implies the dominance of $T_{1}^{*}$ in the local energy conversion. Subsequently, the differences between the results from D2Q25A9H3 and D2Q361A37F for $T_{2}^{*}$ mainly concentrate near the surface of the liquid bridge, as displayed in figures~\ref{KNNS3000},~\ref{KNNS4000} and~\ref{KNNS5000}.

Furthermore, the time evolution of $T_{3}^{*}$ is displayed in figure~\ref{sigmaS}. It is observed that
$T_{3}^{*}$ basically remains positive, whose sign is not significantly changed by the high-order contribution $\bm{\sigma}^{(H)}\bm{:}\bm{S}$.
At $t^*=0.3$ (figure~\ref{disp1500}), the magnitude of $T_{3}^{*}$ obtained by D2Q361A37F is comparable to that from D2Q25A9H3.
In comparison, no obvious oscillations are observed for D2Q361A37F near the collision point $x^*=8$.
At $t^*=0.4$ (figure~\ref{disp2000}), due to the rarefaction effect, the high magnitude region of $T_{3}^{*}$ is depleted along with a wider spatial distribution. In figures~\ref{disp2500} and~\ref{disp3000}, compared to D2Q25A9H3, higher viscous dissipation rate can be observed for D2Q361A37F near the collision point. This is reasonable since the enhanced pressure-dilatation coupling effect in the same region converts more free energy to kinetic energy, causing the growth of the viscous dissipation rate. The characteristic of the distribution of $T_{3}^{*}$ is similar to those of $T_{1}^{*}$ and $T_{2}^{*}$, which is also concentrated near the surface of the liquid bridge (figures~\ref{disp4000} and~\ref{disp5000}).

\section{Conclusions and discussions}\label{Section5}

In this paper, in order to investigate the rarefaction effects in head-on collision of two identical droplets, two DUGKS simulations with the Knudsen numbers $Kn=0.052$ and $0.104$ are performed based on a BGK-Boltzmann equation. The main findings and contributions are summarised as follows.

(a) We observe the rarefaction effects during the binary droplet collision by using the Gauss-Hermite quadratures with different degree of precision. The convergent solutions are obtained by gradually increasing the number of discrete particle velocities for both the simulated cases. For the case with $Kn=0.052$, D2Q81A17F is sufficient to obtain the satisfying convergent solution. In contrast, D2Q361A37F is needed to capture the higher-order non-continuum effects for the case with $Kn=0.104$.

(b) We analyse the rarefaction effects on the time evolution of the binary droplet collision event. 
The spatial distribution of the vertical velocity (hence the droplet morphology) and the viscous stress components are found to be influenced by the non-continuum effects during the formation of the liquid bridge. 
The topology of streamlines near the droplets' surface is significantly altered due to the rarefaction effect. For example, a saddle-node pair and a relatively large crown structure are observed for $Kn=0.052$, which is very similar to the flow structure and surface configuration previously observed in droplet spreading and splash on a solid surface associated with the rarefaction effects~\citep{Mandre2012,Sprittles2015,Sprittles2017}. In addition, high-magnitude vorticity concentration in the interdroplet region is observed to be suppressed, which is related to the previous finding that  the rarefaction effect can boost the droplet coalescence~\citep{Gopinath1997,LiJiePRL2016}. The rarefaction effect also promotes the vorticity diffusion around the outer droplet surface, making the vorticity distribution more uniform there.
Moreover, it is observed the the spatial structures of the density field between the droplets are significantly changed by the rarefaction effects.

(c) We provide a detailed analysis for the rarefaction effects on energy conversion and viscous dissipation.
For the present model, we mathematically prove that only two physical mechanisms are responsible for the energy conversion between the kinetic energy and the free energy, namely the pressure-dilatation coupling effect $T_{1}\equiv p\vartheta$ and the interaction between the density gradient and the strain rate tensor $T_{2}\equiv\kappa\bm{\nabla}\rho\bm{\nabla}\rho\bm{:}\bm{S}$, where the nonlocal total pressure $p$ is given in~\eqref{Pressure_tensor}.
Therefore, the final energy conversion rate depends on 
the competition between $T_{1}$ and $T_{2}$ in different flow problems.
For the case with $Kn=0.104$, it is found that $T_{1}$ dominates the energy conversion from the free energy to the kinetic energy, which facilitates the discharge of the interdroplet gas film along the vertical direction and boosts the coalescence of two droplets. We show that this characteristic is enhanced by the rarefaction effect, which is accompanied by the enhancement of the viscous dissipation rate ($T_3\equiv\bm{\sigma}\bm{:}\bm{S}$) near the surface of the liquid bridge.

Extended studies could be carried out to explore the physical mechanisms associated with different dimensionless parameters,
which could provide new insights into the rarefaction effects on binary droplet collision dynamics and outcomes.
The present code can be modified to investigate the rarefaction effects in non-continuum physical phenomena including moving contact line, boiling and cavitation.
For example, when the size of the bubble is comparable to the mean free path of the gas, the rarefaction effects are likely to dominate the local dynamic behaviour of the bubble. For these problems, the kinetic equation could provide a more physical framework to incorporate the higher-order rarefaction effects compared to the modified Navier-Stokes equation.

\section*{Acknowledgements }
This work is supported the Guangdong-Hong Kong-Macao Joint Laboratory for Data-Driven Fluid Mechanics and Engineering Applications in China under grant 2020B1212030001.

\section*{Declaration of interests} 
The authors report no conflict of interest.

%\section*{DATA AVAILABILITY} 
%The data that support the findings of this study are available from the corresponding author
%upon reasonable request.

%\section*{Author ORCIDs} 

\appendix

\bibliographystyle{jfm}
\bibliography{Chenref}

%% End of file `jfm2esam.bib'.

\end{document}